\documentclass{LMCS}

\def\dOi{11(3:1)2015}
\lmcsheading%
{\dOi}
{1--71}
{}
{}
{Jan.~13, 2013}
{Jul.~28, 2015}
{}

\keywords{Computability logic; Interactive computation; Implicit computational complexity;  Game semantics;  Constructive logics; Efficiency  logics}

\ACMCCS{[{\bf Theory of computation}]: Computational complexity and cryptography---Complexity classes; [{\bf Theory of computation}]: Logic---Constructive mathematics}
\amsclass{primary: 03F50; secondary:  03D75; 03D15; 03D20; 68Q10; 68T27; 68T30}

\newcommand{\cltw}{\mbox{\bf CL12}}
 \newcommand{\blank}{{\scriptsize {\sc Blank}}}
 \newcommand{\transition}{{\sc Update Sketch}}
 \newcommand{\call}{{\sc Fetch Symbol}}
 \newcommand{\main}{{\sc Make History}}

\newcommand{\variables}{\mbox{\sc Variables}}
\newcommand{\constants}{\mbox{\sc Constants}}
\newcommand{\valuations}{\mbox{\sc Valuations}}
\newcommand{\st}{\mbox{\raisebox{-0.05cm}{$\circ$}\hspace{-0.13cm}\raisebox{0.16cm}{\tiny $\mid$}\hspace{2pt}}}  
\newcommand{\sti}{\mbox{\raisebox{-0.02cm}
{\scriptsize $\circ$}\hspace{-0.121cm}\raisebox{0.08cm}{\tiny $.$}\hspace{-0.079cm}\raisebox{0.10cm}
{\tiny $.$}\hspace{-0.079cm}\raisebox{0.12cm}{\tiny $.$}\hspace{-0.085cm}\raisebox{0.14cm}
{\tiny $.$}\hspace{-0.079cm}\raisebox{0.16cm}{\tiny $.$}\hspace{1pt}}}
\newcommand{\intimpl}{\mbox{\hspace{2pt}$\circ$\hspace{-0.14cm} \raisebox{-0.043cm}{\Large --}\hspace{2pt}}}

\newcommand{\adai}{\mbox{$\sqcap$}}      
\newcommand{\successor}{\mbox{\hspace{1pt}\boldmath $'$}}
\newcommand{\plus}{\mbox{\hspace{1pt}\raisebox{0.05cm}{\tiny\boldmath $+$}\hspace{1pt}}}
\newcommand{\mult}{\mbox{\hspace{1pt}\raisebox{0.05cm}{\tiny\boldmath $\times$}\hspace{1pt}}}
\newcommand{\equals}{\mbox{\hspace{1pt}\raisebox{0.05cm}{\tiny\boldmath $=$}\hspace{1pt}}}
\newcommand{\notequals}{\mbox{\hspace{1pt}\raisebox{0.05cm}{\tiny\boldmath $\not=$}\hspace{1pt}}}
\newcommand{\elz}[1]{\mbox{$\parallel\hspace{-3pt} #1 \hspace{-3pt}\parallel$}} 
\newcommand{\elzi}[1]{\mbox{\scriptsize $\parallel\hspace{-3pt} #1 \hspace{-3pt}\parallel$}}
\newcommand{\emptyrun}{\langle\rangle} 
\newcommand{\oo}{\bot}            
\newcommand{\pp}{\top}            
\newcommand{\xx}{\wp}               
\newcommand{\legal}[2]{\mbox{\bf Lr}^{#1}_{#2}} 
\newcommand{\win}[2]{\mbox{\bf Wn}^{#1}_{#2}} 
\newcommand{\seq}[1]{\langle #1 \rangle}           
\newcommand{\pst}{\mbox{\raisebox{-0.01cm}{\scriptsize $\wedge$}\hspace{-4pt}\raisebox{0.16cm}{\tiny $\mid$}\hspace{2pt}}}
\newcommand{\gneg}{\mbox{\small $\neg$}}                 
\newcommand{\mli}{\hspace{2pt}\mbox{\small $\rightarrow$}\hspace{2pt}}                    
\newcommand{\cla}{\mbox{$\forall$}}      
\newcommand{\cle}{\mbox{$\exists$}}        
\newcommand{\mld}{\hspace{2pt}\mbox{\small $\vee$}\hspace{2pt}}     
\newcommand{\mlc}{\hspace{2pt}\mbox{\small $\wedge$}\hspace{2pt}}   
\newcommand{\mlci}{\hspace{2pt}\mbox{\footnotesize $\wedge$}\hspace{2pt}}   
\newcommand{\ade}{\mbox{\large $\sqcup$}}      
\newcommand{\ada}{\mbox{\large $\sqcap$}}      
\newcommand{\add}{\hspace{2pt}\mbox{\small $\sqcup$}\hspace{2pt}}                 
\newcommand{\adc}{\hspace{2pt}\mbox{\small $\sqcap$}\hspace{2pt}} 
\newcommand{\adci}{\hspace{2pt}\mbox{\footnotesize $\sqcap$}\hspace{2pt}}           
\newcommand{\clai}{\forall}    
      
\newcommand{\tlg}{\bot}             
\newcommand{\twg}{\top}              
\newcommand{\col}[1]{\mbox{$#1$:}} 
\theoremstyle{italic}\newtheorem{thesis}[thm]{Thesis}
\theoremstyle{plain}\newtheorem{exercise}[thm]{Exercise}
\newenvironment{idea}{ {\noindent\em Proof idea.} }{\  \rule{1.5mm}{1.5mm} \vspace{.15in} }
\usepackage{amsfonts} 
\usepackage{hyperref}

\begin{document}

\title[On the system CL12 of computability logic]{On the system CL12 of computability logic}

\author[G.~Japaridze]{Giorgi Japaridze}
\address{Department of Computing Sciences, Villanova University, 800 Lancaster Avenue, Villanova, PA 19085, USA}	%required
\urladdr{http://www.csc.villanova.edu/$\sim$japaridz/}  
\email{giorgi.japaridze@villanova.edu}  
%\thanks{thanks 1, optional.}	%optional

%% etc.

%% required for running head on odd and even pages, use suitable
%% abbreviations in case of long titles and many authors:

%%%%%%%%%%%%%%%%%%%%%%%%%%%%%%%%%%%%%%%%%%%%%%%%%%%%%%%%%%%%%%%%%%%%%%%%%%%

%% the abstract has to PRECEDE the command \maketitle:
%% be sure not to issue the \maketitle command twice!

\begin{abstract}
  \noindent {\em Computability logic} (CoL) is  a  long-term project  for  redeveloping logic on the basis of a constructive game semantics, with games seen as abstract models of interactive computational problems. 
Among the  fragments of CoL successfully axiomatized so far is {\bf CL12} --- a conservative extension of classical first-order logic, whose language augments that of classical logic with the so called {\em choice} (``constructive'') sorts of quantifiers and connectives.  This system has already found fruitful applications as a logical basis for constructive and complexity-bound versions of Peano arithmetic, such as arithmetics for polynomial time computability, polynomial space computability, and beyond. 
The present paper introduces a third, indispensable complexity measure for interactive computations termed {\em amplitude complexity}, and establishes the adequacy (soundness/completeness) of {\bf CL12} and the associated Logical Consequence mechanism with respect to (simultaneously) $A$ amplitude, $S$ space and $T$ time computability  under certain minimal conditions on the  triples  $(A,S,T)$ of function classes.  This  result very substantially broadens the potential application areas of {\bf CL12}, even when time and/or space complexity is the only concern. It would be sufficient to point out that, for instance, now {\bf CL12} can be reliably used as a logical basis of systems for logarithmic space or exponential time computabilities --- something that the earlier-known  crude adequacy results for {\bf CL12} were too weak to allow us to do. This paper is self-contained, and targets readers with no prior familiarity with the subject. 
\end{abstract}

\maketitle

\tableofcontents

\section{Introduction}\label{intr}
%\marginpar{intr}

{\bf Computability logic} ({\bf CoL}),\label{0col} introduced in \cite{Jap03,Japic,Japfin}, is a semantically conceived open-ended framework and long-term research project for redeveloping logic as a formal theory of {\em computability}. That is as opposed to the more traditional view of logic as a formal theory of {\em truth}.  
The main pursuit of the  project is to provide ever more expressive and powerful formal tools for systematically telling what can be computed and how,  just like classical logic is a systematic tool for finding what is true. 

Computational problems in CoL are understood in the most general, {\em interactive} sense. Interactive computational problems, in turn, are defined as  {\em games}\label{0game3} played by a {\em machine}\label{0machine3} ($\pp$) against its {\em environment}\label{0environment3} ($\oo$), with computability meaning 
existence of a machine, called a {\em solution},\label{0solution3} that always wins the game. As the name ``machine'' suggests, $\pp$ is a player that always follows determined, algorithmic strategies. On the other hand, there are no restrictions on the possible strategies of $\oo$, which represents a capricious user or the blind forces of nature. Classical propositions and predicates are seen as special sorts of games, called {\em elementary}.\label{0elgame3} These are moveless games automatically won by $\pp$ (and hence lost by $\oo$) when true, and automatically lost by $\pp$ (and hence  won by $\oo$) when false. The approach induces a rich and still-expanding collection of logical operators, standing for various basic operations on games. Those include all operators of classical logic, conservatively generalized from elementary games to all games. This makes CoL a conservative extension of classical first-order logic. As such, the former is dramatically more expressive than the latter, containing a whole zoo of non-classical operators, some reminiscent of those of intuitionistic or linear logic,  and some having no close relatives in the earlier literature. In a series of recent publications \cite{BauerLMCS,Cirq,Japdeep,lmcs,taming1,taming2,XuIGPL,XuIf}, the expressive power of CoL has been further lifted to a qualitatively new level through generalizing formulas to so called {\em cirquents}\label{0cirquent} --- graph-style constructs allowing to explicitly account for {\em sharing} subcomponents between different components.

Since the formalism of CoL is inordinately expressive and, in fact, open-ended, attempts to axiomatize this semantically conceived logic can be reasonably expected to succeed only when focused on various limited yet interesting fragments rather than the whole logic. Recent years have seen rapid and sustained progress in this direction  (\cite{BauerTOCL},\cite{Japtocl1}-\cite{Cirq},\cite{Japtcs}-\cite{Japfour},\cite{Japtowards,Japtoggling,taming1,taming2,Ver,XuIGPL}), and this trend is likely to continue in the near future. 
 
Among the fragments of CoL successfully axiomatized so far is {\bf CL12},\label{0cl12a} to which the present paper is exclusively devoted. This is  a sequent calculus system. Every sequent\label{0sequent} in it looks like 
\[E_1,\ldots,E_n\intimpl F,\]
where $E_1,\ldots, E_n$ ($n\geq 0$) and $F$ are formulas. The language in which formulas are written is that of classical first-order logic with equality, function symbols and constants, augmented  with the so called {\em choice} (``constructive'') operators. Namely, formulas are built from atoms in the standard way using the propositional connectives $\gneg,\mlc,\mld,\adc,\add$ and quantifiers $\cla,\cle,\ada,\ade$. Below we give a brief intuitive characterization of  these logical operators of $\cltw$ as operations on games. It should be noted that, apparently for the exception of $\cla$ and $\cle$, the basic ideas of those game  operations, in one form or another, had surfaced well before CoL was officially introduced, in studies of game semantics by various authors, such as Lorenzen \cite{Lor61}, Hintikka \cite{Hintikka73}, Blass \cite{Bla92}, Japaridze \cite{JapLL,JapND,Jap02} and others. Connections with Girard's   \cite{Gir87} linear logic should also be easily noticeable. Since this is not a survey paper, we refer the reader to \cite{Jap03} for discussions of similarities and differences between the above-mentioned approaches and that of ours. It should also be noted that the overall logical vocabulary of CoL is much wider than the above, including operators for various additional natural operations on games, such as sequential operators (conjunction, disjunction, quantifiers and recurrences) studied in \cite{Japseq,lmcs}, toggling operators (conjunction, disjunction, quantifiers and recurrences) studied in \cite{Japtoggling,lmcs,qu}, parallel quantifiers and recurrences (\cite{Japfour,Japfin,lmcs,separating,Ver,Xure2}), various flavors of branching recurrences 
(\cite{Japfour,Japfin,Japtoggling,separating,face,qu,Ver,Xure1}), and more. Those additional operators will not be discussed here as they are not relevant to what the present paper is focused on.

{\em Negation} $\gneg$\label{0gneg1} can be characterized as a role switch operation: the game $\gneg A$ is the same from the point of view of a given player as what $A$ is from the point of view of the other player. That is, the machine's  moves and wins become those of the environment, and vice versa. For instance, if {\em Chess} is the game of chess\footnote{Modified so that ties are ruled out.} as seen by the white player, then $\gneg${\em Chess} is the same game as seen by the black player. 

Next, $A\mlc B$ and $A\mld B$,\label{0mlc1}\label{0mld1} called {\em parallel conjunction} and {\em parallel disjunction}, respectively, are games playing which means playing both $A$ and $B$ simultaneously. In $A\mlc B$, the machine is considered to be the winner if it wins in both components, while in $A\mld B$ winning in just one component is sufficient. In contrast, {\em choice conjunction } $A\adc B$\label{0adc1}\label{0add1} (resp. {\em choice disjunction} $A\add B$) is a game where the environment (resp. machine) has to choose, at the very beginning, one of the two components, after which the play continues according to the rules of the chosen component, with the failure to make an initial choice resulting in a loss for the corresponding player. To appreciate the difference, compare $\gneg\mbox{\em Chess}\mld  \mbox{\em Chess}$ \ with\  $\gneg\mbox{\em Chess}\hspace{1pt}\add \mbox{\em Chess}$. The former is a two-board game, where the machine plays black on the left board and white on the right board. It is very easily won by the machine by just mimicking on either board the moves made by its adversary on the other board. On the other hand, $\gneg\mbox{\em Chess}\hspace{1pt}\add \mbox{\em Chess}$ is not at all easy to win. Here the machine has to choose between playing black or white, after which the game continues as the chosen one-board game.  Generally, the principle $\gneg P\mld P$ is valid in CoL (in the sense of being ``always winnable'' by a machine) while $\gneg P\add P$ is not.

 With the set of (canonical) {\em constants} throughout CoL being the set $\{0,1,10,11,100,\ldots\}$ of binary numerals for natural numbers,  the {\em choice universal quantification} $\ada xA(x)$\label{0ada} can now be defined as the infinite choice conjunction $A(0)\adc A(1)\adc A(10)\adc A(11)\adc A(100)\adc \ldots$, and the {\em choice existential quantification} $\ade xA(x)$\label{0ade}  defined as the infinite choice disjunction $A(0)\add A(1)\add A(10)\add A(11)\add A(100)\add\ldots$. So, for instance, where $f$ is a unary function, $\ada x\ade y (y=f(x))$ is a game in which the first move is by the environment, consisting in choosing a particular constant $m$ for $x$. Such a move, which intuitively can be seen as the question ``What is the value of $f$ at $m$?'' by the environment, brings the game down to $\ade y(y=f(m))$. Now, in this game/position, the machine is obligated to make a move (otherwise it loses). Such a move should be choosing a constant $n$ for $y$, which further brings the game down to $n=f(m)$. The latter is an elementary game with no further moves, won by the machine if $n=f(m)$, i.e. if it correctly answered the question asked by the environment. From this explanation one can see that $\ada x\ade y(y=f(x))$, in fact, represents the problem of computing $f$, with  the machine having an (algorithmic) winning strategy for  $\ada x\ade y(y=f(x))$ iff $f$ is computable in the standard sense. Similarly, where $p$ is a unary predicate, $\ada x(p(x)\add\gneg p(x))$ is (represents) the problem of deciding $p$. 

Next, in the {\em blind universal quantification} $\cla xA(x)$\label{0cla} (resp. {\em blind existential quantification} $\cle xA(x)$),\label{0cle} no value for $x$ is specified/chosen by either player. In order to win, the machine needs to play $A(x)$ ``blindly'' in a way that guarantees a win for every (resp. at least one) possible value of $x$ from the universe of discourse. To compare the blind sorts of quantifiers with their choice counterparts, note that $\ada x(\mbox{\em Even}(x)\add\mbox{\em Odd}(x))$ is a game easily won by the machine, while, on the other hand, $\cla x(\mbox{\em Even}(x)\add\mbox{\em Odd}(x))$ is impossible to win: this is a one-move-deep game where only the machine has a move; such a move should consist in choosing one of the two $\add$-disjuncts 
$\mbox{\em Even}(x)$ or $\mbox{\em Odd}(x)$;  in order to win, the chosen disjunct should be true for every possible value of $x$, which is an unsatisfiable condition as long as the universe contains both even and odd numbers.   

Finally, $A_1,\ldots,A_n\intimpl B$\label{0intimpl00} can be characterized as the problem of {\em reducing}  $B$ to $A_1,\ldots,A_n$.
Several reduction\label{0reduction} operations emerge naturally within the framework of CoL (cf. \cite{Japjsl,Japfour,Japfin,Japtowards}), including $\mli$\label{0mli1} defined by $A\mli B=\gneg A\mld B$. Among those, $\intimpl$ stands out as the weakest, most general sort of algorithmic reduction. A play of  $A_1,\ldots,A_n\intimpl B$ proceeds, in a parallel fashion, in all of its components. However, in the antecedental components $A_1,\ldots,A_n$, the roles of the two players are switched. That is, from the machine's perspective, they are $\gneg A_1,\ldots,\gneg A_n$ rather than $A_1,\ldots,A_n$.  The machine is considered the winner if it wins in $B$ as long as its adversary wins in each of the components of the antecedent. This game is thus similar to 
$\gneg A_1\mld \ldots\mld \gneg A_n\mld B$, i.e. $A_1\mlc\ldots\mlc A_n\mli B$ (DeMorgan's laws continue to hold in CoL). There is, however, a crucial difference between $A_1\mlc\ldots\mlc A_n\mli B$ and $A_1,\ldots,A_n\intimpl B$. Namely, in the former, each of the antecedental games can be played only once, while, in the latter, they can be played and replayed any number of times at the machine's discretion. Furthermore, at any time, the machine is allowed to split/fork any already reached position of any $A_i$ and thus create several threads continuing from that position (rather than restart $A_i$ from the very beginning). This way, from the machine's perspective, $A_1,\ldots,A_n$ are computational {\em resources}\label{0resource1} that can be used and reused, in the strongest algorithmic sense possible, in the process of playing/solving $B$. In other words, for the machine, solving $A_1,\ldots,A_n\intimpl B$ means solving $B$ while the environment providing (interactive) oracles for $A_1,\ldots,A_n$. It is therefore no surprise that $\intimpl$ turns out to be a conservative generalization of the well-known concept of Turing reduction from traditional, input-output sorts of problems to all interactive problems. Namely, when $A_1,\ldots,A_n,B$ are ``traditional'' kinds of problems such as computing a function or deciding a predicate, $\pp$ has a winning strategy (algorithmic solution) for $A_1,\ldots,A_n\intimpl B$ if and only if $B$ is Turing reducible to $A_1,\ldots,A_n$. 

As promised earlier, the semantical meanings of $\gneg,\mlc,\mld,\cla,\cle$ (and hence $\mli$ as well) are exactly classical when these operations are applied to elementary (moveless) games. For instance, when $A$ and $B$ are elementary, then so is $A\mlc B$, which is automatically won by the machine, i.e. true, iff so are both $A$ and $B$. Furthermore, when all games $A_1,\ldots,A_n,B$ are elementary, $A_1,\ldots,A_n\intimpl B$ can be seen to be equivalent to $A_1\mlc\ldots\mlc A_n\mli B$. It is this fact that eventually makes $\cltw$ a conservative extension of classical first-order logic.

The system $\cltw$ was proven in \cite{lbcs} to be sound and complete in the  sense that a sequent is $\cltw$-provable  if and only if it has a uniform (``purely logical'') solution, i.e. algorithmic strategy of $\pp$ that wins the game/sequent under any interpretation of its nonlogical components such as predicate and function letters. Furthermore, such a strategy can be effectively extracted from a proof of the sequent. Both soundness and completeness, in fact, were  shown in \cite{lbcs} to hold in a significantly stronger sense. Namely, for the completeness part, it was shown that if a sequent $S$ is not provable in $\cltw$, then there is simply no strategy --- whether algorithmic or non-algorithmic --- that wins $S$ under every interpretation. As for the soundness part, it was shown that the strategies extracted from $\cltw$-proofs run in polynomial time and polynomial space. 
What the time and space complexity concepts exactly mean in the context of interactive problems represented by the formulas and sequents of $\cltw$ will be seen later in Section \ref{s7}. For now, it would be sufficient to note that those are natural conservative generalizations of the usual complexity-theoretic concepts.  

While the above adequacy theorem of \cite{lbcs} establishes the completeness of $\cltw$ in an extreme --- strongest possible --- sense,  
the soundness part, as it turns out, can be significantly sharpened. Among the results of the present paper (Theorem \ref{feb9c}) is showing that the strategies extracted from $\cltw$-proofs, in fact, run in linear time and constant space, essentially meaning that such strategies are as efficient as they could possibly be. Also, a conceptual novelty of the present contribution is introducing  (Definition \ref{deftcs}) a third kind of a complexity measure for interactive computations, termed {\em amplitude complexity}.\label{0ampl3} The latter is concerned with the sizes of $\pp$'s moves relative the sizes of the moves made by its adversary. As it happens, in terms of amplitude complexity, $\cltw$ is again as efficient as it could possibly be. Namely, strategies extracted from $\cltw$-proofs run in identity (non-size-increasing) amplitude. Amplitude complexity proves itself to be an indispensable measure when it comes to interactive computation, interesting not only in its own right but also as a means for analyzing time and space complexities at a much finer level than previously possible. 

$\cltw$ induces a rule of inference that we call {\em Logical Consequence}.\label{0logcon3} The latter allows us to jump to conclusion $F$ from premises $E_1,\ldots,E_n$ whenever $\cltw$ proves the sequent $E_1,\ldots,E_n\intimpl F$. The adequacy of $\cltw$ extends to the adequacy of this rule. Namely,  a formula $F$ is a logical consequence of formulas $E_1,\ldots,E_n$ (i.e., the former follows from the latter by Logical Consequence) if and only if a solution  for $F$ can be extracted from solutions for $E_1,\ldots,E_n$ in a purely logical, i.e. interpretation-independent, way. Furthermore, as shown in \cite{lbcs}, whenever $\Omega$ is a class of functions containing all polynomial functions and closed under composition, Logical Consequence preserves both $\Omega$ time and $\Omega$ space computabilities. So, for instance, if $F$ is a logical consequence of $E_1,\ldots,E_n$ and, under a given interpretation of nonlogical symbols, each $E_i$ has a polynomial time solution, then so does $F$, and such a solution for $F$ can be extracted from the solutions for $E_1,\ldots,E_n$. The same holds for space instead of time, as well as for any bigger classes of functions closed under composition, such as the classes of elementary or  primitive recursive functions. 

The adequacy result established in \cite{lbcs} opened a whole new area of applications for $\cltw$: this logic can be used as an appealing alternative to classical or intuitionistic logics as a logical basis for complexity-oriented applied formal theories, such as Peano arithmetic. 
 The papers \cite{Japtowards,cla4,cla8,cla5} constructed the series {\bf CLA1}-{\bf CLA10}\label{0clar}  of CoL-based --- more specifically, $\cltw$-based --- versions of arithmetic, generically named ``{\em clarithmetics}''.\label{0clarithmetic} Below we briefly discuss four of those: {\bf CLA4}, {\bf CLA5}, {\bf CLA6} and {\bf CLA7}. 

All of the above clarithmetical theories have the same language --- the language of classical first-order Peano arithmetic augmented with the choice operators $\adc,\add,\ada,\ade$. They are $\cltw$-based in the sense that the sole logical rule of inference\footnote{This includes logical axioms as special cases of logical rules of inference.}  of all those systems is  Logical Consequence. The set of nonlogical axioms of all those systems consists of the ordinary Peano axioms (including the induction axiom scheme restricted to the ordinary, $\adc,\add,\ada,\ade$-free formulas) plus the single extra-Peano axiom $\ada x\ade y(y=x\plus 1)$, expressing the computability of the successor function. The only exception is {\bf CLA4}, which has $\ada x\ade y(y=2x)$ as an additional extra-Peano axiom (this axiom is derivable/redundant in all other systems). Finally, the only nonlogical rule of inference of all systems is (constructive) induction. It is exactly the induction rule where the systems differ from each other. 

The induction rule of {\bf CLA4} is 
\[\frac{F(0)\hspace{20pt}\ada x(F(x)\mli F(2x))\hspace{20pt}\ada x(F(x)\mli F(2x\plus 1))}{\ada xF(x)},\]
where $F(x)$ is any {\em polynomially bounded} formula, meaning a formula where every subformula $\ada y E(y)$ looks like $\ada y (|y|\leq t\mli G(y))$ and every subformula   $\ade y E(y)$ looks like $\ade y (|y|\leq t\mlc G(y))$, where $t$ is any $0,\successor,\plus,\mult$-combination ($\successor$ stands for the successor function, i.e., $a\successor$ is interpreted as $a\plus 1$) of $|z_1|,\ldots,|z_n|$, where $z_1,\ldots,z_n$ are any variables different from $y$ and not bound by $\cla$ or $\cle$ within $F(x)$,\footnote{The condition ``not bound by $\cla$ or $\cle$ within $F(x)$'' is missing on page 1330 of \cite{cla4}, which is a technical error.}  and $|u|$\label{0|1} is a (pseudo)term for the function ``the size of the binary representation of $x$'' (an integer approximation of the base $2$ logarithm of $u$). This is, in fact, nothing but an adaptation of Buss's \cite{Buss} PIND axiom to the environment in which {\bf CLA4} operates. Note that, however, the boundedness requirement applies only to the choice quantifiers $\ada,\ade$ and not the blind quantifiers $\cla,\cle$. This clear, natural and (as expected) very beneficial separation of constructive and non-constructive operators was  metaphorically characterized in \cite{cla4} as ``giving Caesar what belongs to Caesar and God what belongs to God''. As shown in \cite{cla4}, {\bf CLA4} is sound and complete with respect to polynomial time computability. Sound in the sense that there is an effective procedure that takes any {\bf CLA4}-proof of any formula $F$ and constructs a polynomial time solution for the problem represented by $F$ under the standard arithmetical meanings of its nonlogical symbols $0,\successor,\plus,\mult$. And complete in the sense that any arithmetical problem (meaning the  problem represented by some formula of the language of {\bf CLA4}) with a polynomial time solution is represented by some theorem of {\bf CLA4}.  

The induction rule of {\bf CLA5} is 
\begin{equation}\label{ei}
\frac{F(0)\hspace{20pt}\ada x(F(x)\mli F(x\plus 1))}{\ada xF(x)},\end{equation}
where $F(x)$, again, is any polynomially bounded formula. As proven in \cite{cla5}, {\bf CLA5} is sound and complete in the same sense as {\bf CLA4}, but with respect to polynomial space (rather than polynomial time) computability. 

Next, the induction rule of {\bf CLA6} is the same (\ref{ei}), but with the weaker requirement that $F(x)$ be an exponentially (rather than polynomially) bounded formula. Here ``{\em exponentially bounded}\hspace{1pt}'' means that every subformula $\ada y E(y)$ of $F(x)$ looks like $\ada y (|y|\leq t\mli G(y))$ and every subformula   $\ade y E(y)$ looks like $\ade y (|y|\leq t\mlc G(y))$, where $t$ is any $0,\successor,\plus,\mult$-combination of $z_1,\ldots,z_n$ (rather than $|z_1|,\ldots,|z_n|$ as earlier), where $z_1,\ldots,z_n$ are any variables different from $y$ and not bound by $\cla$ or $\cle$ within $F(x)$. It was proven in \cite{cla5} that {\bf CLA6} is sound and complete in the same sense as {\bf CLA4}, but with respect to elementary recursive time (which can be seen to coincide with elementary recursive space) computability. 

Finally, the induction rule of {\bf CLA7} is also the same (\ref{ei}), but with no restrictions  on $F(x)$ whatsoever. This system was shown in \cite{cla5} to be sound and complete in the same sense as {\bf CLA4}, but with respect to primitive recursive time (= space) computability. 

Efficiency-oriented  systems of clarithmetic, such as {\bf CLA4}, can be seen as  programming languages where ``programming'' simply means theorem-proving. The soundness of the underlying system guarantees that any proof that can be written will be translatable into a program that runs efficiently and indeed is a solution of the problem expressed by the target formula of the proof. Note that the problem of verifying whether a program meets its specification, which is generally undecidable, is fully neutralized here: the ``specification'' is nothing but the target formula of the proof, and the  proof itself, while  encoding an efficient program, also automatically serves as a verification of the correctness of that program. Furthermore,  every step/formula of the proof can be viewed as its own (best possible)  ``comment''. In a more ambitious and, at this point, somewhat fictional perspective, after developing reasonable theorem-provers, CoL-based  efficiency-oriented systems can be seen as declarative programming languages in an extreme sense, where human ``programming'' just means writing a formula expressing the problem whose efficient solution is sought for systematic usage in the future. That is, a program simply coincides with its specification. The compiler's job would be finding a proof (the hard part) and translating it into a machine-language code (the easy part). The process of compiling could thus take long but, once compiled, the program would run fast ever after. 

Beginning from the first study of (the system $I\Delta_0$ of) {\em bounded arithmetic} by Parikh \cite{Parikh} in 1971, 
%and, especially, since  Buss's  \cite{Buss}  seminal contribution  to and systematic advancement of this area in 1986, 
numerous complexity-oriented systems  have been studied in the literature. 
 A notable advantage of CoL-based  systems over the other systems with similar aspirations, which typically happen to be inherently weak systems,  is having actually or potentially unlimited strength, with the latter including the   full  expressive and deductive power of 
classical logic and Peano arithmetic. In view of the above-outlined potential applications, the importance of this feature is obvious: the stronger a system, the better the chances that a proof/program will be found for a declarative, non-preprocessed, ad hoc specification of the goal. Among the other appealing features of clarithmetics is  being semantically meaningful in the full generality of their languages, scalable, and easy to understand in 
their own rights. Syntactically they also tend to be remarkably simple, as we had a chance to see from the above description of {\bf CLA4}--{\bf CLA7}. To achieve adequacy, alternative approaches  often need to do a serious amount of pushing and shoving, including extending the language of arithmetic through symbols for functions expressible in the kind old Peano arithmetic but no longer adequately expressible after  the latter has been ``tampered with'', and introducing a few tens of new axioms to compensate for the loss of deductive power when switching from Peano arithmetic to weaker versions of it. Needless to say  that the alternative systems typically understand and deal with computational problems in the narrow sense of functions,  while clarithmetics deal with a very general class of interactive computational problems.

Having said the above, the most potentially consequential result of the present paper (Theorems \ref{feb9dt} and \ref{feb9ds}) is sharpening  the earlier-mentioned preservation theorem 
for Logical Consequence. As we remember, the  preservation theorem of \cite{lbcs} states that, as long as $\Omega$ is a class 
of functions closed under composition and containing all polynomial functions, Logical Consequence preserves $\Omega$ time and $\Omega$ space computabilities. With these conditions, the extent of applicability of $\cltw$ (of Logical Consequence, that is) as a basis for complexity-oriented systems 
was far from being fully revealed. For instance, the old form of the preservation theorem did not guarantee the soundness of $\cltw$-based theories for sublinear (logarithmic, polylogarithmic etc.)
 space computabilities, as the class of sublinear functions does not contain all polynomial functions; or, it did not guarantee the soundness of 
$\cltw$-based theories for exponential time (or space) complexity, because the class of exponential functions is not closed under composition; 
etc. The present paper strengthens the above preservation theorem 
 by removing the ``closed under composition'' and ``containing all polynomial functions'' requirements on $\Omega$, replacing them with much weaker and finer conditions, bringing into play amplitude complexity along with time and space complexities (as it turns out, the latter can very substantially depend on the former, so that no advanced studies of interactive complexity can avoid explicitly dealing with amplitude complexity). This new adequacy result for Logical 
Consequence dramatically broadens the applicability of $\cltw$ as a basis for complexity-oriented applied theories.

This paper is self-contained and does not assume any prior familiarity with CoL.  To that end, Sections \ref{cg}, \ref{nncg}, \ref{icp} (in part), \ref{ss6} and  \ref{ss8} merely serve the purpose of reintroducing the relevant parts of CoL, and can be more or less safely omitted by readers already well familiar with the subject.\footnote{It should be acknowledged that over 90\% of the contents of those sections are borrowed from some earlier papers on CoL, mostly \cite{lbcs}. Already about 30 articles have been written in the course of developing CoL, and producing 30 original introductions and explanations of the same concepts --- or trying to make them {\em look} original --- is neither technically nor ethically feasible.}

\section{Constant games and ``propositional'' operations}\label{cg}
%\marginpar{cg}

In this and the subsequent few sections we present
 definitions of the basic relevant concepts. A reader aspiring to get additional insights is recommended to consult the first 10 sections of \cite{Japfin}, which provide a tutorial-style introduction to the subject. It should  be however noted that the definitions of  game operations given in the present section are different from --- yet equivalent to --- the ``canonical'' definitions of the same operations found in \cite{Japfin}.  The same applies to some later definitions as well. 

As we already know, computational problems in CoL are understood as games between two players: {\bf Machine},\label{0machine} symbolically $\pp$,\label{0pp} and {\bf Environment},\label{0environment} symbolically $\oo$.\label{0oo} These names will not always be capitalized, and they may take articles. 
A  {\bf move}\label{0move} means any finite string over the standard keyboard alphabet. 
A {\bf labeled move} ({\bf labmove})\label{0labmove} is a move prefixed with $\pp$ or $\oo$, with such a prefix ({\bf label})\label{0label} indicating which player has made the move. 
A {\bf run}\label{0run} is a (finite or infinite) sequence of labmoves, and a {\bf position}\label{0position} is a finite run.

We will be using the letters $\Phi,\Gamma,\Delta$  for runs, and  $\alpha,\beta,\gamma,\delta$ for moves. The letter $\xx$\label{0xx} will always be a variable for players, and \[\gneg{\xx}\label{0pneg}\]  will mean ``$\xx$'s adversary'' (``the other player'').
Runs will be often delimited by ``$\langle$" and ``$\rangle$", with $\emptyrun$ thus denoting the {\bf empty run}.\label{0emptyrun} The meaning of an expression such as $\seq{\Phi,\xx\alpha,\Gamma}$ must be clear: this is the result of appending to the position $\seq{\Phi}$ 
the labmove $\seq{\xx\alpha}$ and then the run $\seq{\Gamma}$.

The following is a formal definition of  constant games, combined with some less formal conventions regarding the usage of certain terminology.

\begin{defi}\label{game}
%\marginpar{game}
 A {\bf constant game}\label{0constantgame} is a pair $A= (\legal{A}{},\win{A}{})$, where:

1. $\legal{A}{}$\label{0lr} is a set of runs  satisfying the condition that a (finite or infinite) run is in $\legal{A}{}$ iff all of its nonempty finite  initial
segments are in $\legal{A}{}$.\footnote{This condition implies that the empty run $\emptyrun$, having no nonempty initial segments, is always among the elements of  $\legal{A}{}$.} The elements of $\legal{A}{}$ are
said to be {\bf legal runs}\label{0legrun} of $A$, and all other runs are said to be {\bf illegal}.\label{0illegrun} We say that $\alpha$ is a {\bf legal move}\label{0legmove} for $\xx$ in a position $\Phi$ of $A$ iff $\seq{\Phi,\xx\alpha}\in\legal{A}{}$; otherwise 
$\alpha$ is {\bf illegal}.\label{0illegmove} When the last move of the shortest illegal initial segment of $\Gamma$  is $\xx$-labeled, we say that $\Gamma$ is a {\bf $\xx$-illegal}\label{0pillegal} run of $A$. 

2. $\win{A}{}$\label{0wn}  is a function that sends every run $\Gamma$ to one of the players $\pp$ or $\oo$, satisfying the condition that if $\Gamma$ is a $\xx$-illegal run of $A$, then $\win{A}{}\seq{\Gamma}= \gneg{\xx}$. When $\win{A}{}\seq{\Gamma}= \xx$, we say that $\Gamma$ is a {\bf $\xx$-won}\label{0wonrun} (or {\bf won by $\xx$}) run of $A$; otherwise $\Gamma$ is {\bf lost}\label{0lostrun} by $\xx$. Thus, an illegal run is always lost by the player who has made the first illegal move in it.  
\end{defi}

A constant game $A$ is said to be {\bf elementary}\label{0elemgamea} iff $\legal{A}{}=\{\emptyrun\}$, i.e., $A$ does not have any nonempty legal runs. There are exactly two elementary constant games, for which we use the same symbols as for the two players. One is  $\twg$\label{0twg} with $\win{\twg}{}\emptyrun=\pp$, and the other is $\tlg$\label{0tlg} with $\win{\tlg}{}\emptyrun=\oo$. Standard true sentences, such as ``snow is white'' or ``$0\equals 0$'', are understood as the game $\twg$, and false sentences, such as ``snow is black'' or ``$0\equals 1$'', as the game $\tlg$. Correspondingly, the two games $\twg$ and $\tlg$ may be referred to as {\bf propositions}.\label{0proposition}

The operation of {\bf prefixation}\label{0prefixation} takes two arguments: a constant game $A$ and a position $\Phi$ 
 that must 
be a legal position of $A$ (otherwise the operation is undefined), and returns the game $\seq{\Phi}A$.
Intuitively, $\seq{\Phi}A$ is the game playing which means playing $A$ starting (continuing) from position $\Phi$. 
That is, $\seq{\Phi}A$ is the game to which $A$ {\bf evolves}   (will be ``{\bf brought down}") after the moves of $\Phi$ have been made.  Here is a definition:

\begin{defi}\label{prfx}
%\marginpar{prfx}
Let $A$ be a constant game and $\Phi$ a legal position of $A$. The game 
$\seq{\Phi}A$\label{0pr} is defined by: 
\begin{itemize}
\item $\legal{\seq{\Phi}A}{}= \{\Gamma\ |\ \seq{\Phi,\Gamma}\in\legal{A}{}\}$;
\item $\win{\seq{\Phi}A}{}\seq{\Gamma}= \win{A}{}\seq{\Phi,\Gamma}$.
\end{itemize}
\end{defi}

\begin{conv}\label{poscon}
%\marginpar{poscon}
A terminological convention important to remember is that we often identify a legal position\label{0posa} $\Phi$ of a game $A$ with the game $\seq{\Phi}A$. So, for instance, we may say that the move $1$ by $\oo$ brings the game $B_0\adc B_1$ down to the position $B_1$. Strictly speaking, $B_1$ is not a position but a game, and what {\em is} a position is $\seq{\oo 1}$, which we here identified with the game $B_1=\seq{\oo 1}(B_0\adc B_1)$.
\end{conv}

Note that, in order to define the $\legal{}{}$ component of   a constant game $A$,  it is sufficient  to specify what the {\bf initial legal (lab)moves}\label{0ilm} --- i.e., the elements of  $\{\xx\alpha\ |\ \seq{\xx\alpha}\in\legal{A}{}\}$ --- are, and to 
what game the game $A$ is  brought down after such an initial legal labmove $\xx\alpha$ is made;   
then, the set of legal runs of $A$ will be uniquely defined  (this can be seen to hold even in recursive definitions of game operations as in clauses 1 and 3 of Definition \ref{op} below). Similarly, note that defining the $\win{}{}$ component only for the legal runs of $A$ would be sufficient, for then it uniquely extends to all runs. With these observations in mind, we define the operations $\gneg,\mlc,\mld,\adc,\add$  as follows:

\begin{defi}\label{op} Let $A$, $B$, $A_0,A_1,\ldots$ be   constant games, and $n$  a positive integer.\vspace{9pt}
%\marginpar{op}

\noindent 1. $\gneg A$\label{0gneg2} ({\bf negation}) is defined by: 
\begin{quote}\begin{description}
\item[(i)] $\seq{\xx\alpha}\in\legal{\gneg A}{}$ iff $\seq{\gneg{\xx}\alpha}\in\legal{A}{}$. Such an initial legal labmove $\xx\alpha$ brings the game down to 
$\gneg (\seq{\gneg{\xx}\alpha}A)$.
\item[(ii)] Whenever $\Gamma$ is a legal run of $\gneg A$, $\win{\gneg A}{}\seq{\Gamma} = \pp$ iff $\win{A}{}\seq{\gneg\Gamma} =\oo$. Here \[\gneg\Gamma\label{0rneg}\] means $\Gamma$ with each label $\xx$ changed to $\gneg\xx$.\vspace{5pt}
\end{description}\end{quote}

\noindent 2. $A_0\adc\ldots\adc  A_n$\label{0adc2} ({\bf choice conjunction}) is defined by: 
\begin{quote}\begin{description}
\item[(i)] $\seq{\xx\alpha}\in\legal{A_0\adci\ldots\adci  A_n}{}$ iff $\xx= \oo$ and $\alpha= i  \in\{0,\ldots,n\}$.\footnote{Here and in the other clauses of this definition,  the number $i$ is identified with its binary representation.  It should be pointed out that the binary representation of zero in the present case is meant to be the  symbol ``$0$''. This is as opposed to our later (Section \ref{nncg}) treatment of constants, where the ``official'' binary representation of zero is agreed to be the empty bit string.}  Such an initial legal labmove $\oo i$ brings the game down to 
$A_i$.
\item[(ii)] Whenever $\Gamma$ is a legal run of $A_0\adc\ldots\adc  A_n$, $\win{A_0\adci\ldots\adci  A_n}{}\seq{\Gamma} = \oo$ iff $\Gamma$ looks like $\seq{\oo i,\Delta}$ ($i  \in\{0,\ldots,n\}$) and $\win{A_i}{}\seq{\Delta} = \oo$.\vspace{5pt} 
\end{description}\end{quote}

\noindent 3. $A_0\mlc\ldots\mlc A_n$\label{0mlc2} ({\bf parallel conjunction}) is defined by: 
\begin{quote}\begin{description}
\item[(i)] $\seq{\xx\alpha}\in\legal{A_0\mlci\ldots\mlci  A_n}{}$ iff $\alpha= i.\beta$, where $i\in\{0,\ldots,n\}$ and $\seq{\xx\beta}\in\legal{A_i}{}$. Such an initial legal labmove $\xx i.\beta$ brings the game down to  
\[ A_0\mlc\ldots\mlc A_{i-1}\mlc \seq{\xx\beta}A_i\mlc A_{i\plus 1}\mlc\ldots\mlc A_n.\] 
\item[(ii)] Whenever $\Gamma$ is a legal run of $A_0\mlc\ldots\mlc A_n$, $\win{A_0\mlci\ldots\mlci  A_n}{}\seq{\Gamma}= \pp$ iff, for each $i\in\{0,\ldots,n\}$,  $\win{A_i}{}\seq{\Gamma^{i.}}= \pp$. Here \[\Gamma^{i.}\label{0apr2}\] means the result of removing, from $\Gamma$, all labmoves except those that look like $\xx i.\alpha$, and then further changing each such (remaining) $\xx i.\alpha$ to $\xx \alpha$.\footnote{Intuitively, $\Gamma^{i.}$ is the run played in the $A_i$ component. The present condition thus means that $\pp$ wins a $\mlci$-conjunction of games iff it wins in each conjunct.} \vspace{5pt}  
\end{description}\end{quote}

\noindent 4. $A_0\add\ldots\add A_n$\label{0add2} ({\bf choice disjunction}) and $A_0\mld\ldots\mld  A_n$
\label{0mld2} ({\bf parallel disjunction}) 
are defined exactly as $A_0\adc\ldots\adc A_n$ and $A_0\mlc\ldots\mlc  A_n$, respectively, only with ``$\pp$" and ``$\oo$" interchanged.\vspace{7pt} 

\noindent 5. $A\mli B$\label{imli2} ({\bf strict reduction})\label{0mli2} is treated as an abbreviation of $(\gneg A)\mld B$.
\end{defi}

We also agree that, when $k=1$,   $A_1\adc\ldots\adc A_k$ simply means $A_1$, and so do $A_1\add\ldots\add A_k$, $A_1\mlc\ldots\mlc A_k$ and $A_1\mld\ldots\mld A_k$. We further agree that, when the set $\{A_1,\ldots , A_k\}$ is empty ($k=0$, that is), both  $A_1\adc\ldots\adc A_k$ and  $A_1\mlc\ldots\mlc A_k$ mean $\pp$, while both  $A_1\add\ldots\add A_k$ and  $A_1\mld\ldots\mld A_k$ mean $\oo$.

\begin{exa}
 The game 
\[(0\equals 0\adc 0\equals 1)\mli(10\equals 11\adc 10\equals 10)\ ,\]
 i.e. 
\[\gneg (0\equals 0\adc 0\equals 1)\mld(10\equals 11\adc 10\equals 10),\]
 has thirteen legal runs, which are: 
\begin{description}
\item[1] $\seq{}$. It is won by $\pp$, because $\pp$ is the winner in the right $\mld$-disjunct (consequent).
\item[2] $\seq{\pp 0.0}$. (The labmove of) this run brings the game down to $\gneg 0\equals 0\mld(10\equals 11\adc 10\equals 10)$, and $\pp$ is the winner for the same reason as in the previous case.
\item[3] $\seq{\pp 0.1}$. It brings the game down to $\gneg 0\equals 1\mld(10\equals 11\adc 10\equals 10)$, and $\pp$ is the winner because it wins in both $\mld$-disjuncts. 
\item[4] $\seq{\oo 1.0}$. It brings the game down to $\gneg(0\equals 0\adc 0\equals 1)\mld 10\equals 11$.  $\pp$ loses as it loses in both $\mld$-disjuncts. 
\item[5] $\seq{\oo 1.1}$. It brings the game down to $\gneg (0\equals 0\adc 0\equals 1)\mld 10\equals 10$.  $\pp$ wins as it wins in the right $\mld$-disjunct. 
\item[6-7] $\seq{\pp 0.0,\oo 1.0}$ and $\seq{\oo 1.0, \pp 0.0}$. Both bring the game down to the false $\gneg 0\equals 0 \mld$ $10\equals 11$, and both are therefore lost by  $\pp$. 
\item[8-9] $\seq{\pp 0.1,\oo 1.0}$ and $\seq{\oo 1.0, \pp 0.1}$. Both bring the game down to the true $\gneg 0\equals 1 \mld$ $10\equals 11$, which makes  $\pp$ the winner.
\item[10-11] $\seq{\pp 0.0,\oo 1.1}$ and $\seq{\oo 1.1, \pp 0.0}$. Both bring the game down to the true $\gneg 0\equals 0 \mld$ $10\equals 10$, so $\pp$ wins.
\item[12-13] $\seq{\pp 0.1,\oo 1.1}$ and $\seq{\oo 1.1, \pp 0.1}$. Both bring the game down to the true $\gneg 0\equals 1 \mld$ $10\equals 10$, so $\pp$ wins.
\end{description}
\end{exa}

\noindent Later we will be using some relaxed informal jargon already established in CoL for describing runs and strategies, referring to moves via the intuitive meanings of their effects on the game. For instance, the initial labmove $\pp 0.0$ in a play of the game $p\adc q\mli r$ we can characterize as ``$\pp$ made the move $0.0$''. Remembering the meaning of the prefix ``$0.$'' of this move, we may as well say that ``$\pp$ made the move $0$ in the antecedent''. Further remembering the effect of such a move on the antecedent, we may just as well say ``$\pp$ chose $p$ (or the left $\adc$-conjunct) in the antecedent''. We may also 
 say something like ``$\pp$ (made the move that) brought the game down to $p\mli r$'', or ``$\pp$ (made the move that) brought the antecedent down to $p$''. 

To define the   operation $\st$ in the style of Definition \ref{op}, we need some preliminaries.  What we call a  {\bf tree of games}\label{0treeofgames} is a structure defined inductively as an element of the smallest set satisfying the following conditions:
\begin{itemize}
\item Every constant game $A$ is a tree of games. The one-element sequence $\seq{A}$ is said to be the {\bf yield}\label{0yield} of such a tree, and the {\bf address}\label{0address} of $A$ in this tree is the empty bit string. 
\item Whenever $\mathcal A$ is a tree of games with yield $\seq{A_1,\ldots,A_m}$ and $\mathcal B$ is a  tree of games with yield $\seq{B_1,\ldots,B_n}$, the pair ${\mathcal A}\circ{\mathcal B}$\label{0circ} is a tree of games with {\bf yield} $\seq{A_1,\ldots,A_m,$ $B_1,\ldots,B_n}$. The {\bf address} of each $A_i$  in this tree is $0w$, where $w$ is the address of $A_i$ in $\mathcal A$. Similarly,  the address of each $B_i$ is $1w$, where $w$ is the address of $B_i$ in $\mathcal B$.
\end{itemize}

\begin{exa} Where $A,B,C,D$ are constant games, $(A\circ B)\circ(C\circ(A\circ D))$ is a tree of games with yield $\seq{A,B,C,A,D}$. The address of the first $A$ of the yield, to which we may as well refer as the first {\bf leaf}\label{0leaf} of the tree, is $00$; the address of the second leaf $B$ is $01$; the address of the third leaf $C$ is $10$; the address of the fourth leaf $A$ is $110$; and the address of the fifth leaf $D$ is $111$.
\end{exa}

Note that $\circ$ is not an operation on games, but just a symbol used instead of the more common comma to separate the two parts of a pair. And a tree of games itself is not a game, but a collection of games arranged into a certain structure, just as a sequence of games is not a game but a collection of games arranged as a list. 

For bit strings $u$ and $w$, we will write $u\preceq w$\label{0preceq} to indicate that $u$ is a (not necessarily proper) {\bf prefix} (initial segment) of $w$.

\begin{defi}\label{op1} 
%\marginpar{op1}
Let  $A_1,\ldots,A_n$ ($n\geq 1$) be   constant games, and $\mathcal T$ be a tree of games with yield $\seq{A_1,\ldots,A_n}$. Let $w_1,\ldots,w_n$ be the addresses of $A_1,\ldots,A_n$ in $\mathcal T$, respectively. The game   $\st {\mathcal T}$ (the {\bf branching recurrence}\label{0st} of $\mathcal T$) is defined by:
\begin{description}
\item[(i)] $\seq{\xx\alpha}\in\legal{\sti {\mathcal T}}{}$ iff one of the following conditions is satisfied: 
\begin{enumerate}
\item $\xx\alpha=\xx u.\beta$, where $u\preceq w_i$ for at least one
  $i\in\{1,\ldots,n\}$ and, for each  $i$ with $u\preceq w_i$,
  $\seq{\xx\beta}\in\legal{A_i}{}$. We call such a (lab)move a {\bf
    nonreplicative}\label{0nonrep} (lab)move. It brings the game down
  to  $$\st {\mathcal T}'\ ,$$ where ${\mathcal T}'$ is the result of replacing $A_i$ by $\seq{ \xx\beta}A_i$ in $\mathcal T$ for each $i$ with $u\preceq w_i$. If here $u$ is $w_i$ (rather than a proper prefix of such) for one of $i\in\{1,\ldots,n\}$, we say that the  move $\xx u.\beta$ is {\bf focused}.\label{0focused} Otherwise it is {\bf unfocused}.\label{0unfocused} 
\item $\xx\alpha=\oo \col{w_i}$, where $i\in\{1,\ldots,n\}$. We call such a (lab)move a {\bf replicative}\label{0rep} (lab)move. It brings the game down to  $\st {\mathcal T}'$, where ${\mathcal T}'$ is the result of replacing $A_i$ by $(A_i\circ A_i)$ in $\mathcal T$.
\end{enumerate}
\item[(ii)] Whenever $\Gamma$ is a   legal run of $\st {\mathcal T}$,   $\win{\sti {\mathcal T}}{}\seq{\Gamma}=\pp$ iff, for each $i\in\{1,\ldots,n\}$ and every infinite bit string $v$ with $w_i\preceq v$, we have $\win{A_i}{}\seq{\Gamma^{\preceq v}}=\pp$. Here \[\Gamma^{\preceq v}\label{0susu}\] means the result of deleting, from $\Gamma$, all labmoves except those that look like $\xx u.\alpha$ for some bit string $u$ with $u\preceq v$, and then further changing each such (remaining) labmove  $\xx u.\alpha$ to $\xx \alpha$.\footnote{Intuitively, $\Gamma^{\preceq v}$ is the run played in one of the multiple ``copies'' of $A_i$ that have been generated in the play, with $v$ acting as a (perhaps longer than necessary yet meaningful) ``address'' of that copy.}
\end{description}
\end{defi}

\begin{exa} \label{ex37}
%\marginpar{ex37} 
 Let \[G\ =\    p\add(q\adc (r\adc (s\add t))) ,\] where $p,q,r,s,t $ are constant elementary games. And let  
\[\Gamma\ = \ \seq{ \oo \col{},\ \pp .1, \ \oo 0.0,\ \oo 1.1,\  \oo \col{1},\ \oo 10.0,\ \oo 11.1,\ \pp 11.0}.\]  Then $\Gamma$ is a legal run of $\st G$. Below we trace, step by step, the effects of its moves on $\st G$.

The 1st (lab)move $\oo \col{}$ means that $\oo$ replicates the (only) leaf of the tree, with the address of that leaf being the empty bit string. This move 
 brings the game down to --- in the sense that $\seq{\oo \col{}}\st G$ is --- the following game:
\[\st \Bigl(\bigl(p\add(q\adc (r\adc (s\add t)))\bigr)\circ \bigl(p\add(q\adc (r\adc (s\add t)))\bigr)\Bigr).\]

The 2nd move $\pp .1$ means choosing the second $\add$-disjunct $q\adc (r\adc (s\add t))$ in {\em both} leaves of the tree. This is so because the addresses of those leaves are $0$ and $1$, and the empty bit string --- seen between ``$\pp$'' and ``$.1$'' in $\pp .1$ --- is an initial segment of both addresses. The effect of this unfocused move is the same as the effect of the two consecutive focused moves $\pp 0.1$ and $\pp 1.1$ (in whatever order) would be, but $\pp$ might have its reasons for having made an unfocused move. Among such reasons could be that $\pp$ did not notice $\oo$'s initial move (or the latter arrived late over the asynchronous network) and thought that the position was still $G$, in which case making the moves $\pp 0.1$ and $\pp 1.1$ would be simply illegal. Note also that the ultimate effect of the move $\pp .1$ on the game would remain the same as it is now even if this move was made before the replicative move $\oo\col{}$. It is CoL's striving to achieve this sort of flexibility and asynchronous-communication-friendliness  that has determined our seemingly ``strange'' choice of trees rather than sequences as the underlying structures for $\st$-games. Any attempt to deal with sequences instead of trees would encounter the problem of violating what CoL calls the  {\em static} (speed-independent) property of games, defined later in Section \ref{nncg}.

 Anyway, the position resulting from the second move of $\Gamma$ is    
\[\st \Bigl(\bigl( q\adc (r\adc (s\add t))\bigr)\circ \bigl(q\adc (r\adc (s\add t))\bigr)\Bigr).\]
The effect of the 3rd move $ \oo 0.0$ is choosing the left $\adc$-conjunct $q$ in the left ($0$-addressed) leaf of the tree, which results in 
\[\st \Bigl(q\circ \bigl(q\adc (r\adc (s\add t))\bigr)\Bigr).\]
Similarly, the 4th move $\oo 1.1$ chooses the right $\adc$-conjunct in the right leaf of the tree, resulting in 
\[\st \Bigl(q\circ \bigl(r\adc (s\add t)\bigr)\Bigr).\]
The 5th move $\oo \col{1}$ replicates  the right leaf, bringing the game down to  
\[\st \Bigl(q\circ \bigl((r\adc (s\add t))\circ (r\adc (s\add t))\bigr)\Bigr).\]
The 6th move $\oo 10.0$ chooses the left $\adc$-conjunct in the second ($00$-addressed) leaf, and, similarly, the 7th move $\oo 11.1$ chooses the right $\adc$-conjunct in the third ($11$-addressed) leaf. These two moves bring the game down to   
\[\st \Bigl(q\circ \bigl(r\circ (s\add t)\bigr)\Bigr).\]
The last, 8th move $\pp 11.0$ chooses the left $\add$-disjunct of the third leaf, and the final position is 
\[\st \bigl(q\circ (r\circ s)\bigr).\]

According to clause (ii) of Definition \ref{op1}, $\Gamma$ is a $\pp$-won run of $\st G$ iff, for any infinite bit string $v$, $\Gamma^{\preceq v}$ is a $\pp$-won run of $G$. Observe that for any infinite --- or, ``sufficiently long'' finite --- bit string $v$, $\Gamma^{\preceq v}$ is either $\seq{\pp 1,\oo 0}$ (if $v=0\ldots$) or $\seq{\pp 1,\oo 1, \oo 0}$ (if $v=10\ldots$) or $\seq{\pp 1,\oo 1, \oo 1, \pp 0}$ (if $v=11\ldots$).  We also have $\seq{\pp 1,\oo 0}G=q$,
$\seq{\pp 1,\oo 1, \oo 0}G=r$ and  $\seq{\pp 1,\oo 1, \oo 1, \pp 0}G=s$. So it is no accident that we see $q,r,s$ at the leaves in the final position. Correspondingly, the game is won iff each one of these three propositions is true. 

The cases where $\oo$ makes infinitely many replications in a run $\Gamma$ of a game $\st H$ and hence the ``eventual tree'' is infinite are similar, with the only difference that the ``addresses'' of the ``leaves'' of such a ``tree'', corresponding to different plays of  $H$, may be infinite bit strings.  But, again, the overall game $\st H$ will be won by $\pp$ iff all of those plays --- all $\Gamma^{\preceq v}$ where $v$ is an infinite bit string, that is --- are $\pp$-won plays of $H$. 
\end{exa}

\begin{defi}\label{opp}
%\marginpar{opp}
 Let $B$,  $A_1,\ldots,A_n$ ($n\geq 0$) be   constant games. We define 
$$A_1,\ldots,A_n\intimpl B$$ --- let us call it the (generalized) {\bf Turing reduction}\label{0intimpl} of $B$ to $A_1,\ldots,A_n$ --- as the game $$\st A_1\mlc\ldots\mlc\st A_n\mli B.$$
\end{defi}

\section{Not-necessarily-constant games and ``quantifier'' operations}\label{nncg}
%\marginpar{nncg}

Constant games can be seen as generalized propositions: while propositions in classical logic are just elements 
of $\{\twg,\tlg\}$, constant games are functions from runs to $\{\twg,\tlg\}$.  Our concept of a (not-necessarily-constant) game defined in this section generalizes that of a constant game in the same sense as the classical concept of a predicate generalizes that of a proposition. 

Throughout CoL, we have a fixed set   $\variables$
of expressions called {\bf variables},\label{0variable}
\[\variables\ =\ \{\mbox{\em var}_1,\mbox{\em var}_2,\mbox{\em var}_3,\ldots\}.\label{0variablee}\]  
Each variable is thus the string $\mbox{\em var}_i$\label{0vari} for some positive decimal numeral $i$.  However, we seldom write variables in this form and, instead, usually use  
 the letters (metavariables) $r,s,t,u,v,w,x,y,z$ for them.   We also have another fixed set $\constants$  of expressions called {\bf canonical constants},\label{0constant} or simply {\bf constants}. Specifically,  
\[\constants\ =\ \{\epsilon,1,10,11,100,101,110,111,1000,\ldots\}.\label{0consntme}\] Constants are thus  {\bf binary numerals}\label{0binnum} --- the strings matching the regular expression $\epsilon\cup 1(0\cup 1)^*$, where $\epsilon$\label{0epsilon} is the empty string.  We will be often identifying such strings --- by some rather innocent abuse of concepts --- with the natural numbers represented by them in the standard binary notation, and vice versa. Note that $\epsilon$ represents $0$. For this reason, following tradition, we shall usually write $0$ instead of $\epsilon$, keeping in mind that, in such cases, the length $|0|$ of the string $0$ should (still) be seen to be $0$ rather than $1$. We will be mostly using $a,b,c,d$ as metavariables for constants. 

A {\bf universe}\label{0universe} (of discourse) is a pair $U=(\mbox{\em Dom}, \mbox{\em Den})$, where 
$\mbox{\em Dom}$, called the {\bf domain}\label{xxdomain1} of $U$, is a nonempty set, and $ \mbox{\em Den}$, called 
the {\bf denotation}\label{0nf} of $U$,  is a (total) function of type 
$\constants\rightarrow \mbox{\em Dom}$.  The elements of $\mbox{\em Dom}$ will be referred to as the {\bf individuals}\label{0individual} of $U$. The intuitive meaning of $d=\mbox{\em Den}(c)$ is that the individual
$d$ is the {\bf denotat}\label{0denotat} of the constant $c$ and thus $c$ is a {\bf name}\label{0name} of $d$. So, the function 
$\mbox{\em Nam}$ from $\mbox{\em Dom}$ to the powerset of $\constants$ satisfying the condition $c\in \mbox{\em Nam}(d)\Leftrightarrow d=\mbox{\em Den}(c)$ can be called 
the {\bf naming}\label{0naming} of $U$. 
Of course, whenever convenient, a universe can be characterized in terms of its naming rather than denotation.  

The {\bf ideal universe}\label{0ideal} is the universe whose domain is $\constants$ and whose denotation is the identity function on that domain.  
All earlier papers on CoL, except \cite{cla4,lbcs,cla8,cla5}, dealt only with the ideal universe. This was for simplicity considerations, yielding no loss of generality as no results relied on the assumption that the underlying universe was ideal. Our present treatment, however, for both technical and philosophical reasons, {\em does} call for the above-defined, more general, concept of a universe --- a universe where some individuals may have  unique names, some  have many names, and some have no names at all.\footnote{Further generalizations are possible if and when a need arises. Namely, one may depart from our present assumption that the set of constants is infinite and/or fixed, as long as there is a fixed constant --- say, $0$ --- that belongs to every possible set of constants ever considered. No results of this or any earlier papers on CoL would be in any way affected by doing so.} Note that real-world universes are typically not ideal: not all people living or staying in the United States have social security numbers;  most stars and planets of the Galaxy have no names at all, while some  have several names (Morning Star = Evening Star = Venus); etc. A natural example of a non-ideal universe from the world of mathematics would be the one whose domain is the set  of real numbers, only some of whose elements have names, such as $5$, $1/3$, $\sqrt{2}$ or $\pi$. Generally, even if the set of constants was not fixed, no universe with an uncountable domain would  be ``ideal''  for the simple reason that  there can  only be countably many names. This is so because names, by their very nature and purpose, have to be finite objects. Observe also that many  properties of common interest, such as computability or decidability, are usually sensitive to how objects (individuals) are named, as they deal with the names of those objects rather than the objects themselves. For instance, strictly speaking, computing a function $f(x)$ means the ability to tell, after seeing a (the) name of an arbitrary object $\mathfrak{a}$, to produce a (the) name of the object $\mathfrak{b}$ with $\mathfrak{b}=f(\mathfrak{a})$.  Similarly, an algorithm that decides a predicate $p(x)$ on a set $S$, strictly speaking, takes not elements of $S$ --- which may be abstract objects such as numbers or graphs --- but rather names of those elements (such as binary numerals or codes). It is not hard to come up with a nonstandard naming of the natural numbers through binary numerals where the predicate ``$x$ is even'' is undecidable.  On the other hand, for any undecidable arithmetical predicate $p(x)$, one can come up with a naming  such that $p(x)$ becomes decidable --- for instance, one that assigns even-length names to all $\mathfrak{a}$ with $p(\mathfrak{a})$    and assigns odd-length names to all $\mathfrak{a}$ with $\gneg p(\mathfrak{a})$. Classical logic exclusively deals with individuals of a universe without a need for also considering names for them, as it is not concerned with decidability or computability. CoL, on the other hand, with its computational semantics, inherently calls for being more careful about differentiating between individuals and their names, and hence for explicitly considering universes in the form $(\mbox{\em Dom}, \mbox{\em Den})$ rather than just $\mbox{\em Dom}$  as classical logic does.

By a {\bf valuation}\label{0valuation} on a universe $U=(\mbox{\em Dom},\mbox{\em Den})$, or a {\bf $U$-valuation},  we mean 
a (total) function $e$ of type  $\variables \rightarrow \mbox{\em Dom}$. For the purposes of this definition, the $\mbox{\em Den}$ component of $U$ is thus irrelevant and, whenever convenient, we may talk about valuations in terms of just domains instead of universes. 
%For technical convenience, we uniquely extend every $U$-valuation $e$ to constants as well, by stipulating that, for any constant $c$, $e(c)=\mbox{\em Den}(c)$. 
The set of all $U$-valuations will be denoted by $\valuations(U)$.\label{0valu} When a universe $U$ (or its domain) is fixed,  irrelevant or clear from the context, we may omit an explicit reference to it and simply say ``valuation''. References to $U$ can be similarly omitted when talking about individuals, denotats, names or some later-defined concepts such as those of a game or a function. 

In the above terms, a classical predicate $p$ can be understood as 
a function that sends each valuation $e$ to a proposition, i.e., to a constant predicate.   Similarly, what we call a game sends valuations to constant games: 

\begin{defi}\label{ngame}
%\marginpar{ngame}
Let $U$ be a universe. A {\bf game on $U$}\label{0game} is a (total) function $A$ from $U$-valuations   to constant games. For a valuation $e$, we write $e[A]$\label{0ea} rather than $A(e)$ to denote the value of $A$ at $e$. Such a constant game $e[A]$ is said to be an {\bf instance}\label{0instance} of $A$. 
For readability, we usually write $\legal{A}{e}$\label{0lre} and $\win{A}{e}$\label{0wne} instead of $\legal{e[A]}{}$ and $\win{e[A]}{}$.
\end{defi}

Just as it is the case with propositions versus predicates, constant games in the sense of Definition \ref{game} will
be thought of as special, constant cases of games in the sense of Definition \ref{ngame}. Namely, in the context of a given universe $U$, each constant game $A'$ is the game $A$ (on $U$) such that, for every valuation $e$,
$e[A]= A' $. From now on we will no longer distinguish between such $A$ and $A' $, so that, if $A$ is a constant game,
it is its own instance, with $A= e[A]$ for every valuation $e$.   
 
Where $n$ is a natural number, we say that a game $A$ is {\bf $n$-ary}\label{0arg} iff there are $n$ variables such that, for any two valuations $e_1$ and $e_2$ that agree on all those variables, we have $e_1[A]= e_2[A]$. Note that, if $A$ is $n$-ary, then it is also $(n+1)$-ary, $(n+2)$-ary, etc. A game that is $n$-ary for some $n$ is said to be {\bf finitary}.\label{0finitary} The present paper is going to exclusively deal with finitary games and, for this reason, we agree that, from now on, when we say ``game'', we always mean ``finitary game''.  

For a variable $x$ and valuations  $e_1,e_2$, we write $e_1\equiv_x e_2$\label{0equiv} to mean that the two valuations  agree on all variables  other than (perhaps) $x$.

We say that a game $A$ {\bf depends}\label{0depend} on a variable $x$ iff there are two valuations  $e_1,e_2$ with $e_1\equiv_x e_2$  such that $e_1[A]\not= e_2[A]$. An $n$-ary game thus depends on at most $n$ variables. And constant games are nothing but $0$-ary games, i.e., games that do not depend on any variables. 

We say that a (not necessarily constant) game $A$ is {\bf elementary}\label{0elemgameb} iff so are all of its instances.

Just as constant games are generalized propositions, games can be seen as generalized predicates. Namely, in the context of a given universe $U$, we will view each predicate $p$ on $U$'s domain  as  the elementary game such that, for every valuation $e$,
$\win{p}{e}\emptyrun=\pp$ iff $p$ is true at $e$.  
And vice versa: every elementary game $p$ will be viewed as the  predicate on $U$'s domain which is true at a given valuation $e$ iff  $\win{p}{e}\emptyrun=\pp$.   
Thus, for us, ``predicate'' and ``elementary game'' are synonyms. Accordingly,  any standard terminological or notational conventions familiar from the literature for predicates also apply to them  viewed as elementary games. 
%And most of those conventions naturally extend to all games. 

There are two different yet not always clearly differentiated understandings of predicates in the literature. One --- more common --- understanding is {\em extensional},\label{0euop} 
according to which an $n$-ary predicate is a set of $n$-tuples of individuals (that satisfy $p$). The other 
understanding --- let us call it {\em subextensional}\label{0iuop} --- sees a predicate $p$ as a set of valuations (at which $p$ is true). Note that, unlike the extensional understanding, the subextensional understanding is variable-sensitive: it, for instance,  sees $\mbox{\em var}_1\geq 100$ and $\mbox{\em var}_2\geq 100$ as different predicates, because one depends on $\mbox{\em var}_1$ while the other depends on $\mbox{\em var}_2$ instead. Our understanding of predicates and, more generally, our understanding of games, is clearly subextensional. We agree on the following  understanding of functions in the same subextensional style:

\begin{defi}\label{nfun}
%\marginpar{nfun}
Let $U=(\mbox{\em Dom},\mbox{\em Den})$ be a universe. A {\bf function on $U$}\label{0fun} is a mapping $f$ from $\valuations(U)$   to $\mbox{\em Dom}$. For a valuation $e$, we write $e[f]$\label{0ef} rather than $f(e)$ to denote the value of $f$ at  $e$.
% For each variable or constant $w$, $w$ will (also) stand for the function that sends each $U$-valuation $e$ to (the value of) $e[w]$.
\end{defi} 

Much of our terminology for games naturally extends to functions as well. Namely, where $n$ is a natural number, we say that a function $f$ is {\bf $n$-ary}\label{0argf} iff there are $\leq n$ variables such that, for any two valuations $e_1$ and $e_2$ that agree on all those variables, we have $e_1[f]= e_2[f]$. A  function that is $n$-ary for some $n$ is said to be {\bf finitary}.\label{0finitaryf} The present paper  exclusively deals with finitary functions, and we agree that, from now on, ``function'' always means ``finitary function''.  
We say that a function $f$ {\bf depends}\label{0dependf} on a variable $x$ iff there are two valuations  $e_1,e_2$ with $e_1\equiv_x e_2$  such that $e_1[f]\not= e_2[f]$. {\bf Constant functions}\label{0constantfunction} are nothing but $0$-ary functions, i.e. functions   that do not depend on any variables. 

\begin{conv}\label{wuding}
%\marginpar{wuding}
In the context of a given universe $U=(\mbox{\em Dom},\mbox{\em Den})$, all individuals, constants  and variables
will be (simultaneously) understood as (denoting) functions on $U$. Namely: 
\begin{enumerate}[label=\arabic*.]
\item An individual $a$ is the (constant) function that sends every valuation $e$ to $a$. 

\item A constant $c$ is the (constant) function that sends every valuation $e$ to $\mbox{\em Den}(c)$.

\item A variable $x$ is the (not-necessarily-constant) function that
  sends every valuation $e$ to $e(x)$.
\end{enumerate}
\end{conv}

\begin{defi}\label{sov}
%\marginpar{sov}
Let $W$ be a game or a function on a universe $U$, $x_1,\ldots,x_n$ be pairwise distinct variables, and $f_1,\ldots,f_n$ be  functions on $U$. 
The result of {\bf substituting $f_1,\ldots,f_n$ for $x_1,\ldots,x_n$ in $W$}, denoted by 
$W(f_1/x_1,\ldots,f_n/x_n)$,\label{0dash} is defined as the game on the same universe $U$ such that,  for every valuation $e$, $e[W(f_1/x_1,\ldots,f_n/x_n)]= e'[W]$, where $e'$ is the valuation that sends each $x_i\in\{x_1,\ldots,x_n\}$ to (the value of) $e[f_i]$ and agrees with $e$ on all other variables. 
\end{defi}

Following the standard readability-improving practice established in the literature for predicates and functions, we will often fix pairwise distinct  variables $x_1,\ldots,x_n$ for a game or function $W$ and write $W$ as $W(x_1,\ldots,x_n)$. 
Representing $W$ in this form  sets a context in which we can write $W(f_1,\ldots,f_n)$ to mean the same as the more clumsy expression $W(f_1/x_1,\ldots,f_n/x_n)$. 

Just as the Boolean operations straightforwardly extend from propositions to all predicates, our operations 
$\gneg,\mlc,\mld,\mli,\adc,\add,\st,\intimpl$ extend from constant games to all games, with each operation taking one or more games on a given (common-for-all-arguments) universe and returning a game on the same universe. This is done by simply stipulating that $e[\ldots]$ commutes with all of those operations: $\gneg A$ is 
the game  such that, for every valuation $e$, $e[\gneg A]=\gneg e[A]$; $A\adc B$ is the game such that,
for every $e$, $e[A\adc B]= e[A]\adc e[B]$; etc. So does the operation of prefixation: provided that $\Phi$ is a legal position of every instance of $A$,  $\seq{\Phi}A$ is  understood as the unique game such that, for every $e$, $e[\seq{\Phi}A]= \seq{\Phi}e[A]$. 

\begin{defi}\label{bq}
%\marginpar{bq} 
Below  $A(x)$ is any game on a universe $U$. On the same universe:\vspace{9pt}

\noindent 1. The game $\ada xA(x)$ ({\bf choice universal quantification})\label{0ada2} is defined by stipulating that, for every $U$-valuation $e$, we have: 
\begin{quote}\begin{description}
\item[(i)] $\seq{\xx\alpha}\in\legal{\adai xA(x)}{e}$ iff $\xx= \oo$ and $\alpha$ is $\#c$\label{0pound} for some constant $c$.   Such an initial legal labmove $\oo \#c$ brings the game $e[\ada xA(x)]$ down to 
$e[A(c)]$.
\item[(ii)] Whenever $\Gamma$ is a legal run of $\ada x A(x)$, $\win{\adai xA(x)}{e}\seq{\Gamma} = \oo$ iff $\Gamma$ looks like $\seq{\oo \#c,\Delta}$   and $\win{A(c)}{e}\seq{\Delta} = \oo$.\vspace{5pt} 
\end{description}\end{quote}

\noindent 2. The game $\ade x A(x)$ ({\bf choice existential quantification})\label{0ade2}  is defined in exactly the same way, only with $\pp$ and $\oo$ interchanged.  
\end{defi}

Thus, every initial legal move of $\ada xA(x)$ or $\ade xA(x)$ is the string $\#c$ for some $c\in\constants$,  which in our informal language we may refer to as ``the constant chosen (by the corresponding player) for $x$''. Note that $\ada x A(x)$ is nothing but the infinite choice conjunction $A(0)\adc A(1)\adc \ldots$, with the only technical difference that, in the former, the initial, ``choice'' move should be prefixed with the symbol ``$\#$''. Similarly, $\ade xA(x)$ is essentially nothing but the infinite choice disjunction $A(0)\add A(1)\add\ldots$.

 We say that a game $A$ is  {\bf unistructural}\label{0unistructural} iff, for any two valuations $e_1$ and $e_2$,   we have $\legal{A}{e_1}= \legal{A}{e_2}$. Of course, all constant or elementary games are unistructural. It can also be easily seen that all our game operations preserve the unistructural property of games. For the purposes of the present paper, considering only  unistructural games is sufficient.

We define the remaining operations $\cla$ and $\cle$ only for unistructural games.

\begin{defi}\label{op5} Below  $A(x)$ is any  unistructural game on a universe $U$. On the same universe:\vspace{9pt}
%\marginpar{op5}

\noindent 1. The game $\cla x A(x)$ ({\bf blind universal quantification})\label{0cla2} is defined by stipulating that, for every $U$-valuation $e$,  we have: 
\begin{quote}\begin{description}
\item[(i)] $\seq{\xx\alpha}\in\legal{\clai x A(x)}{e}$ iff $\seq{\xx\alpha}\in\legal{A(x)}{e}$. Such an initial legal labmove $\xx\alpha$ brings the game $e[\cla x A(x)]$ down to 
$e[\cla x\seq{\xx\alpha}A(x)]$.
\item[(ii)] Whenever $\Gamma$ is a legal run of $e[\cla x A(x)]$,  $\win{\clai x A(x)}{e}\seq{\Gamma}= \pp$ iff, for every valuation $g$ with $g\equiv_x e$, $\win{A(x)}{g}\seq{\Gamma}= \pp$. \vspace{5pt}
\end{description}\end{quote}
\noindent 2. The game $\cle x A(x)$ ({\bf blind existential quantification})\label{0cle2}  is defined in exactly the same way, only with $\pp$ and $\oo$ interchanged.  
\end{defi}

\begin{exa}\label{may14}
%\marginpar{may14}
Let $G$ be the following game on the ideal universe, with the predicates {\em Even} and {\em Odd} having their expected meanings: 
\[\cla y\Bigl(\mbox{\em Even}(y)\add \mbox{\em Odd}(y) \mli \ada   x\bigl(\mbox{\em Even}(x\plus y)\add \mbox{\em Odd}(x\plus y)\bigr)\Bigr).\]
 Then the sequence 
$\seq{\oo 1.\#11,\ \oo 0.0,\ \pp 1.1}$ 
is a legal run of $G$, the effects of the moves of which are shown below:
\[\begin{array}{ll}
G:  & \cla y\Bigl(\mbox{\em Even}(y)\add \mbox{\em Odd}(y) \mli \ada   x\bigl(\mbox{\em Even}(x\plus y)\add \mbox{\em Odd}(x\plus y)\bigr)\Bigr)\\
\seq{\oo 1.\#11}G:  & \cla y\bigl(\mbox{\em Even}(y)\add \mbox{\em Odd}(y) \mli \mbox{\em Even}(11\plus y)\add \mbox{\em Odd}(11\plus y)\bigr)\\
\seq{\oo 1.\#11, \oo 0.0}G: &  \cla y\bigl(\mbox{\em Even}(y) \mli \mbox{\em Even}(11\plus y)\add \mbox{\em Odd}(11\plus y)\bigr)\\
\seq{\oo 1.\#11, \oo 0.0,\pp 1.1}G: & \cla y\bigl(\mbox{\em Even}(y) \mli \mbox{\em Odd}(11\plus y)\bigr)
\end{array}\]
The play hits (ends as) the true proposition $\cla y\bigl(\mbox{\em Even}(y) \mli \mbox{\em Odd}(11\plus y)\bigr)$ and hence is won by $\pp$. 
\end{exa}

\begin{exa}\label{newexample}
The \ sequence \ \ $\langle\pp 0.1.\col{},\ \oo 1.\#10,\ \pp 0.1.0.\#10,\ \pp 0.1.0.\#10,\ \oo 0.1.0.\#100,$ $\pp 0.1.1.\#100,$ \  $\pp 0.1.1.\#10,\ \oo 0.1.1.\#1000,\ \pp 1.\#1000\rangle$
is a legal run of the game 
\begin{equation}\label{mar7}
\cla x \bigl(x^3\equals (x\mult x)\mult x\bigr),\ \ada x\ada y \ade z (z\equals x\mult y) \ \intimpl \ \ada x\ade y(y\equals x^3).
\end{equation}
Below we see how the game evolves according to the scenario of this run: 

\

\[\begin{array}{ll}
& \cla x \bigl(x^3\equals (x\mult x)\mult x\bigr),\ \ada x\ada y \ade z (z\equals x\mult y) \ \intimpl \ \ada x\ade y(y\equals x^3)\\
& \\
\mbox{$\pp 0.1.\col{}$ yields} & \cla x \bigl(x^3\equals (x\mult x)\mult x\bigr),\ \ada x\ada y \ade z (z\equals x\mult y)\circ\ada x\ada y \ade z (z\equals x\mult y)  \ \intimpl\\
& \hspace{15pt}  \ada x\ade y(y\equals x^3)\\
& \\
 \mbox{$\oo 1.\#10$ yields} & \cla x \bigl(x^3\equals (x\mult x)\mult x\bigr),\ \ada x\ada y \ade z (z\equals x\mult y)\circ\ada x\ada y \ade z (z\equals x\mult y) \ \intimpl  \\
& \hspace{15pt}   \ade y(y\equals 10^3)\\
& \\
\mbox{$\pp 0.1.0.\#10$ yields} & \cla x \bigl(x^3\equals (x\mult x)\mult x\bigr),\ \ada y \ade z (z\equals 10\mult y)\circ\ada x\ada y \ade z (z\equals x\mult y)  \ \intimpl\\
& \hspace{15pt}  \ade y(y\equals 10^3)\\
& \\
 \mbox{$\pp 0.1.0.\#10$ yields} & \cla x \bigl(x^3\equals (x\mult x)\mult x\bigr),\ \ade z (z\equals 10\mult 10)\circ\ada x\ada y \ade z (z\equals x\mult y)  \ \intimpl\\
& \hspace{15pt}  \ \ade y(y\equals 10^3)\\
& \\
\mbox{$\oo 0.1.0.\#100$ yields} & \cla x \bigl(x^3\equals (x\mult x)\mult x\bigr),\ (100\equals 10\mult 10)\circ\ada x\ada y \ade z (z\equals x\mult y)  \ \intimpl\\
& \hspace{15pt}  \ade y(y\equals 10^3)\\
& \\
\mbox{$\pp 0.1.1.\#100$ yields} & \cla x \bigl(x^3\equals (x\mult x)\mult x\bigr),\ (100\equals 10\mult 10)\circ\ada y \ade z (z\equals 100\mult y)  \ \intimpl\\
& \hspace{15pt}   \ade y(y\equals 10^3)\\
& \\
\mbox{$\pp 0.1.1.\#10$ yields} & \cla x \bigl(x^3\equals (x\mult x)\mult x\bigr),\ (100\equals 10\mult 10)\circ \ade z (z\equals 100\mult 10)  \ \intimpl \ \ade y(y\equals 10^3)\\
& \\
\mbox{$\oo 0.1.1.\#1000$ yields} & \cla x \bigl(x^3\equals (x\mult x)\mult x\bigr),\ (100\equals 10\mult 10)\circ (1000\equals 100\mult 10)  \ \intimpl \ \ade y(y\equals 10^3)\\
& \\
 \mbox{$\pp 1.\#1000$ yields} &\cla x \bigl(x^3\equals (x\mult x)\mult x\bigr),\ (100\equals 10\mult 10)\circ (1000\equals 100\mult 10)  \ \intimpl \ 1000\equals 10^3\\
& 
\end{array}\]
The play hits a true proposition and hence is won by $\pp$. Note that here, unlike the case in the previous example, $\top$ is the winner no matter what universe we consider and what the meanings of the expressions $x\times y$ and $x^3$ are. In fact, $\pp$ has a ``purely logical'' winning strategy in this game, in the sense that the strategy is successful  regardless of whether things have their standard arithmetical meanings or some other meanings. This follows from the promised soundness of {\bf CL12} and the fact --- illustrated later in Example \ref{ecube} --- that (\ref{mar7}) is provable in {\bf CL12}. Such ``purely logical'' strategies in CoL are called {\em uniform solutions} (see Section \ref{ss6}).
\end{exa}

We close this section by defining a couple of additional concepts that are  central in CoL, even if nothing in the present paper directly or indirectly relies on them. 
For either player $\xx$,  a run $\Phi$ is said to be a {\bf $\xx$-delay}\label{0delay} of a run $\Gamma$ iff (1) 
 for both players $\xx'\in\{\top,\bot\}$, the subsequence of $\wp'$-labeled moves of $\Phi$ is the same as that of $\Gamma$, and (2)
for any $n,k\geq 1$, if the $n$th $\xx$-labeled move is made later than (is to the right of) the $k$th $\gneg\wp$-labeled move in $\Gamma$, then so is it in $\Phi$.  For instance, $\seq{\oo\alpha_1,\pp\beta_1,\oo\alpha_2,\pp\beta_2}$ is a $\pp$-delay of  $\seq{\pp\beta_1,\oo\alpha_1,\pp\beta_2,\oo\alpha_2}$. Next, 
we say that a run is {\bf $\xx$-legal}\label{0plegal} iff it is not $\xx$-illegal. 
Now, we say that a constant game  $A$ is {\bf static}\label{0static} iff, for either player $\wp$, whenever a run $\Phi$ is a $\wp$-delay of 
a run $\Gamma$, we have: (i) if $\Gamma$ is a $\wp$-legal run of $A$, then so is $\Phi$, and 
(ii) if $\Gamma$ is a $\wp$-won run of $A$, then so is $\Phi$. This concept extends to all games by stipulating that a not-necessarily-constant game is static iff all of its instances are so.  

Static games form a very wide natural subclass of all games. Intuitively, such games are interactive tasks where the relative speeds of the players are irrelevant, as it never hurts a player to postpone making moves. In other words, static games are games that are contests of intellect rather than contests of speed. And one of the theses that CoL philosophically relies on is that static games present an adequate formal counterpart of our intuitive concept of ``pure'', speed-independent interactive computational problems. Correspondingly, CoL restricts its attention (more specifically, possible interpretations of the atoms of its formal language) to static games. Every elementary game is trivially static, and the class of static games turns out to be closed under all game operations studied in CoL. This can be seen to immediately imply that all games expressible in the language of the later-defined logic $\cltw$  are static (as well as finitary and unistructural).

\section{Interactive machines}\label{icp}
%\marginpar{icp}

In traditional game-semantical approaches, including those of Lorenzen \cite{Lor61}, Hintikka \cite{Hintikka73} and Blass's \cite{Bla92}, players' strategies are understood as {\em functions} --- typically as functions from interaction histories (positions) to moves, or sometimes (Abramsky and Jagadeesan \cite{Abr94}) as functions that only look at the latest move of the history. This {\em strategies-as-functions} approach, however, is generally inapplicable in the context of CoL, 
whose relaxed semantics, in striving to get rid of  ``bureaucratic pollutants'' and only deal with the remaining true essence of games,  does not impose any regulations on which player can or should move in a given situation. Here, in many cases, either player may have (legal) moves, and then it is unclear whether the next move should be the one prescribed by $\pp$'s strategy function or the one prescribed by the strategy function of $\oo$. For a game semantics whose ambition is to provide a comprehensive, natural and direct tool for modeling interaction, the strategies-as-functions approach would be less than adequate, even if technically possible. This is so for the simple reason that  the strategies that real computers follow are not functions. If the strategy of your personal computer was a function from the history of interaction with you, then its performance would keep noticeably worsening due to the need to read the continuously lengthening --- and, in fact, practically infinite --- interaction history every time before responding. Fully ignoring that history and looking only at your latest keystroke in the spirit of \cite{Abr94} is also certainly not what your computer does, either. The advantages of our approach thus become especially appreciable when one tries to bring complexity theory into interactive computation: hardly (m)any really meaningful and interesting complexity-theoretic concepts can be defined for games (particularly, games that may last long) with the strategies-as-functions approach.  

In CoL, ($\pp$'s effective) strategies are defined in terms of interactive machines, where computation is one continuous process interspersed with --- and influenced by --- multiple ``input'' (environment's moves) and ``output'' (machine's moves) events. Of several, seemingly rather different yet equivalent,  machine models of interactive computation studied in CoL, this paper only employs the most basic, {\bf HPM}\label{0HPM} (``Hard-Play Machine'') model. 
 
An HPM  is   a Turing machine with the additional capability of making moves. The adversary can also move at any time, and such moves are the only nondeterministic events from the machine's perspective. Along with one or more ordinary read/write {\bf work tapes},\footnote{This is the first time in the literature on CoL that multiple-work-tape HPMs are considered. All earlier HPMs had a single work tape. Just as in the traditional theory of computation, as long as we are willing to ignore certain ``small'' polynomial time differences in computation (game-playing) efficiency, 
 nothing depends on how many --- one or more --- work tapes are allowed.}\label{0worktape} the machine has an additional, read-only   tape called  the  {\bf run tape}.\label{0runtape}\footnote{Together with the work and run tapes, the HPMs from the earlier literature on CoL also had an additional tape called the {\em valuation tape}. The latter becomes redundant in our present treatment due to the fact that we are exclusively interested in constant games or finitary games identified with their (constant) $\adc$-closures.} The latter, serving as a dynamic input, at any time  spells the ``current position'' of the play. Its role is to make the evolving run fully visible to the machine. 
In these terms,  an  algorithmic solution ($\pp$'s winning strategy) for a given constant game $A$ is understood as an HPM $\mathcal M$ such that,  no matter how the environment acts during its interaction with $\mathcal M$ (what moves it makes and when),  the run incrementally spelled on the run tape is a $\pp$-won run of $A$.   
As for $\oo$'s strategies, there is no need to define them: all possible behaviors by $\oo$ are accounted for by the different possible nondeterministic updates  of the run tape of an HPM. 

In the above outline, we described HPMs in a relaxed fashion, without being specific about details  such as, say, how, exactly, moves are made by the machine, how many moves either player can make at once, what happens if both players attempt to move ``simultaneously'', etc. As it happens, all reasonable design choices yield the same class of winnable games as long as we only consider static games, including all games expressible in the language of logic $\cltw$. 
 Such games are not necessarily constant but, due to being finitary, they can and will be thought of to be constant by identifying them with their $\ada$-closures. Correspondingly, in this paper, we use the 
 term ``{\bf computational problem}'',\label{0compp} or simply ``{\bf problem}'', as a synonym of ``constant static game''.

While design choices are largely unimportant and ``negotiable'', we still want to agree on some technical details for clarity.   
Just like an ordinary Turing machine, an HPM has a finite set of {\bf states},\label{0state} one of which has the special status of being the {\bf start state}.\label{0startstate} There are no accept, reject, or halt states, but there are specially designated states called  {\bf move states}.\label{0movestate} For simplicity, we do not allow the start state to (also) be a move state.  
Each tape of the machine has a beginning but no end, and is divided into infinitely many {\bf cells},\label{0cell} arranged in the left-to-right order. At any time, each cell  contains one symbol from a certain fixed finite set of {\bf tape symbols}.\label{0tapesymbol} The {\bf blank} symbol\label{0blank} \blank,  as well as $\pp$ and $\oo$, are among the tape symbols. 
We also assume that these three symbols  are not among the symbols that the body of any (legal or illegal) move can ever contain, i.e. not among the symbols of the keyboard alphabet of which, as we agreed in Section \ref{cg}, all moves are composed.    
Each tape has its own {\bf scanning head},\label{0sch} at any given time looking (located) at one of the cells of the tape. 

For technical purposes, we additionally assume the (physical or imaginary) presence of a {\bf buffer}.\label{0buffer}\footnote{The earlier HPM models did not include buffers. The reason for a modification in the present style will be explained in Section \ref{s7}.}
The size of the latter is unlimited and, at any time, it contains some (possibly empty) finite string over the keyboard alphabet.
The function of the buffer is to let the machine construct a (``large'') move piece-by-piece before officially making such a move.

A transition from one {\bf computation step}\label{0cc}  (``{\bf clock cycle}'', ``{\bf time}'')   to another happens according to the fixed {\bf transition function}\label{0tf} of the machine. The latter, depending on the current state and the symbols seen by the  scanning heads on the corresponding tapes, deterministically prescribes: (1) the next state that the machine should assume; (2) the tape symbol by which the old symbol should be overwritten in the current cell   (the cell currently scanned by the  head), for each work tape individually; (3) the (finite, possibly empty) string over the keyboard alphabet that should be appended to the content of the buffer;
%\footnote{Unless the next state is a move state, in which case, as will be seen shortly, the above string is directly appended to the move that the machine is making, without being first put into the buffer.} 
and (4)  the (not necessarily the same) direction --- stay put, one cell to the left, or one cell to the right --- in which each  scanning head should move. It is stipulated that  when the head of a tape is looking at the first (leftmost) cell, an attempt to move to the left results in staying put. The same happens when the head tries to move to the right while looking at a cell containing  \blank. We additionally assume that \blank\ can never be written\footnote{I.e., \blank\ can never replace a non-blank symbol.} by the machine on its work tapes.  
 Note that, in view of the above conditions, at any time, all cells of a work tape found to the right of a blank cell are also blank. The same holds for the run tape, of course. 
%For simplicity we further require that, whenever the machine transitions to a move state, the string added to the buffer on that transition should be empty. Finally, we assume that the start state is not a move state.  

When the machine starts working, it is in its start state, all scanning heads are looking at the leftmost cells of the corresponding tapes, all work tapes are blank (i.e., all their cells contain \blank), the run tape does not contain any $\pp$-labeled moves (but it may contain some $\oo$-labeled moves, signifying that Environment has made those moves ``at the very beginning''), and the buffer is empty.  
Whenever 
the machine enters a move state, the move $\alpha$  written in the buffer by that time, in the $\pp$-prefixed form $\pp\alpha$,   is automatically appended  to the contents of the run tape, and the buffer is simultaneously  emptied. Here we assume that $\alpha$ includes (as a suffix) the string $\gamma$ that the machine tried to put into the buffer on that transition. For terminological and conceptual convenience, we think of this situation as that  $\gamma$ is actually written into the buffer before it, immediately after that (on the same transition), migrates from the buffer to the run tape together with the rest of $\alpha$.  
Also, on every transition,  any finite sequence $\oo\beta_1,\ldots,\oo\beta_m$ of $\oo$-labeled moves may be nondeterministically appended to the contents of the run tape. If the above two events happen on the same clock cycle, then both $\pp\alpha$ and  $\oo\beta_1,\ldots,\oo\beta_m$ will be appended to the contents of the run tape, where $\pp\alpha$ can (nondeterministically) go before, after or anywhere in the middle of $\oo\beta_1\ldots\oo\beta_m$. Whenever we say that a move (by either player) was {\bf made} {\em at time $c$}, or {\em on the transition to} step $c$, we mean that $c$ is the clock cycle on which that move (first) appeared on the run tape. We agree that the count of clock cycles starts from $0$ rather than $1$.

In the future, when describing the work of a machine, we may use the jargon ``{\bf retire}''.\label{x16} What will be meant by retiring is going into an infinite loop that makes no moves, puts nothing into the buffer, and does not reposition the scanning heads.  Retiring thus achieves the same effect as halting would achieve if this was technically allowed.

A {\bf configuration}\label{0configuration} is a full description of the situation in the machine at some given computation step. It consists of records of the (``current'') contents of the work and run tapes, the content of the buffer, the location of each 
scanning head, and the state of the machine. 
A {\bf computation branch}\label{0cb} of the machine is an infinite sequence $C_0,C_1,C_2,\ldots$ of configurations, where $C_0$ is an {\bf initial configuration}\label{0initial configuration} (one described at the beginning of the preceding paragraph), and every $C_{i+ 1}$ is a configuration that could have legally followed (again,  in the sense explained earlier) $C_i$ according to the transition function of the machine. In less formal contexts, we may say ``{\bf play}''\label{0play} instead of ``computation branch''.  
For a computation branch $B$, the {\bf run spelled by $B$}\label{0run spelled by a computation branch} is the run $\Gamma$ incrementally spelled on the run tape in the corresponding scenario of interaction. 
We say that such a $\Gamma$ is {\bf a run generated by} \label{0rgb} the machine.

We say that a given HPM $\mathcal M$ {\bf wins}\label{0win} ({\bf computes},\label{0cg} {\bf solves}) a given constant game $A$  --- and write ${\mathcal M}\models A$\label{0models} --- iff  every run generated by $\mathcal M$  is a $\pp$-won run of $A$. We say that $A$ is {\bf computable}\label{0computable} iff there is an HPM $\mathcal M$ with ${\mathcal M}\models A$; such an HPM is said to be an (algorithmic) {\bf solution},\label{0sol} or {\bf winning strategy}, for $A$.  

\section{Interactive complexity}\label{s7}
%\marginpar{s7}

At present, the theory of interactive computation is far from being well developed, and even more so is the corresponding complexity theory. The studies of interactive computation in the context of complexity, while having been going on since long ago, have been relatively scattered and ad hoc:  more often than not,  interaction 
has  been used for better understanding certain complexity issues for traditional, non-interactive problems  rather than being treated as an object of systematic studies 
in its own rights (examples would be alternating computation \cite{Chandra}, or   interactive proof systems and Arthur-Merlin games \cite{Goldwasser,Babai}). 
 As if complexity theory was not ``complex'' enough already, taking it to the interactive level would most certainly generate a by an order of magnitude greater diversity of species from the complexity zoo. The paper \cite{lbcs} made the first 
 attempt to bring complexity concepts into CoL.  The present paper significantly refines that attempt. The main novelty, as mentioned in Section \ref{intr}, is throwing amplitude complexity (see below) into the mix. It should be however pointed out that our way of measuring the complexity of  strategies is merely one out of a huge and interesting potential variety of complexity measures meaningful and useful in the interactive context.

For the following two definitions, we assume the presence of some fixed function that sends every move $\alpha$ to a natural number called the {\em magnitude}\label{0magnitudenew} of $\alpha$. 

\begin{defi}\label{newdef}
%\marginpar{newdef}
 In the context of a given computation branch (play) of a given HPM $\mathcal M$:
\begin{enumerate}
\item By the 
{\bf background}\label{0background} of a clock cycle  $c$ we mean the greatest of the magnitudes of Environment's moves made before time $c$, or is $0$ if there are no such moves. 

\item By  the {\bf timecost}\label{0timecost} of a cycle $c$ we mean  $c- d$, where $d$ is the greatest cycle with $d< c$ on which a move was made by Environment, or is $0$ if there is no such cycle. 

\item By the {\bf spacecost}\label{0spacecost} of a  cycle $c$ we mean the maximum number of cells in which $\mathcal M$ has ever  written anything on any (any {\em one}) work tape before time $c$.\footnote{Here, of course, ``written anything'' means ``written anything new'', i.e., ``overwritten an old symbol by a new one'', for, technically, not writing simply means writing/repeating the old symbol of the cell. Note that we only look at the ``worst-case'' work tape, deeming the numbers of cells written on all other   (if any) work tapes irrelevant. Also note that we only count the numbers of {\em cells} ever written rather the numbers of {\em writes} --- that is, we only count the first write in a given cell, with all subsequent writes contributing nothing extra to the spacecost. Further note that the number of cells ever {\em visited} (rather than {\em written}) by the scanning head of the worst-case work tape of $\mathcal M$ before time $c$ is either the same as the spacecost of $c$, or exceeds the latter by $1$. This is so because, as we agreed, a scanning head can never move to the right while looking at a blank cell. Finally, note that the spacecost of the first step/cycle of a computation branch is always $0$, and the spacecost of every subsequent step is nothing but the number of non-blank cells (which form a contiguous block) on the worst-case work tape of $\mathcal M$ at the preceding step.}

\item If $\mathcal M$ makes a move on a cycle $c$, then the {\bf background} (resp. {\bf timecost}, resp. {\bf spacecost}) of that move\footnote{As easily understood, here and later in similar contexts, including Definition \ref{deftcs}, a ``move'' means a move not (just) as a {\em string}, but (also) as an {\em event}, namely, the event of $\mathcal M$ moving at time $c$.} means the background (resp. timecost, resp. spacecost) of $c$. 
\end{enumerate}
\end{defi}

\noindent Throughout this paper, an $n$-ary ($n\geq 0$)  {\bf arithmetical function}\label{0aff} means a total function from $n$-tuples of natural numbers to natural numbers. As always, ``unary'' is a synonym of ``$1$-ary''. 

Where $\mathcal M$ is an HPM and $A$ is a constant game, a {\bf $\oo$-legal play\label{0lpob} of $A$ by $\mathcal M$} means a computation branch $B$ of $\mathcal M$ such that the run spelled by $B$ is a $\oo$-legal run of $A$.

\begin{defi}\label{deftcs}
%\marginpar{deftcs}
Let $\mathcal M$ be an HPM, $h$ a unary arithmetical  function,  and $A$ a constant game.  We say that: 
\begin{enumerate}[label=\arabic*.]
\item {\bf $\mathcal M$ plays $A$ in amplitude\label{0amplitude} $h$} iff,  in every $\oo$-legal play of $A$ by $\mathcal M$, whenever   $\mathcal M$ makes a move $\alpha$, the magnitude of $\alpha$  does not exceed   $h(\ell)$,  where $\ell$ is the background of $\alpha$; 

\item {\bf $\mathcal M$ plays $A$ in space\label{0space} $h$} iff,  in every $\oo$-legal play of $A$ by $\mathcal M$,  the spacecost of any given clock cycle $c$ does not exceed  $h(\ell)$,   where $\ell$ is the background of $c$; 

\item {\bf $\mathcal M$ plays $A$ in time\label{0time} $h$} iff, in every $\oo$-legal play of $A$ by $\mathcal M$, whenever   $\mathcal M$ makes a move $\alpha$,   the timecost of $\alpha$ does not exceed  $h(\ell)$,  where $\ell$ is the background of $\alpha$.  
\end{enumerate}
\end{defi}

\noindent When  a game $A$ is fixed or clear from the context, we will usually omit a reference to it and simply say that a given machine ${\mathcal M}$ {\bf plays}, or {\bf runs}, in time (space, amplitude) $h$, or that $h$ is a {\bf bound}\label{0bound} for the time (space, amplitude) complexity of $\mathcal M$. 

Note that each clause of Definition \ref{deftcs}   only looks at $\oo$-legal plays of $A$. An intuitive reason for this arrangement is that the machine is expected to ``act  properly'' only as long as its adversary plays legally; once  Environment makes an illegal move, the (confused) machine is no longer responsible for anything.

Amplitude complexity is to keep track of (set bound on) the magnitudes of $\pp$'s moves relative to the magnitudes of $\oo$'s moves. For instance,  assume ``magnitude'' means ``size'', and consider the problem $\ada x\ade y(y=2x)$. This is about computing the function $2x$, the traditional sort of an input-output problem. A machine that solves it would play in amplitude $\lambda\ell.\ell+1$. This is because the size of $\pp$'s move (``output'') will exceed the size of $\oo$'s move (``input'') by $1$.\footnote{Unless, of course, the input is $\#0$, in which case the output is also $\#0$. But, as the other sorts of complexities, amplitude complexity is an ``at most'' kind of a measure.} 
Next, our time complexity concept can be seen to be in the spirit of what is usually called {\em response time}. The latter generally does not and should not depend on the length of the preceding interaction history. On the other hand, it is not and should not  merely be a function of the adversary's last move, either. A   similar characterization applies to our concept of space complexity. All three complexity measures are equally  meaningful whether it be in the context of ``short-lasting'' games (such as the ones represented by the formulas  of the later-defined logic $\cltw$) or  the context of games that may have ``very long'' and even infinitely long legal runs. It is worthwhile to note one substantial difference between  amplitude complexity on one hand, and  time and space complexities on the other hand:  the former, unlike the latter,  only depends on {\em what} runs the machine generates, regardless of  {\em how} those runs are generated.    

In \cite{lbcs}, amplitude complexity was not explicitly considered. Also, the definitions of space and time complexities looked at all runs generated by the machine rather than only at $\oo$-legal runs. Also, timecost was measured as the time elapsed since the last move by either player rather than only by Environment. Finally, as mentioned, \cite{lbcs}, just like all other earlier papers on CoL, only dealt with single-work-tape HPMs. 
Ignoring these minor  and (mostly) inconsequential  differences, and with magnitude understood as symbolwise length, our present time complexity concept is, in fact, no different from that introduced in \cite{lbcs}. The same is not quite the case for space complexity though. In the version of the underlying HPM model employed in \cite{lbcs} and all other earlier papers on CoL, there was no buffer, and the machine had to construct the moves that it made on its work tape. We have revised that technical detail because the old arrangement did not offer sufficient flexibility. Namely, under the approach of \cite{lbcs}, it was impossible to meaningfully talk about  cases where amplitude complexity exceeds space complexity --- for instance,   natural and frequently occurring combinations such as logarithmic space with polynomial amplitude.\footnote{By the way, the situation is similar with Turing machines in the traditional theory of computation. For instance, a ``canonical'', single-tape Turing machine is not fit for defining sublinear space complexity, for which reason such a machine is usually modified through including, in addition to the work tape, two separate tapes for input and output. Such an input tape is reminiscent of our run tape, and output tape is reminiscent of our buffer.} It should be noted that, while the concept of amplitude complexity was absent in \cite{lbcs}, due to the above-mentioned choice of the underlying HPM model, amplitude complexity was automatically bounded by space complexity there.

Definitions \ref{newdef} and \ref{deftcs} are relative to the choice of how the magnitude of a move is exactly measured. More than one reasonable choice is possible here. The most straightforward and scalable measure would be to understand the magnitude of a move $\alpha$ as the ``raw'' length of the string $\alpha$. However, we here prefer a more refined measure. Namely, we agree that:

\begin{conv}\label{decc10a} 
%\marginpar{decc10a}
By the {\bf magnitude}\label{0magnitude} of a move $\alpha$ we mean the number $m$ defined as follows:
\begin{itemize}
  \item if $\alpha$ does not contain the symbol $\#$, then $m=0$;
  \item otherwise $m$ is $|c|$\label{0|2} (the bitwise length of $c$), where $c$ is the greatest constant such that $\#c$ is a substring of $\alpha$.\footnote{Such a $c$ does exist. As we remember, the constant $0$, as a bit string,  is empty. Thus,  due to the omnipresence of the empty string, it is impossible for an occurrence of $\#$ not to be ``followed'' by {\em some}, even if empty, constant.}
\end{itemize}
\end{conv}

\noindent The reason for our preference of the above magnitude measure over the raw lengths of moves is purely technical, and otherwise it has no effect on asymptotic (or even much finer) analysis of complexity. The point is that, in games represented by formulas of the language of $\cltw$,  the only essentially varying parameter that affects move sizes is the length $|c|$ of a constant $c$ chosen for a variable $x$ in a  $\ada x E$ or $\ade x E$ component, and we want to focus on that parameter only. For instance, consider the game (represented by) $\ade x P(x)\mli (\pp\mlc \ade xP(x))$. If the environment chooses $c$ for $x$ in the antecedent, the raw size of the corresponding move $0.\#c$ will be $|c|+3$ rather than $|c|$, with $3$ being the overhead imposed by the technical prefix ``$0.\#$''. It is natural to want to ignore that overhead.  Imagine further that the machine responded to the above move by choosing the same constant $c$ for $x$ in (the second conjunct of) the consequent. The raw size $|c|+5$ of the corresponding actual  move $1.1.\#c$ happens to be  greater than that of   the adversary's move $0.\#c$, despite the fact that the machine made ``the same'' --- namely, ``non-size-increasing'' ---  choice. We do not want to let such necessary yet rather arbitrary 
 ``overhead bureaucracy'' anyhow interfere with and annoy us in our complexity analysis.

Let $A$ be a constant game, $h$ a unary arithmetical function, and $\mathcal M$ an HPM. We say that {\bf $\mathcal M$ wins} ({\bf computes}, {\bf solves}) {\bf $A$ in time $h$}, or that {\bf $\mathcal M$ is an $h$ time solution for $A$},  iff $\mathcal M$ plays $A$ in time $h$ and ${\mathcal M}\models A$. 
We say that $A$ is {\bf computable} ({\bf solvable}) {\bf in time $h$} iff it has an $h$ time solution. Similarly for ``{\bf space}'' or ``{\bf amplitude}'' instead of ``time''. 

When we say {\bf polynomial time}, it is to be understood as ``time $h$ for some polynomial function $h$''. Similarly for {\bf polynomial space}, {\bf polynomial amplitude}, {\bf logarithmic space}, etc. More generally, using the asymptotic 
``Big-O''\label{xbigo} notation, where $g$ is a unary arithmetical function, ``time (space, amplitude) $O(g)$'' should be understood as ``time (space, amplitude) $h$ for some function $h$ with $h\in O(g)$''.

\section{Language of \texorpdfstring{$\cltw$}{CL12} and its semantics}\label{ss6}
%\marginpar{ss6}

Logic $\cltw$ will be axiomatically constructed in Section \ref{ss8}.\label{0cl12b} The present section is merely devoted to its {\em language}. The building blocks of the formulas of the latter are:

\begin{itemize} 
\item {\bf Nonlogical predicate letters},\label{0predicatelettern} for which we use $p,q$  as metavariables. With each predicate letter is associated a fixed nonnegative integer called its {\bf arity}.\label{0arp} We assume that, for any $n$, there are countably infinitely many $n$-ary predicate letters.   
\item {\bf Function letters},\label{0fl} for which we use $f,g$ as metavariables. Again, each function letter comes with a fixed {\bf arity}, and  we assume that, for any $n$, there are countably infinitely many $n$-ary function letters.  
\item The binary {\bf logical predicate letter}\label{0predicateletterl} $\equals $.
\item {\bf Variables} and {\bf constants}.\label{0variable2}\label{0constant2}  These are the same as the ones fixed in \mbox{Section \ref{nncg}} --- that is, the elements of $\variables$ and $\constants$, respectively. 
\item Logical connectives and quantifiers $\twg,\tlg,\gneg,\mlc,\mld,\cla,\cle,\adc,\add,\ada,\ade$. 
\end{itemize}

\noindent {\bf Terms},\label{0term} for which in this paper we use $\tau $  as a metavariable,  are built from variables, constants and function letters in the standard way.  An {\bf atomic formula}\label{0af} is $p(\tau_1,\ldots,\tau_n)$, where $p$ is an $n$-ary predicate letter and the $\tau_i$ are terms. 
When the arity of $p$ is $0$, we write $p$ instead of $p()$. Also, we write $\tau_1\equals \tau_2$ instead of $\equals (\tau_1,\tau_2)$, and $\tau_1\notequals \tau_2$ instead of $\gneg (\tau_1\equals \tau_2)$. 
{\bf Formulas}\label{0formula} are built from atomic formulas, propositional connectives $\twg,\tlg$ (nullary), $\gneg$ (unary), $\mlc,\mld,\adc,\add$ (binary), variables and quantifiers $\cla,\cle,\ada,\ade$  in the standard way, with the exception that, officially, $\gneg$ is only allowed to be applied to atomic formulas. The definitions of {\em free} and {\em bound} occurrences of variables are standard  (with $\ada,\ade$ acting as quantifiers along with $\cla,\cle$). A formula with no free occurrences of variables is said to be {\bf closed}.\label{0closedformula}

Note that, conceptually, $\twg$ and $\tlg$ do not count as atoms. For us, atoms are formulas containing no logical operators. The formulas $\twg$ and $\tlg$ do not qualify because they {\em are} ($0$-ary) logical operators themselves.

$\gneg E$, where $E$ is not atomic, will be understood as a standard abbreviation: 
$\gneg\twg=\tlg$, $\gneg\gneg E= E$, $\gneg(A\mlc B)= \gneg A\mld \gneg B$, $\gneg \ada xE= \ade x\gneg E$, etc. And $E\mli F$ will be understood as an abbreviation of $\gneg E\mld F$.

Parentheses will often be omitted --- as we just did in the preceding paragraph --- if there is no danger of ambiguity. When omitting parentheses, we assume that $\gneg$ and the quantifiers have the highest precedence, and $\mli$ has the lowest precedence. 
An expression $E_1\mlc\ldots\mlc E_n$, where $n\geq 2$, is to be understood as $E_1\mlc(E_2\mlc (\ldots\mlc(E_{n-1}\mlc E_n)\ldots)$. Sometimes we can write this  expression for an unspecified $n\geq 0$ (rather than $n\geq 2$). Such a formula, in the case of $n= 1$, should be understood as simply $E_1$. Similarly for $\mld,\adc,\add$.  As for the case of $n=0$, the nullary $\mlc$ and $\adc$ should be understood as $\twg$ while the nullary $\mld$ and $\add$ as $\tlg$.

Sometimes a formula $F$ will be represented as $F(s_1,\ldots,s_n)$, where the $s_i$ are variables. 
When doing so, we do not necessarily mean that each  $s_i$ has a free occurrence in $F$, or that every variable occurring free in $F$ is among $s_1,\ldots,s_n$. However, it {\em will}  always be assumed (usually only implicitly) that the $s_i$ are pairwise distinct, and have no bound occurrences in $F$.  In the context set by the above representation, $F(\tau_1,\ldots,\tau_n)$ will mean the result of replacing, in $F$, each  occurrence of each $s_i$   by term $\tau_i$. When writing $F(\tau_1,\ldots,\tau_n)$, it will always be assumed (again, usually only implicitly) that the terms $\tau_1,\ldots,\tau_n$ contain no variables that have bound occurrences in $F$, so that there are no unsafe collisions of variables when doing replacements.  

Similar --- well established in the literature --- notational conventions apply to terms.

A {\bf sequent}\label{0sequent2} is an expression $E_1,\ldots,E_n\intimpl F$, where $E_1,\ldots,E_n$ ($n\geq 0$) and $F$ are formulas. Here $E_1,\ldots,E_n$ is said to be the {\bf antecedent}\label{0antecedent} of the sequent, and $F$ said to be the {\bf succedent}.\label{0succedent} 

By a {\bf free} (resp. {\bf bound}) {\bf variable} of a sequent we shall mean a variable that has a free (resp. bound) occurrence in one of the formulas of the sequent. A {\bf closed sequent} is one with no free variables.\label{0closedsequent} For safety and simplicity, throughout the rest of this paper we assume that the sets of all free and bound variables of any sequent that we ever consider --- unless strictly implied otherwise by the context ---  are disjoint.   This restriction, of course, does not yield any loss of expressive power as variables can always be renamed so as to satisfy this condition. 

An {\bf interpretation}\label{0int}\footnote{The concept of an interpretation in CoL is usually more general than the present one. Interpretations in our present sense are called  {\bf perfect}. But here we omit the word ``perfect'' as we do not consider any nonperfect interpretations, anyway.} is a function $^*$ that:
\begin{itemize}
  \item Sends the word ``$\mbox{\em Universe}$'' to a universe $\mbox{\em Universe}^*$,\label{00unhg} called the {\bf universe} of $^*$. The {\bf domain}\label{000dom} and {\bf denotation}\label{0OIKOL} of $^*$ are understood as those of $\mbox{\em Universe}^*$, and are denoted by  $\mbox{\em Domain}^*$\label{0xcs} and $\mbox{\em Denotation}^*$,\label{000den} respectively. 
\item  Sends every $n$-ary function letter $f$ to an $n$-ary function $f^*(var_1,\ldots,var_n)$ on $\mbox{\em Universe}^*$.
\item  Sends every nonlogical $n$-ary  predicate letter $p$ to an $n$-ary predicate $p^*(var_1,\ldots,var_n)$ on $\mbox{\em Universe}^*$.
% which does not depend on any variables other than $s_1,\ldots,s_n$. 
 \end{itemize}

\noindent We uniquely extend the above $^*$ to  a  mapping that sends each term $\tau$ to a function $\tau^*$, and each formula or sequent $S$ to a game $S^*$ over $\mbox{\em Universe}^*$, by stipulating that: 
\begin{enumerate}
  \item Where $c$ is a constant, $c^*$ is $\mbox{\em Denotation}^*(c)$  (with the individual $\mbox{\em Denotation}^*(c)$ understood as a function according to Convention \ref{wuding}).
\item Where $s$ is a variable, $s^*$ is $s$ (with $s$ understood as a function according to Convention \ref{wuding}). 
\item Where $f$ is an $n$-ary function letter and $\tau_1,\ldots,\tau_n$ are terms, $\bigl(f(\tau_1,\ldots,\tau_n)\bigr)^*$ is  
$f^*(\tau_{1}^{*},\ldots,\tau_{n}^{*})$.
% i.e. $f^*(\tau_{1}^{*}/var_1,\ldots,\tau_{n}^{*}/var_n)$ 
\item Where   $\tau_1$ and $\tau_2$ are terms, $(\tau_1\equals \tau_2)^*$ is $\tau_{1}^{*}\equals \tau_{2}^{*}$. 
\item Where \ $p$ \ is \ an \ $n$-ary \ nonlogical  \ predicate \ letter  \ and \ $\tau_1,\ldots,\tau_n$ \ are \ terms, $\bigl(p(\tau_1,\ldots,\tau_n)\bigr)^*$ is $p^*(\tau_{1}^{*},\ldots,\tau_{n}^{*})$.
% i.e. $p^*(\tau_{1}^{*}/var_1,\ldots,\tau_{n}^{*}/var_n)$. 
\item $^{*}$ commutes with all logical operators, seeing them as the corresponding game operations: $\tlg^*$ is $\tlg$,  $(E_1\mlc\ldots\mlc E_n)^{*}$ is $E^{*}_{1}\mlc \ldots\mlc E^{*}_n$, $(\ada x E)^{*}$ is $\ada x(E^{*})$, etc.
\item Similarly, \ \ $^*$ \ sees \ the \ sequent \ symbol \ \ $\intimpl$ \ \ as \ the \ same-name \ game \ operation:  \mbox{$(E_1,\ldots,E_n\intimpl F)^*$ is $E_{1}^{*},\ldots,E_{n}^{*}\intimpl F^{*}$.}  
\end{enumerate}

\noindent  When $O$ is a function letter, a predicate letter, a term, a formula or a sequent and $O^* =  W$, we say that $^*$ {\bf interprets}\label{0interpret} $O$ as $W$. We can also refer to such a $W$ as 
``{\bf $O$ under interpretation $^*$}''.

When \ a \ given \ formula \ is \ represented \ as \ $F(x_1,\ldots,x_n)$, \ we \ will \ typically \ write \ $F^*(x_1,\ldots,x_n)$ instead of 
$\bigl(F(x_1,\ldots,x_n)\bigr)^*$. A similar practice will be used for terms as well.

We agree that, for a sequent or formula $S$, an interpretation $^*$ and an HPM $\mathcal M$, whenever we say that $\mathcal M$ is a {\bf  solution} of $S^*$ or write ${\mathcal M}\models S^*$, we mean that   $\mathcal M$ is a  solution of the (constant) game $\ada x_1\ldots\ada x_n (S^*)$, where $x_1,\ldots,x_n$ are exactly the free variables of $S$, listed according to their lexicographic order. We call the above game the {\bf $\ada$-closure}\label{0adaclosure} of $S^*$, and denote it by 
$\ada S^*$. 

Note that, for any given sequent or formula $S$, the $\legal{}{}$ component of the game $\ada S^*$ does not depend on the interpretation $^*$. Hence we can safely 
say ``legal run of $\ada S$'' --- or even just ``legal run of $S$''\label{026} --- without indicating an interpretation applied to $S$.

We say that an HPM $\mathcal M$ is a (polynomial time, $h$ space, etc.) {\bf uniform} (or {\bf logical}) {\bf solution}\label{0uniformsolution}  of a sequent $X$ iff, for any interpretation $^*$, ${\mathcal M}$ is a (polynomial time, $h$ space, etc.) solution of $X^*$. 

Intuitively, a uniform  solution is   a ``purely logical'' solution. ``Logical'' in the sense that it does not depend on the universe and the meanings of the nonlogical symbols (predicate and function letters) ---  does not depend on a (the) interpretation $^*$, that is. It is exactly these kinds of   solutions that we are interested in when seeing CoL as a logical basis for applied theories or knowledge base systems. As a universal-utility tool, CoL (or a CoL-based compiler) would have no knowledge of the meanings of those nonlogical symbols (the meanings that will be changing from application to application and from theory to theory), other than what is explicitly  given by the target formula and the axioms or the knowledge base   of the system.

\section{Axiomatics of \texorpdfstring{$\cltw$}{CL12}}\label{ss8}
%\marginpar{ss8}

The purpose of the system $\cltw$\label{cl12c} that we deductively construct in this section is to axiomatize the set of   sequents with logical solutions. 
 Our formulation of the system relies on the terminology and notation explained below.

\begin{enumerate}
\item A {\bf surface occurrence}\label{0surface occurrence} of a subformula is an occurrence that is 
not in the scope of any choice operators ($\adc,\add,\ada$ and/or $\ade$). 
\item A formula not containing choice operators --- i.e., a formula of the language of classical first order logic --- is said to be {\bf elementary}.\label{0elformula} 
\item A sequent is {\bf elementary}\label{0elsequent} iff all of its formulas are so. 
\item The {\bf elementarization}\label{0elz} \[\elz{F}\] of a formula $F$ is the result of replacing
in $F$ all $\add$- and $\ade$-subformulas by $\tlg$, and all $\adc$- and $\ada$-subformulas by $\twg$. Note that $\elz{F}$ is (indeed) an elementary formula.
\item The {\bf elementarization}\label{0elz2} $\elz{G_1,\ldots,G_n\intimpl F}$ of a sequent 
$G_1,\ldots,G_n\intimpl F$ is the elementary formula \[\elz{G_1}\mlc\ldots\mlc \elz{G_n}\mli \elz{F}.\] 
\item A sequent  is said to be {\bf stable}\label{0stable} iff its elementarization is classically valid; otherwise it is {\bf unstable}.\label{0unstable}  By ``classical validity'', in view of G\"{o}del's completeness theorem,  we mean provability in some fixed standard version of classical first-order calculus with constants, function letters and $\equals$, where $\equals$ is treated as the logical {\em identity} predicate (so that, say, $x\equals x$, $x\equals y\mli (E(x)\mli E(y))$, etc. are provable).
\item We will be using the notation \[F[E]\label{0fe}\] to mean a formula $F$ together with some (single) fixed  surface occurrence of a subformula $E$. Using this notation sets a context, in which $F[H]$ will mean the result of replacing in $F[E]$ the (fixed) occurrence of $E$ by $H$.  Note that here we are talking about some {\em occurrence} of $E$. Only that occurrence gets replaced when moving from $F[E]$ to $F[H]$, even if the formula also had some other occurrences of $E$.
\item By a {\bf rule}\label{0rule} (of inference) in this section we mean a binary relation $\mathbb{Y}{\mathcal R} X$, where $\mathbb{Y}=\seq{Y_1,\ldots,Y_n}$ is a finite sequence of sequents and $X$ is a sequent. {\bf Instances}\label{0ruleinstance} of such a relation are schematically written as 
\[\frac{Y_1,\ldots,Y_n}{X},\] 
where $Y_1,\ldots,Y_n$ are called the {\bf premises},\label{0premise} and $X$ is  called the {\bf conclusion}.\label{0ruleconclusion} Whenever $\mathbb{Y}{\mathcal R}X$ holds, we say that $X$ {\bf follows} from $\mathbb{Y}$ by $\mathcal R$.  
\item Expressions such as $\vec{G},\vec{K},\ldots$ will usually stand for finite sequences of formulas. The standard meaning of an expression such as $\vec{G},F,\vec{K}$  should also be clear. 
\end{enumerate}

\begin{center}
\begin{picture}(100,30)

\put(0,10){\bf THE RULES OF $\cltw$}

\end{picture}
\end{center}

$\cltw$ has the six rules listed below, with the following additional conditions/explanations: 
\begin{enumerate}
\item In $\add$-Choose and $\adc$-Choose, $i\in\{0,1\}$.
\item In $\ade$-Choose and $\ada$-Choose,  $\mathfrak{t}$ is either a constant or a variable with no bound occurrences in the premise, and $H(\mathfrak{t})$ is the result of replacing by $\mathfrak{t}$ all free occurrences of $x$ in $H(x)$ (rather than vice versa).
\end{enumerate}
\begin{center}
\begin{picture}(287,70)

\put(14,50){\bf $\add$-Choose}\label{0choose}
\put(12,30){$\vec{G}\ \intimpl\  F[H_i]$}
\put(0,22){\line(1,0){78}}
\put(0,8){$\vec{G}\ \intimpl \ F[H_0\add H_1]$}

\put(232,50){\bf $\adc$-Choose}
\put(212,30){$\vec{G},\ E[H_i],\ \vec{K}\  \intimpl \ F$}
\put(200,22){\line(1,0){113}}
\put(200,8){$\vec{G},\ E[H_0\adc H_1], \ \vec{K}\ \intimpl\ F$}

\end{picture}
\end{center}

\begin{center}
\begin{picture}(287,70)

\put(231,50){\bf $\ada$-Choose}
\put(210,30){$\vec{G},\ E[H(\mathfrak{t})],\ \vec{K}\ \intimpl\ F$}
\put(200,22){\line(1,0){113}}
\put(200,8){$\vec{G},\ E[\ada xH(x)],\ \vec{K}\ \intimpl\ F$}

\put(16,50){\bf $\ade$-Choose}
\put(08,30){$\vec{G}\ \intimpl\  F[H(\mathfrak{t})]$}
\put(0,22){\line(1,0){78}}
\put(0,8){$\vec{G}\ \intimpl \ F[\ade x H(x)]$}

\end{picture}
\end{center}

\begin{center}
\begin{picture}(74,70)

\put(12,50){\bf Replicate}\label{0replicate}
\put(8,8){$\vec{G},E,\vec{K}\intimpl F$}
\put(0,22){\line(1,0){69}}
\put(0,30){$\vec{G},E,\vec{K},E\intimpl F$}
\end{picture}
\end{center}

\begin{center}
\begin{picture}(300,70)
\put(140,50){\bf Wait}\label{0wait}
\put(0,30){$Y_1,\ldots,Y_n$}
\put(0,22){\line(1,0){43}}
\put(55,20){($n\geq 0$), where all of the following five conditions are satisfied:}
\put(17,8){$X$}
\end{picture}
\end{center}

\begin{enumerate}
\item {\bf $\adc$-Condition:}  Whenever $X$ has the form $\vec{G}\intimpl F[H_0\adc H_1]$, both of the sequents $\vec{G}\intimpl F[H_0]$ and 
$\vec{G}\intimpl F[H_1]$ are among $Y_1,\ldots,Y_n$.
\item {\bf $\add$-Condition:} Whenever $X$ has the form $\vec{G},E[H_0\add H_1],\vec{K}\intimpl F$, both of the sequents $\vec{G},E[H_0],\vec{K}\intimpl F$ and 
$\vec{G},E[H_1],\vec{K}\intimpl F$ are among $Y_1,\ldots,Y_n$.
\item {\bf $\ada$-Condition:} Whenever $X$ has the form $\vec{G}\intimpl F[\ada xH(x)]$, for some variable $y$ not occurring in $X$, the sequent  $\vec{G}\intimpl F[H(y)]$ is among  $Y_1,\ldots,Y_n$. Here and below, $H(y)$ is the result of replacing by $y$ all free occurrences of $x$ in $H(x)$ (rather than vice versa).
\item {\bf $\ade$-Condition:} Whenever $X$ has the form $\vec{G},E[\ade xH(x)],\vec{K}\intimpl F$, for some variable $y$ not occurring in $X$, the sequent  $\vec{G},E[H(y)],\vec{K}\intimpl F$ is among  $Y_1,\ldots,Y_n$.
\item {\bf Stability condition:} $X$ is stable.
\end{enumerate}

As will be seen in Section \ref{ssoundness}, each rule --- seen bottom-up --- encodes an action that a winning strategy should take in a corresponding situation, and the name of each rule is suggestive of that action. For instance, Wait (indeed) prescribes the strategy to wait till the adversary moves. This explains why we have called  ``Replicate'' the rule which otherwise is nothing but what is commonly known as Contraction.   

A {\bf $\cltw$-proof} of a sequent $X$ is a sequence $X_1,\ldots,X_n$ of sequents, with $X_n=X$, such that, each $X_i$ follows  by one of the rules of $\cltw$ from some (possibly empty in the case of Wait, and certainly empty in the case of $i=1$) set $\mathcal P$ of premises such that ${\mathcal P}\subseteq \{X_1,\ldots, X_{i-1}\}$.
When a $\cltw$-proof of $X$ exists, we say that $X$ is {\bf provable} in $\cltw$, and write $\cltw\vdash X$.\label{0stopor}

   A {\bf $\cltw$-proof} of a formula $F$ will be understood as a  $\cltw$-proof of the empty-antecedent sequent $\intimpl F$. Accordingly, $\cltw\vdash F$ means $\cltw\vdash\intimpl F$.

\begin{fact}\label{fce}
%\marginpar{fce}
$\cltw$ is a conservative extension of classical logic. That is, an elementary sequent {\em $E_1,\ldots,E_n\intimpl F$} is provable in $\cltw$ iff the formula $E_1\mlc\ldots\mlc E_n\mli F$ is valid in the classical sense.
\end{fact}

\begin{proof} Assume $E_1,\ldots,E_n,F$ are elementary formulas. If $E_1\mlc\ldots\mlc E_n\mli F$ is classically valid, then $E_1,\ldots,E_n\intimpl F$ follows   from the empty set of premises by Wait. And if $E_1\mlc\ldots\mlc E_n\mli F$ is not classically valid, then $E_1,\ldots,E_n\intimpl F$  cannot be the conclusion of any of the rules of $\cltw$ except Replicate. However, applying (bottom-up) Replicate does not take us any closer to finding a proof of the sequent, as the premise still remains an unstable elementary sequent.  
\end{proof}

$\cltw$ can also be seen to be a conservative extension of the earlier known logic {\bf CL3} studied in \cite{Japtcs}.\footnote{Essentially the same logic, called {\bf L}, was in fact known as early as in \cite{Jap02}.} The latter is nothing but the empty-antecedent fragment of $\cltw$ without function letters and identity. 

\begin{exa}\label{ecube} In this example, $\mult$ is a binary function letter and $^3$ is a unary function letter. We write $x\mult y$ and $x^3$ instead of $\mult(x,y)$ and $^3(x)$, respectively. The following sequence of sequents is a $\cltw$-proof of the sequent (\ref{mar7}) from Example \ref{newexample}. It may be worth observing that the strategy used by $\top$ in that example, in a sense, ``follows'' our present proof step-by-step in the bottom-up direction. And this is no accident: as we are going to see in the course of proving the soundness of $\cltw$, every $\cltw$-proof rather directly encodes a winning strategy.\vspace{7pt}

\noindent 1. $\cla x \bigl(x^3\equals (x\mult x)\mult x\bigr),\    t\equals s\mult s, \  r\equals t\mult s \ \intimpl \  r\equals s^3$  \ \ {Wait: (no premises)} \vspace{3pt}

\noindent 2. $\cla x \bigl(x^3\equals (x\mult x)\mult x\bigr),\    t\equals s\mult s, \  r\equals t\mult s \ \intimpl \  \ade y(y\equals s^3)$  \ \   {$\ade$-Choose: 1}\vspace{3pt}

\noindent 3. $\cla x \bigl(x^3\equals (x\mult x)\mult x\bigr),\    t\equals s\mult s, \  \ade z (z\equals t\mult s) \ \intimpl \  \ade y(y\equals s^3)$  \ \ {Wait: 2} \vspace{3pt}

\noindent 4. $\cla x \bigl(x^3\equals (x\mult x)\mult x\bigr),\    t\equals s\mult s, \ \ada y \ade z (z\equals t\mult y) \ \intimpl \  \ade y(y\equals s^3)$  \ \ {$\ada$-Choose: 3}\vspace{3pt}

\noindent 5. $\cla x \bigl(x^3\equals (x\mult x)\mult x\bigr),\    t\equals s\mult s, \ \ada x\ada y \ade z (z\equals x\mult y) \ \intimpl \  \ade y(y\equals s^3)$  \ \ {$\ada$-Choose: 4}\vspace{3pt}

\noindent 6. $\cla x \bigl(x^3\equals (x\mult x)\mult x\bigr),\    \ade z (z\equals s\mult s), \ \ada x\ada y \ade z (z\equals x\mult y) \ \intimpl \  \ade y(y\equals s^3)$  \ \ {Wait: 5}\vspace{3pt}

\noindent 7. $\cla x \bigl(x^3\equals (x\mult x)\mult x\bigr),\  \ada y \ade z (z\equals s\mult y), \ \ada x\ada y \ade z (z\equals x\mult y) \ \intimpl \  \ade y(y\equals s^3)$  \ \ { $\ada$-Choose: 6}\vspace{3pt}

\noindent 8.  $\cla x \bigl(x^3\equals (x\mult x)\mult x\bigr),\ \ada x\ada y \ade z (z\equals x\mult y), \ \ada x\ada y \ade z (z\equals x\mult y) \ \intimpl \  \ade y(y\equals s^3)$ \ \ { $\ada$-Choose: 7}\vspace{3pt}

\noindent 9.  $\cla x \bigl(x^3\equals (x\mult x)\mult x\bigr),\ \ada x\ada y \ade z (z\equals x\mult y) \ \intimpl \ \ade y(y\equals s^3)$ \ \ {Replicate: 8}\vspace{3pt}

\noindent 10. $\cla x \bigl(x^3\equals (x\mult x)\mult x\bigr),\ \ada x\ada y \ade z (z\equals x\mult y) \ \intimpl \ \ada x\ade y(y\equals x^3)$ \ \ { Wait: 9}
\end{exa}

\begin{exa}\label{j28a}
%\marginpar{j28a}
The \ formula \ $\cla x\hspace{1pt}p(x)\mli\ada x\hspace{1pt}p(x)$ \ is \ provable \ in \ $\cltw$. \ It \ follows  \ from $\cla x\hspace{1pt}p(x)\mli p(y)$ by Wait. The latter, in turn, follows by Wait from the empty set of premises. 

On the other hand, the formula $\ada x\hspace{1pt}p(x)\mli\cla x\hspace{1pt}p(x)$, i.e. $\ade x\gneg p(x)\mld \cla x\hspace{1pt}p(x)$, in not provable. Indeed, its elementarization is $\tlg\mld \cla x\hspace{1pt}p(x)$, which is not classically valid.  Hence $\ade x\gneg p(x)\mld \cla x\hspace{1pt}p(x)$ cannot be derived by Wait. Replicate can also be dismissed for obvious reasons. This leaves us with $\ade$-Choose. But if $\ade x\gneg p(x)\mld \cla x\hspace{1pt}p(x)$ is derived  by $\ade$-Choose, then the premise should be $\gneg p(\mathfrak{t})\mld \cla x\hspace{1pt}p(x)$ for some variable or constant $\mathfrak{t}$. The latter, however, is a classically non-valid elementary formula and hence, by Fact \ref{fce}, is not provable. 
\end{exa}

\begin{exa}\label{j28b}
%\marginpar{j28b} 
The formula $\ada x\ade y\bigl(p(x)\mli p(y)\bigr)$ is provable in $\cltw$ as follows:\vspace{7pt}

\noindent 1. $\begin{array}{l}
p(s)\mli p(s)
\end{array}$  \ \ Wait:\vspace{3pt}

\noindent 2. $\begin{array}{l}
\ade y\bigl(p(s)\mli p(y)\bigr)
\end{array}$  \ \ $\ade$-Choose: 1\vspace{3pt}

\noindent 3. $\begin{array}{l}
\ada x\ade y\bigl(p(x)\mli p(y)\bigr)
\end{array}$  \ \ Wait: 2\vspace{7pt}

On the other hand, the formula $\ade y\ada x \bigl(p(x)\mli p(y)\bigr)$ can be seen to be unprovable, even though its classical counterpart $\cle y\cla x \bigl(p(x)\mli p(y)\bigr)$ is a classically valid elementary formula and hence provable in $\cltw$. 
\end{exa}

\begin{exa}\label{j28c}
%\marginpar{j28c} 
While the formula $\cla x\cle y \bigl(y\equals f(x)\bigr)$  is classically valid and hence provable in $\cltw$, its constructive counterpart 
$\ada x\ade y \bigl(y\equals f(x)\bigr)$ can be easily seen to  be unprovable. This is no surprise. In view of the expected soundness of $\cltw$,  provability  of $\ada x\ade y \bigl(y\equals f(x)\bigr)$ would imply that every function $f$ is computable (furthermore, computable in a ``uniform'' way), which, of course, is not the case.  
\end{exa}

\begin{exercise}\label{feb1a}{\em 
%\marginpar{feb1a}
To see the resource-consciousness of $\cltw$, show that it does not prove $p\adc q\mli (p\adc q)\mlc (p\adc q)$, even though this formula has the form $F\mli F\mlc F$ of a classical tautology. Then show that, in contrast, $\cltw$ proves the {\em sequent} $p\adc q\intimpl (p\adc q)\mlc (p\adc q)$ because, unlike the antecedent of a $\mli$-combination, the antecedent of a $\intimpl$-combination is reusable (trough Replicate).
} 
\end{exercise}

\begin{exercise}\label{feb1ae}{\em 
%\marginpar{feb1ae}
Show that $\cltw\vdash \ade x\ada y\hspace{2pt} p(x,y)\intimpl \ade x\bigl(\ada y\hspace{2pt}p(x,y)\mlc \ada y\hspace{2pt}p(x,y)\bigr)$. Then observe that, on the other hand,  $\cltw$ does not prove any of the formulas 
\[\begin{array}{rcl}
\ade x\ada y\hspace{2pt} p(x,y) & \mli & \ade x\bigl(\ada y\hspace{2pt}p(x,y)\mlc \ada y\hspace{2pt}p(x,y)\bigr);\\
\ade x\ada y\hspace{2pt} p(x,y)\ \mlc \ \ade x\ada y\hspace{2pt} p(x,y) & \mli & \ade x\bigl(\ada y\hspace{2pt}p(x,y)\mlc \ada y\hspace{2pt}p(x,y)\bigr);\\
\ade x\ada y\hspace{2pt} p(x,y)\ \mlc\ \ade x\ada y\hspace{2pt} p(x,y)\ \mlc\ \ade x\ada y\hspace{2pt} p(x,y) & \mli & \ade x\bigl(\ada y\hspace{2pt}p(x,y)\mlc \ada y\hspace{2pt}p(x,y)\bigr);\\
 & \ldots & 
\end{array}\]
Intuitively, this contrast is due to the fact that, even though  both $\st A$ and $\pst A = A\mlc A\mlc\ldots$\label{0pst} are resources allowing to reuse $A$ any number of times, the ``branching'' form of reusage offered by $\st A$ is substantially stronger than the ``parallel'' form of reusage offered by $\pst A$.  $\st \ade x\ada y\hspace{2pt} p(x,y)\mli \ade x\bigl(\ada y\hspace{2pt}p(x,y)\mlc \ada y\hspace{2pt}p(x,y)\bigr)$  is  a  valid  principle  of CoL while $\pst \ade x\ada y\hspace{2pt} p(x,y)$ $\mli \ade x\bigl(\ada y\hspace{2pt}p(x,y)\mlc \ada y\hspace{2pt}p(x,y)\bigr)$ is not.  
}
\end{exercise}

\section{Soundness and completeness of \texorpdfstring{$\cltw$}{CL12}}\label{ssoundness}
%\marginpar{ssoundness}

%\begin{conv}\label{ell}
%\marginpar{ell}
Throughout this paper,  the letter $\ell$\label{0ell} is used as a variable for the {\em background parameter}\label{0bp}  in terms of which complexity is measured. When writing $f(\ell)$, for whatever arithmetical function $f$,
in some cases  it is to be understood as the quantity $f(\ell)$, while in some other cases   
 as the function $\lambda\ell.f(\ell)$. We typically do not explicitly use the lambda operator, and whether $f(\ell)$ is meant to stand for the function $\lambda \ell.f(\ell)$ or for the value of that function at the argument $\ell$  should be determined based on the context.  
%\end{conv}

 We say that a computation branch (play) of a given HPM $\mathcal M$  is 
{\bf provident}\label{0prvb} iff there are infinitely many configurations in it with empty buffer contents.  Intuitively, this means that there are no ``unfinished moves'': every (nonempty) move that $\mathcal M$ has started to construct in its buffer was eventually made in the play. 

We define the {\bf native magnitude}\label{0nm} of a sequent $X$ as the smallest integer $\mathfrak{c}\geq 0$ such that  $X$ contains no constant $c$ with $|c|>\mathfrak{c}$. Thus, if $X$ contains no constants, or no constants other than $0$, its native magnitude is $0$ (remember from Section \ref{nncg} that, when denoting a constant, ``$0$'' stands for the empty bit string $\epsilon$ and hence its size is $0$ rather than $1$). 

 By a {\bf minimal amplitude \label{0minamp} logical solution} of a sequent $X$ we shall mean a  $\max\{\ell,\mathfrak{c}\}$ amplitude  logical solution of $X$, where $\mathfrak{c}$ is the native magnitude of $X$.    Of course, $\max\{\ell,\mathfrak{c}\}=O(\ell)$,  simply meaning linear amplitude. However, unlike  space or time complexities, we may not always be willing to settle for merely asymptotic analysis when it comes to amplitude complexity. 

\begin{defi}\label{wb}
%\marginpar{wb}
We say that a logical solution $\mathcal M$ of a sequent $X$  is {\bf well behaved}\label{0wbs}  iff  the following five conditions are satisfied:
\begin{enumerate} 
\item There is an integer $\mathfrak{r}$ such that, in every $\oo$-legal play of $X$\hspace{1pt}\footnote{Remember our convention (page \pageref{026}), according to which we may unambiguously say ``legal play/run of $X$'' even if, strictly speaking,  $X$ is not a game, but rather  turns into one only after some interpretation $^*$ is applied to it.} by $\mathcal M$, at most $\mathfrak{r}$ replicative moves are made in the antecedent. 
\item There is an integer $\mathfrak{k}$ such that, in every $\oo$-legal play of $X$ by $\mathcal M$,  the run-tape head of $\mathcal M$ revisits any given labmove $\xx\alpha$  at most $\mathfrak{k}$ times. Here by {\bf revisiting}\label{0revisit} such a $\xx\alpha$ we mean entering the cell containing the last symbol of this string in the right-to-left direction; that is, coming back to the labmove after it has already been fully scanned.  
\item In every $\oo$-legal play of $X$ by $\mathcal M$,  every move that $\mathcal M$ makes in one of the components of the antecedent of $X$ is focused. 
\item Every $\oo$-legal play of $X$ by $\mathcal M$ is provident.
\item $\mathcal M$ has only one work tape.
\end{enumerate}
\end{defi}

\begin{thm}\label{feb9c}
%\marginpar{feb9c}
For any sequent $X$,
the following conditions are equivalent: 
\begin{enumerate}[label=(\roman*)]
\item $\cltw\vdash X$.
\item  $X$ has a logical solution.
\item $X$ has a well behaved, minimal amplitude, constant space and linear time logical solution $\mathcal M$. 
\end{enumerate}
Furthermore, the implication $(i)\Rightarrow (iii)$ holds in the strong sense that there is an effective  procedure 
that takes an arbitrary $\cltw$-proof $\mathbb{P}$ of an arbitrary sequent $X$ and constructs an HPM \ ${\mathcal M}$ satisfying   $(iii)$.    
\end{thm}

\begin{rem}\label{efrem}
%\marginpar{efrem}
In the ``Furthermore'' clause of the above theorem, ``effective'' can be strengthened to ``efficient''. Indeed, let  ${\mathcal M}'$ be an HPM that, using the procedure whose existence is claimed in the ``Furthermore'' clause, constructs $\mathcal M$ and then, simulating $\mathcal M$, plays just as $\mathcal M$ would play. It can be seen with some thought that such an ${\mathcal M}'$ (as opposed to $\mathcal M$ itself) is constructed from $\mathbb{P}$ in linear time. It inherits, from $\mathcal M$, the property of being a well-behaved, minimal amplitude logical solution of $X$. With properly arranged details of simulation,  the space and time complexities of ${\mathcal M}'$, asymptotically, can also be easily achieved to be the same as those of ${\mathcal M}$.  Thus ${\mathcal M}'$, just like $\mathcal M$, satisfies condition of $(iii)$ of the theorem. 

A similar comment applies to the later-stated Theorems \ref{feb9dt} and \ref{feb9ds}. 
\end{rem}

We begin our proof of Theorem \ref{feb9c} by noting that the implication $(iii)\Rightarrow (ii)$ is trivial. The implication $(ii)\Rightarrow (i)$ was proven in \cite{lbcs}, and that proof is reproduced in Appendix \ref{scompleteness}  of this paper.\footnote{As mentioned, the underlying HPM model dealt with in \cite{lbcs} was not exactly the same as the present one; yet the above proof goes through in either case without a need for any readjustments whatsoever.} So, here we only need to verify the implication $(i)\Rightarrow (iii)$, in the strong sense of the ``Furthermore'' clause of the theorem. This is what the entire rest of the present section is devoted to, for the exception of its last subsection.  

Our proof of $(i)\Rightarrow (iii)$ proceeds by induction on the length of (the number of sequents involved in) a $\cltw$-proof of $X$ and, as such, is nothing but a combination of six cases, corresponding to the six rules of $\cltw$ by which the final sequent $X$ could have been derived from its premises (if any).

In each case, our efforts will be focused on showing how to construct a single-work-tape  HPM $\mathcal M$ --- a  logical solution of the conclusion --- from an arbitrary instance 
\[\frac{Y_1,\ldots,Y_n}{X}\] ($n\geq 0$) of the rule and arbitrary HPMs ${\mathcal N}_1,\ldots,{\mathcal N}_n$ ---  well behaved, minimal amplitude, constant space and linear time logical solutions of the premises, that exist according to the induction hypothesis.  It will be immediately clear from our description of $\mathcal M$  that it is constructed effectively (thus taking care of the ``Furthermore'' claim of the theorem), and that the work of $\mathcal M$ in no way depends on an interpretation $^*$ applied to the sequents involved, so that the solution is logical. 

 Since an interpretation $^*$ is typically irrelevant in such proofs, we will usually omit mentioning it and write simply $S$ where, strictly speaking, $S^*$ is meant. That is, we identify formulas or sequents with the games into which they turn once an interpretation is applied to them. Accordingly, in  contexts  where $S^*$ has to be understood as $\ada S^*$ anyway (e.g., when talking about computability of $S^*$), we may further omit ``$\ada $''  and write $S$ instead of $\ada S$. 

In all cases, whether it be the description of $\mathcal M$'s work or our further analysis of its play of the target sequent $X$, we shall implicitly rely on what in \cite{Japtcs,Japfin} is called the {\bf Clean Environment Assumption},\label{0cea} according to which $\mathcal M$'s  adversary  never makes illegal moves. That is,  all plays of $X$ by $\mathcal M$ are $\oo$-legal. Such an assumption is legitimate because, once Environment makes an illegal move in a given play, $\mathcal M$ is an automatic winner regardless of how it works afterwards, trivially satisfying the required or relevant well-behavedness and complexity conditions because the latter only look at $\oo$-legal runs.

\subsection{\texorpdfstring{$\add$}{sup}-Choose}\label{sep10aa} 
\[\frac{\vec{G}\ \intimpl\  F[H_i]}{\vec{G}\ \intimpl \ F[H_0\add H_1]}\]

\noindent Assume (induction hypothesis)  that  $\mathcal N$ is a well behaved, minimal amplitude, constant space and linear time logical solution of the premise $\vec{G} \intimpl  F[H_i]$ ($i\in\{0,1\}$).  We want to construct  a well behaved, minimal amplitude, constant space and linear time logical solution $\mathcal M$ of the conclusion $\vec{G} \intimpl  F[H_0\add H_1]$. 

 The basic idea for the strategy followed by $\mathcal M$ is very simple: $\add$-Choose 
 most directly encodes an action that $\mathcal M$ should perform in order to successfully solve the conclusion.  Namely, $\mathcal M$ should choose $H_i$ in the $H_0\add H_1$ component  and then continue playing like $\mathcal N$. $\mathcal M$ wins because the above initial move --- call it $\beta_{initial}$ --- brings the conclusion down to the premise, and $\mathcal N$ wins the latter. 

In more detail, this is how $\mathcal M$ works. Let $k$ be the number of free variables in $\vec{G} \intimpl  F[H_0\add H_1]$. 
At the beginning of the play, $\mathcal M$ waits till Environment chooses constants  for all of those free variables. That is, until $k$  $\oo$-labeled moves appear on the run tape. If Environment never makes $k$ moves, $\mathcal M$ wins. Besides, in this case, $\mathcal M$ will not be billed for any time, because it did not make any moves. Nor will it be billed for any space, because it does not write anything on the work tape while waiting. Suppose now Environment does make $k$ moves. Then $\mathcal M$ goes ahead and makes the above-mentioned move $\beta_{initial}$ signifying selecting $H_i$. By this time, $\mathcal M$'s space consumption remains $0$ as in the preceding case. As for the time bill, with a moment's thought we can see that it will be at most proportional to $k\ell$, where $\ell$ is the background of $\beta_{initial}$.\footnote{Note that, if Environment was slow in making moves, $\mathcal M$ is not billed for waiting, because $\mathcal M$'s time counter is reset to $0$ every time Environment moves; thus, $\mathcal M$'s time bill will be the biggest in the scenario where Environment made all $k$ moves at once, right during the first clock cycle. In this case $\mathcal M$ will have to read all $k$ moves, plus spend a constant amount of time on making the move $\beta_{initial}$.} The time consumption so far is linear, that is.

After $\mathcal M$ chooses $H_i$, the main part of its work consists in simulating $\mathcal N$ forever. To do this, at any time $\mathcal M$ needs to remember the ``current'' configuration of $\mathcal N$, which at the beginning is $\mathcal N$'s start configuration in the scenario where $\mathcal N$'s adversary has made all of the initial $k$ moves right at the beginning of the play, and which will be continuously updated to the ``next configuration" by the simulation routine. Maintaining this information requires remembering $\mathcal N$'s ``current'' state, work-tape content, work-tape head location, run-tape content, run-tape head location, and the content of $\mathcal N$'s buffer. The first piece of this information --- $\mathcal N$'s state ---
$\mathcal M$ remembers through its own state memory without utilizing any work-tape space. Next, $\mathcal M$ remembers $\mathcal N$'s work-tape content and work-tape head location by fully synchronizing those with its own work-tape content and work-tape head location. That is, at any time, the work-tape content of $\mathcal M$ is an exact copy of that of $\mathcal N$, and the work-tape head location of $\mathcal M$ also coincides with that of $\mathcal N$.\footnote{In fact, since $\mathcal N$ runs in constant space, $\mathcal M$ could remember these two pieces of information (just like the information on $\mathcal N$'s state) in its state memory rather than on its work tape.}  
There is no need for $\mathcal M$ to remember $\mathcal N$'s run-tape content (which would require linear rather than constant space) on its work tape. That is because all information about the content of the imaginary run tape of $\mathcal N$ resides on $\mathcal M$'s own run tape. $\mathcal M$ does not need to remember on its work tape $\mathcal N$'s run-tape head location either (which otherwise would require a logarithmic rather than constant amount of space). Rather, just as in the case of the work-tape head location, $\mathcal M$ keeps track of this information through correspondingly positioning its own run-tape head, usually (with an exception discussed below) letting it scan the same symbol as $\mathcal N$'s imaginary run-tape head is scanning. Finally,  $\mathcal M$ keeps  the content of its buffer identical to that of $\mathcal N$'s imaginary buffer, and thus consumes no work-tape space on remembering the corresponding piece of information. Of course, in addition to remembering the ``current'' configuration of $\mathcal N$,  $\mathcal M$ also needs to maintain (the never-changing) full information on $\mathcal N$'s transition function. But, again, there is no need to keep this information on the work tape --- it is understood that, instead, it is simply ``hardwired'' into $\mathcal M$'s finite control.

The purpose of 
simulation is to play as $\mathcal N$ would play. Namely, every time $\mathcal N$ goes from one configuration to another,\footnote{And, of course, this happens ``all the time''.} $\mathcal M$ correspondingly updates its representation of $\mathcal N$'s ``current'' configuration using the built-in knowledge of $\mathcal N$'s 
transition function. Among other things, this means that, every time $\mathcal N$ adds some nonempty string to the content of its buffer, $\mathcal M$ puts  the same string into its own buffer. Every time $\mathcal N$ makes a move (enters a move state), $\mathcal M$ does the same. And every time $\mathcal N$ looks up a symbol on its run tape, $\mathcal M$ looks up the same symbol on its own run tape and feeds the corresponding information back to the simulation routine.  Since the magnitude of $\beta_{initial}$ is $0$ and all other moves that $\mathcal M$ makes are also made in the corresponding scenario by $\mathcal N$, $\mathcal M$ runs in precisely the same amplitude as $\mathcal N$ does. Next, note that $\mathcal M$ utilizes  the same  amount  of work-tape space as $\mathcal N$ does. And the time that $\mathcal M$ consumes is also the same (or, rather, ``essentially the same'' as we are going to see below) as the time consumed by $\mathcal N$. So, $\mathcal M$ inherits $\mathcal N$'s constant space and linear time complexities. Observe that the native magnitude $\mathfrak{c}$ of the premise 
$\vec{G} \intimpl  F[H_i]$ 
does not exceed the native magnitude of the conclusion $\vec{G} \intimpl  F[H_0\add H_1]$. Hence, the fact that $\mathcal M$ runs in the same amplitude $\max\{\ell,\mathfrak{c}\}$ as 
$\mathcal N$ does additionally implies that, as desired, $\mathcal M$ is a minimal amplitude logical solution of $\vec{G} \intimpl  F[H_0\add H_1]$.

One detail of simulation suppressed so far requires further attention. In the above explanation we pretended that the content of $\mathcal M$'s run tape (more precisely, of its so-far-scanned portion) was the same as that of the simulated $\mathcal N$'s imaginary run tape. This, however, is not exactly so. The difference between the two contents is that $\mathcal M$'s run tape will have $\pp\beta_{initial}$ as an additional labmove. This is not a serious problem though.  Let us call the portion of $\mathcal M$'s run tape on which $\pp\beta_{initial}$ is written the {\em special zone}. $\mathcal M$ handles the complication caused by the presence of $\pp\beta_{initial}$ by remembering (through its state memory), at any time, the integer $D$ indicating ``how many labmoves away   its run-tape head is from the special zone''. Namely, if the run-tape head is scanning a cell in the special zone, then $D=0$; if it is scanning the labmove written immediately after (resp. before) the special zone, then $D=1$ (resp. $D=-1$); if it is scanning the labmove immediately after (resp. before) the previously mentioned labmove, then $D=2$ (resp. $D=-2$); and so on. 
As long as $D$ is not $0$,  the simulation proceeds in the ``normal" way described in the preceding two paragraphs, with $\mathcal M$'s run-tape head scanning precisely the same symbol as $\mathcal N$'s imaginary run-tape head is scanning, every time moving that head in the same direction as the direction in which $\mathcal N$'s run-tape head moves. An exception occurs when $\mathcal M$'s run-tape head enters the special zone. When this happens, $\mathcal M$ temporarily stops simulation, and continues moving its run-tape head in the same direction (left-to-right or right-to-left) in which it entered the special zone. This continues until $\mathcal M$'s run-tape head exits the special zone, after which the normal simulation resumes.  
Thus, every time $\mathcal M$ crosses the special zone from one end to the other, $\mathcal M$ spends a  constant  amount of additional time. Remembering that $\mathcal N$ is well behaved,\footnote{What is relevant here is condition 2 of Definition \ref{wb}.} this event (crossing the special zone) will only happen a constant number of times. Thus, overall, the additional time consumption of $\mathcal M$ (additional to the quantity established in the preceding paragraph)  is also constant, meaning that, asymptotically, $\mathcal M$'s time complexity is not affected. 
Finally, it is immediately obvious that $\mathcal M$  inherits the   well-behavedness of $\mathcal N$. To summarize, $\mathcal M$ is a well behaved, minimal amplitude, constant space and linear time logical solution of $\vec{G} \intimpl  F[H_0\add H_1]$, as desired.

\subsection{\texorpdfstring{$\adc$}{inf}-Choose}\label{sep10aa2} 
This case is similar to the preceding one.

\subsection{\texorpdfstring{$\ade$}{SUP}-Choose}\label{sep10bb} 
\[\frac{\vec{G}\ \intimpl\  F[H(\mathfrak{t})]}{\vec{G}\ \intimpl \ F[\ade x H(x)]}\]
Taking into account that the choice existential quantifier is nothing but a ``long'' choice disjunction, this case is also rather similar to the case of $\add$-Choose. Assume   that  $\mathcal N$ is a  well behaved, minimal amplitude, constant space and linear time logical solution of  the premise.  We want to  construct a  same kind of a  solution $\mathcal M$ for the conclusion.

First, consider the case of $\mathfrak{t}$ being a constant that occurs (not only in the premise but also) in the conclusion.  We let ${\mathcal M}$ be a machine that works as follows. At the beginning,  ${\mathcal M}$ waits till Environment specifies some constants for all free variables of $\vec{G} \intimpl F[\ade x H(x)]$. For readability, we continue referring to the resulting game as $\vec{G} \intimpl F[\ade x H(x)]$, even though, strictly speaking, it is $e[\vec{G} \intimpl F[\ade x H(x)]]$, where $e$ is a valuation that agrees with the choices that Environment has just made for the free variables of the sequent.     Now $\mathcal M$  makes the move $\beta_{initial}$ that brings $\vec{G} \intimpl F[\ade x H(x)]$ down to $\vec{G} \intimpl F[H(\mathfrak{t})]$. For instance, if $\vec{G} \intimpl F[\ade x H(x)]$ is $\vec{G} \intimpl E\mlc(K\mld \ade x H(x))$ and thus $\vec{G} \intimpl F[H(\mathfrak{t})]$ is $\vec{G} \intimpl E\mlc (K\mld H(\mathfrak{t}))$, then $\beta_{initial}$ is $1.1.1.\#\mathfrak{t}$.  After this move, ${\mathcal M}$ ``turns itself into $\mathcal N$'' in the same fashion as in the proof of the case of $\add$-Choose.   The only difference between the present case and the case of $\add$-Choose worth pointing out is that the magnitude of $\beta_{initial}$ is no longer $0$ --- rather, it is $|\mathfrak{t}|$. But note that, since $\mathfrak{t}$ occurs in the conclusion, 
$|\mathfrak{t}|$ does not exceed the native magnitude $\mathfrak{c}$ of the conclusion. This guarantees that $\mathcal M$ runs in amplitude $\max\{\ell,\mathfrak{c}\}$,  as desired. As for the bounds on time and space complexities, their being as desired is guaranteed by the same reasons as in the case of $\add$-Choose. So is $\mathcal M$'s being well behaved.

Next, consider the case of  $\mathfrak{t}$ being a variable that is among the free variables of the conclusion. It can be handled in a similar way to the above, with the only difference that now $\beta_{initial}$ is a move signifying choosing, for $x$, the constant chosen by Environment for $\mathfrak{t}$. To make such a move, $\mathcal M$ finds on its run tape the corresponding $\oo$-labeled move, and then copies its symbols into its buffer in a single pass (after first putting into the buffer the necessary technical/addressing prefix for the move, of course). This only takes a linear amount of time and a constant (if any) amount of space.   In analyzing the time overhead  imposed  on simulation by the presence of the labmove $\pp\beta_{initial}$ on $\mathcal M$'s run tape, it is worth pointing out that a difference between the present case and the case of $\add$-Choose is that, while the length of the ``special zone'' in the former case was constant, in the present case it is linear. This is however fine, because, as before, the special zone will be crossed only a constant number of times, so that the time overhead remains linear.  As for the amplitude complexity of $\mathcal M$, it also remains as desired, because the magnitude of $\beta_{initial}$ does not exceed its background $\ell$.

The remaining possibility to consider is that of  $\mathfrak{t}$ being either a constant with no occurrence in the conclusion, or a variable with no (free) occurrence  in the conclusion. 
 By induction on the length of the proof of the premise, it is very easy to see that such a proof remains a proof with $\mathfrak{t}$ replaced by $0$ (or any other constant for that matter) everywhere in it. Thus, we may assume that the premise of $\vec{G}\ \intimpl \ F[\ade x H(x)]$ is simply $\vec{G}\ \intimpl\  F[H(0)]$, i.e., that $\mathfrak{t}=0$. Now, this case will be handled exactly as the earlier case of  $\mathfrak{t}$ being a constant that occurs  in the conclusion. When dealing with that case, the assumption that $\mathfrak{t}$ had an occurrence in the conclusion was only needed to guarantee that $|\mathfrak{t}|$ did not exceed the native magnitude of the conclusion. In the present case, i.e. when $\mathfrak{t}=0$, we can no longer assume that $\mathfrak{t}$ occurs in the conclusion. However, $|\mathfrak{t}|$ still does not exceed the native magnitude of the conclusion because $|\mathfrak{t}|=0$.  

\subsection{\texorpdfstring{$\ada$}{INF}-Choose}\label{sep10bb2} 
This case is similar to the preceding one.

\subsection{Replicate}\label{sep10bb3} 
\[\frac{\vec{G},E,\vec{K},E\intimpl F}{\vec{G},E,\vec{K}\intimpl F}\]
Remembering that we agreed to see no distinction between sequents and the games they represent, and disabbreviating $\vec{G}$, $\vec{K}$ and $\intimpl$,   the premise and the conclusion of this rule can be rewritten as the following two games, respectively: 
%\marginpar{fff11a1}
\begin{equation}\label{fff11a1}
\st G_1\mlc\ldots\mlc\st G_m\ \mlc \ \st E\ \mlc \ \st K_{1}\mlc\ldots\mlc \st K_n \ \mlc\ \st E\ \ \mli\ \  F;\end{equation}
%\marginpar{fff11a2}
\begin{equation}\label{fff11a2}
\st G_1\mlc\ldots\mlc\st G_m\ \mlc\ \st E\ \mlc\ \st K_{1}\mlc\ldots\mlc \st K_n\ \ \mli \ \ F.\end{equation}

Assume $\mathcal N$ is a well behaved, minimal amplitude, constant space and linear time logical solution   of the premise. We let a logical solution $\mathcal M$ of the conclusion be a machine that works as follows.

After Environment chooses some constants for all free variables of (\ref{fff11a2}), $\mathcal M$ makes a replicative move in the $\st E$ component of the latter, thus bringing the game down to 
%\marginpar{fff11b}
\begin{equation}\label{fff11b}
\st G_1\mlc\ldots\mlc\st G_m\ \mlc\ \st (E\circ E)\ \mlc\ \st K_{1}\mlc\ldots\mlc \st K_n\ \ \mli \ \ F \end{equation}
(more precisely, it will be not (\ref{fff11b}) but $e[(\ref{fff11b})]$, where $e$ is a valuation that agrees with Environment's choices for the free variables of the sequent.   As we did earlier, however, notationally we ignore this difference). Later the section on $\mathcal M$'s run tape containing this replicative move will be treated as a ``special zone'' in the same way as in the case of $\add$-Choose, so we may safely pretend that it is simply not there.  
  
Now we need to observe that (\ref{fff11b}) is ``essentially the same as'' (\ref{fff11a1}), so that $\mathcal M$, through simulation, can continue playing (\ref{fff11b}) ``essentially as'' $\mathcal N$ would play (\ref{fff11a1}) in the scenario where the adversary of $\mathcal N$ chose the same constants for free variables as the adversary of $\mathcal M$ just did. All that $\mathcal M$ needs to do to account for the minor technical differences between (\ref{fff11b}) and (\ref{fff11a1}) is to make a very simple ``reinterpretation'' of moves. Namely:
\begin{itemize}
\item Any move made within any of the $\st G_i$ or $\st K_i$ components of (\ref{fff11b}) $\mathcal M$ sees exactly as $\mathcal N$ would see the same move in the same component of (\ref{fff11a1}), and vice versa.
\item  Any (focused) move of the form $0.\alpha$ (resp. $1.\alpha$)  made in the $\st (E\circ E)$ component of (\ref{fff11b}) $\mathcal M$ sees as $\mathcal N$ would see the  move $\alpha$ as if it was made in the first (resp. second) $\st E$ component of (\ref{fff11a1}), and vice versa.
\item  Any (unfocused) move of the form $\epsilon.\alpha$  made by Environment in the $\st (E\circ E)$ component of (\ref{fff11b}) $\mathcal M$ sees as $\mathcal N$ would see the move $\alpha$ made twice (but on the same clock cycle) by its adversary: once in  the first $\st E$ component of 
(\ref{fff11a1}), and once in the second $\st E$ component of 
(\ref{fff11a1}).
\end{itemize}

\noindent Due to the third clause above, the difference between the contents of the run tapes of $\mathcal M$ and the simulated $\mathcal N$ in the present case is greater than in the case of $\add$-Choose. This means that the run-tape head of $\mathcal M$ may need to make certain additional back-and-forth journeys to properly maintain simulation. Without going into details, we simply point out that this can be handled in a way similar to the way we handled the presence of the ``special zone'' when discussing $\add$-Choose. In view of the fact that $\mathcal N$ is well behaved, with some thought it can be seen that $\mathcal M$ only needs a constant amount of extra space and a linear amount of extra time for this, so that $\mathcal M$, just like $\mathcal N$, runs in constant space and linear time. $\mathcal M$'s amplitude complexity obviously remains the same as that of $\mathcal N$, which implies that $\mathcal M$ is a minimal amplitude logical solution as required. Finally, it can  be seen that $\mathcal M$ inherits $\mathcal N$'s 
well-behavedness as well.

\subsection{Wait}\label{iwait} 
\[\frac{Y_1,\ldots,Y_n}{X}\]
(where $n\geq 0$ and the $\adc$-, $\add$-, $\ada$-, $\ade$- and Stability conditions are satisfied). 

We shall rely on the following lemma. It   can be verified by a straightforward induction on the complexity of $Z$, which we omit. Remember that $\seq{}$ stands for the empty run, and $\elz{Z}$ for the elementarization of $Z$.

\begin{lem}\label{new1}
%\marginpar{new1}
For any sequent $Z$, valuation $e$ and interpretation $^*$, $\win{Z^*}{e}\emptyrun= \win{\elzi{Z}^*}{e}\emptyrun$.
\end{lem}

Assume ${\mathcal N}_1,\ldots,{\mathcal N}_n$ are well behaved, minimal amplitude, constant space and linear time logical solutions   of $Y_1,\ldots,Y_n$, respectively.  We  
 let ${\mathcal M}$, a  logical solution of $X$,  be a machine that works as follows. 

 At the beginning, as always, $\mathcal M$ waits till Environment chooses some constants for all free variables of the conclusion. Let $e$ be a (the) valuation that agrees with the choices just made by Environment (in the previous cases, we have suppressed the $e$ parameter, but now we prefer to deal with it explicitly). So, the conclusion is now brought down to $e[X]$. After this event,  ${\mathcal M}$ continues waiting until Environment makes one more move.  If such a move is never made, then the run of (the $\ada$-closure of) $X$ generated in the play can be simply seen as the empty run of $e[X]$.  Due  to the Stability condition, $\elz{X}$ is classically valid, meaning that $\win{\elzi{X}}{e}\emptyrun=\pp$.  But then, in view of Lemma \ref{new1}, $\win{X}{e}\emptyrun=\pp$. This makes $\mathcal M$ the winner without having consumed any space or any move-preceding time for which it can be billed.  

 Suppose now Environment makes a move $\alpha$.    With a little thought, one can see that any (legal) move $\alpha$ by Environment brings the game $e[X]$ down to $g[Y_i]$ for a certain valuation $g$   and one of the premises $Y_i$ of the rule. For example, if $X$ is $P \intimpl (E\adc F)\mld  \ada x G(x)$ where $P$ is atomic, then a legal move $\alpha$ by Environment should be either $1.0.0$ or $1.0.1$ or $1.1.\#c$ for some constant $c$. In the case $\alpha=1.0.0$, the above-mentioned premise $Y_i$ will be $P\intimpl E \mld  \ada x G(x)$, and $g$ will be the same as $e$.  In the case $\alpha=1.0.1$,   $Y_i$ will be $P\intimpl F \mld  \ada x G(x)$, and $g$, again, will be the same as $e$. Finally, in the case $\alpha=1.1.\#c$,   $Y_i$ will be $P\intimpl (E\adc F) \mld G(y)$  for a variable $y$  not occurring in $X$, and   $g$ will be the valuation that sends $y$ to $c$'s denotat and agrees with $e$ on all other variables, so that $g[P\intimpl (E\adc F) \mld G(y)]$ is $e[P\intimpl (E\adc F) \mld G(c)]$, with the latter being the game to which $e[X]$ is brought down by the labmove $\oo 1.1.\#c$. 

After the above event, ${\mathcal M}$ does the usual trick of turning itself into --- and continuing playing as --- ${\mathcal N}_i$,  with the only difference that,  if $g\not=e$, the behavior of ${\mathcal N}_i$ should be followed for the scenario where the adversary of the latter, at the very beginning of the play, chose constants for the free variables of $Y_i$ in accordance with $g$ rather than $e$.  

It is left to the reader to convince himself or herself that, with adequately arranged details of simulation, $\mathcal M$ is as desired.

\subsection{On the completeness of \texorpdfstring{$\cltw$}{CL12}}\label{comment}
%\marginpar{comment}

 While CoL takes no interest in nonalgorithmic ``solutions'' of problems, it would still be a pity to let one fact go unobserved. As noted, the completeness proof for $\cltw$ from \cite{lbcs} (namely, the $(ii)\Rightarrow(i)$ part of our present Theorem \ref{feb9c}) will be reproduced in 
Appendix \ref{scompleteness}. An analysis of that proof reveals that virtually nothing in it 
 relies on the fact that a purported HPM solving the problem represented by a $\cltw$-unprovable sequent, whose non-existence is proven, follows an algorithmic strategy. So, the completeness result  can be strengthened by saying that, if $\cltw$ does not prove a sequent $X$, then $X$ does not even have a nonalgorithmic logical solution. Precisely defining the meaning of a ``nonalgorithmic'', or rather ``not-necessarily-algorithmic'' logical solution, is not hard. The most straightforward way to do so would be to simply take our present definition of a logical solution but generalize its underlying model of computation by allowing HPMs to have oracles  --- in the standard sense --- for whatever functions.

As an aside, among the virtues of CoL is that it eliminates the need for many ad hoc inventions such as the just-mentioned oracles. Namely, observe that a problem $A$ is computable  by an HPM with an oracle for a function $f(x)$ if and only if the problem $\ada x\ade y\bigl(y\equals f(x)\bigr)\intimpl A$ is computable in the ordinary sense (i.e., computable by an ordinary HPM without any oracles). So, a CoL-literate person, regardless of his or her aspirations, would never really have to speak in terms of  oracles or nonalgorithmic strategies.   This explains why `{\em CoL takes no interest in nonalgorithmic ``solutions'' of problems}'.

\section{Logical Consequence}\label{slc}
%\marginpar{slc}

\begin{defi}\label{dlc}
%\marginpar{dlc}
We say that a formula $F$ is a {\bf logical consequence}\label{0lcr} of formulas  $E_1,\ldots,E_n$ ($n\geq 0$) iff the sequent $E_1,\ldots,E_n\intimpl F$, in the role of $X$, satisfies any of the (equivalent) conditions (i)-(iii) of Theorem \ref{feb9c}. 
\end{defi}

As noted in Section \ref{intr}, the following rule, which we (also) call {\bf Logical Consequence},\label{0lcl}   will be the only logical rule of inference in $\cltw$-based applied systems:
\[\mbox{\em From $E_1,\ldots,E_n$ conclude $F$ as long as $F$ is a logical consequence of $E_1,\ldots,E_n$}.\]   
 
A reader familiar with  earlier essays on CoL would remember that, philosophically speaking, computational {\em resources}\label{0resource2} are symmetric to computational problems: what is a problem for one player to solve is a resource that the other player can use. Namely, having a problem $A$ as a computational resource intuitively means having the (perhaps externally provided) ability  to successfully solve/win $A$. For instance, as a resource, $\ada x\ade y(y=x^2)$ means the ability to tell the square of any number.   

According to the following thesis, logical consequence lives up to its name. A justification for it, as well as an outline of its significance, was provided in Section \ref{intr}:

\begin{thesis}\label{thesis}
%\marginpar{thesis}
Assume  $E_1,\ldots,E_n$ and $F$ are formulas such that  there is a $^*$-independent (whatever interpretation $^*$) intuitive description and justification of a winning strategy for $F^*$, which relies on the availability and ``recyclability'' --- in the strongest sense possible --- of $E_{1}^{*},\ldots,E_{n}^{*}$ as computational resources.   Then $F$ is a logical consequence of $E_1,\ldots,E_n$. 
\end{thesis}

\begin{exa}\label{intex}
%\marginpar{intex}
Imagine a $\cltw$-based applied formal theory, in which we have already proven two facts: $\cla x\bigl(x^3\equals(x\mult x)\mult x\bigr)$ (the meaning of ``cube'' in terms of multiplication) and $\ada x\ada y\ade z(z\equals x\mult y)$ (the computability of multiplication), and now we want to derive $\ada x\ade y(y\equals x^3)$ (the computability of ``cube''). This is how we can reason to justify $\ada x\ade y(y\equals x^3)$:  
\begin{quote}{\em Consider any $s$ 
(selected by Environment for $x$ in $\ada x\ade y(y\equals x^3)$). We need to find  $s^3$. Using the resource $\ada x\ada y\ade z(z\equals x\mult y)$, we first find the value $t$ of $s\mult s$, and then  the value $r$ of $t\mult s$. According to $\cla x(x^3\equals (x\mult x)\mult x)$, such an $r$ is the sought $s^3$.}
\end{quote}   

\noindent Thesis \ref{thesis} promises that the above intuitive argument will be translatable into a $\cltw$-proof of 
\[\cla x\bigl(x^3\equals(x\mult x)\mult x\bigr),\ \ada x\ada y\ade z(z\equals x\mult y)\ \intimpl \ \ada x\ade y(y\equals x^3) \]
(and hence the succedent will be derivable in the  theory by Logical Consequence as the formulas of the antecedent are already proven). Such a proof indeed exists --- see 
 Example \ref{ecube}.
\end{exa}

While Thesis \ref{thesis} is about the {\em completeness} of Logical Consequence, the following Theorems \ref{feb9dt} and \ref{feb9ds} are about {\em soundness}, establishing that Logical Consequence preserves computability, and does so in a certain very strong sense. They are the most important results of the present paper: as noted in Section 1, it is these two theorems that dramatically broaden the applicability of $\cltw$ (of Logical Consequence, that is) as a logical basis for complexity-oriented applied theories. Theorem \ref{feb9dt} carries good news for the cases where time efficiency is of main concern, and otherwise we are willing to settle for at least linear space.  Theorem \ref{feb9ds} does the same but for the cases where the primary concern is space efficiency --- namely, when we want to keep the latter sublinear.

\section{First preservation theorem}\label{sstm}
%\marginpar{sstm}
We say that a given HPM $\mathcal M$ plays a given constant game $A$  {\bf providently}\label{0prvp} iff all $\oo$-legal plays of $A$ by $\mathcal M$ are provident. 
 Intuitively, 
this means that, as long as the adversary plays $A$ legally, $\mathcal M$   never starts constructing a (nonempty) move in its buffer unless it is going to actually make such a move later.  A {\bf provident solution}\label{0prvs} of a given constant game $A$ means an HPM $\mathcal M$ with ${\mathcal M}\models A$  that plays $A$  providently.

Due to being technical and probably not particularly interesting for most readers,   proofs of the following two Lemmas \ref{sep1} and \ref{sep2} are   postponed to  Appendix   \ref{sapa} and Appendix \ref{sapb}, respectively. Here we only present   brief outlines of the ideas underlying those proofs.

\begin{lem}\label{sep1}
%\marginpar{sep1}
There is an effective procedure that takes an 
 arbitrary HPM $\mathcal N$, together with an arbitrary formula $E$,  and constructs an HPM ${\mathcal Q}$  such that, for any interpretation $^*$   and any unary arithmetical functions 
$\mathfrak{a},\mathfrak{s},\mathfrak{t}$,  if $\mathcal N$ is an   
$\mathfrak{a}$ amplitude, $\mathfrak{s}$ space and $\mathfrak{t}$ time solution of $E^*$, then  
${\mathcal Q}$  is a provident,  $\mathfrak{a}$ amplitude, $O(\mathfrak{s})$ space and $O(\mathfrak{t})$ time solution of $E^*$. Such a $\mathcal Q$ has the same number of work tapes as $\mathcal N$ does.
\end{lem} 

\begin{idea} \label{pid} We let $\mathcal Q$ simulate $\mathcal N$ and play  exactly as $\mathcal N$ does yet in a way that guarantees providence.  Namely,     every time   $\mathcal N$ starts constructing a new move $\alpha$ in its buffer, $\mathcal Q$ waits --- without hurrying to put any symbols of $\alpha$  into its own buffer --- to see if $\mathcal N$ eventually   makes such a move; if and when the simulation shows     that $\mathcal N$  moves, $\mathcal Q$ resimulates  $\mathcal N$'s work, this time actually copying the evolving content of $\alpha$  into its own buffer and entering a move state once  that content is mature.   Through an 
appropriate arrangement of the details of $\mathcal Q$'s work, one can achieve the satisfaction of  all conditions of   the lemma. 
\end{idea}

\begin{lem}\label{sep2} \ 
%\marginpar{sep2}
%Consider any unary arithmetical function $\mathfrak{a}$.  
\begin{enumerate}[label=\arabic*.]
\item Consider any formula $E$. There is a number $c$ such that, for any HPM $\mathcal Q$, interpretation $^*$ and unary arithmetical function $\mathfrak{a}$, if $\mathcal Q$ is a provident and $\mathfrak{a}$-amplitude solution of $E^*$,  then,   at any step $t$ of any $\oo$-legal play of $E$ by ${\mathcal Q}$, the number of symbols in $\mathcal Q$'s buffer does not exceed $\mathfrak{a}(\ell)+c$, where $\ell$ is the background of $t$. 

\item Consider any sequent $X$ with a well behaved and minimal
  amplitude logical solution $\mathcal Q$. There is a number $c$ such
  that, at any step $t$ of any $\oo$-legal play of $X$ by ${\mathcal
    Q}$, the number of symbols in $\mathcal Q$'s buffer does not
  exceed $\ell+c$, where $\ell$ is the background of $t$.
\end{enumerate} 
\end{lem}

\begin{idea} 
Assuming all the conditions of clause 1 of the lemma,  $\mathcal Q$  never constructs an ``oversized'' --- longer that $\mathfrak{a}(\ell)$ plus a constant --- move in its buffer because, due to providence, it would have to actually make such a move sooner or later, in which case  it would violate its  amplitude constraints (unless, of course, by good luck, Environment makes an illegal or ``very long'' move meanwhile). Clause 2 is similar, taking into account that providence is one of the necessary conditions for  being well behaved. \end{idea}

Below and later, when $f$ is a unary arithmetical function and $m$ a natural number, $f^{m}(x)$ means the $m$-fold composition of $f$ with itself applied to $x$, i.e. $f(f(\ldots (x)\ldots))$ with $f$ repeated $m$ times.   On the other hand, $(f(x))^m$ has its usual meaning of the $m$th power of $f(x)$, i.e. $f(x)$ multiplied by itself $m$ times.

\begin{thm}\label{feb9dt}  
%\marginpar{feb9dt}
There is an  effective procedure that takes an arbitrary $\cltw$-proof \ $\mathbb{P}$ of an arbitrary sequent {\em $E_1,\ldots,E_n\intimpl F$}, arbitrary HPMs  ${\mathcal N}_1,\ldots,{\mathcal N}_n$
 and constructs an  HPM \ ${\mathcal M}$ satisfying the condition below:

\begin{quote}
Assume $^*$ is an interpretation and $\mathfrak{a},\mathfrak{s},\mathfrak{t}$ are unary arithmetical functions such that:
\begin{description}
  \item[(i)] For each $i\in\{1,\ldots,n\}$, \ ${\mathcal N}_i$ is an $\mathfrak{a}$ amplitude, $\mathfrak{s}$ space and $\mathfrak{t}$ time solution of $E_{i}^{*}$.
  \item[(ii)] For any $x$,  $\mathfrak{a}(x)\geq \max\{x,\mathfrak{c}\}$, where  $\mathfrak{c}$ is the native magnitude of 
{\em $E_1,\ldots,E_n\intimpl F$}.  
  \item[(iii)] For any $x$,  $\mathfrak{s}(x)\geq x$.
  \item[(iv)] For any $x$,  $\mathfrak{t}(x)\geq  x  $.
\end{description} 
Then there is a number $\mathfrak{b}$ which only depends on $\mathbb{P}$ such that, with $\mathfrak{R}$ abbreviating $\mathfrak{a}^{\mathfrak{b}}(\ell)$,  \ ${\mathcal M}$ is a solution of $F^*$ that runs in amplitude  $\mathfrak{R}$, space  $O\bigl(\mathfrak{s}(\mathfrak{R})\bigr)$ and time   $O\bigl(\mathfrak{t}(\mathfrak{R})\bigr)$.
\end{quote}
\end{thm}

\noindent The rest of the present section is  devoted to a proof of the above theorem. 

 Consider an arbitrary sequent $E_1,\ldots,E_n\intimpl F$ together with a $\cltw$-proof $\mathbb{P}$ of it, and let $\mathfrak{c}$ be the native magnitude of $E_1,\ldots,E_n\intimpl F$. Let $\mathfrak{a},\mathfrak{s},\mathfrak{t}$ be functions satisfying conditions (ii)-(iv) of the theorem.  By Theorem \ref{feb9c}, there is a well behaved logical solution  $\mathcal K$ of $E_1,\ldots,E_n\intimpl F$ which runs in $\max\{\ell,\mathfrak{c}\}$ amplitude, constant space and linear time,  and such a $\mathcal K$ --- fix it --- can be effectively constructed from $\mathbb{P}$. Note that, in view of conditions 
(ii)-(iv) of the present theorem, $\mathcal K$ automatically also runs in amplitude $\mathfrak{a}$, space $O(\mathfrak{s} )$ and time $O(\mathfrak{t} )$. 
Consider an arbitrary interpretation $^*$ (which, as done before, we shall notationally suppress) and  
arbitrary HPMs ${\mathcal N}_1,\ldots,{\mathcal N}_n$ satisfying condition (i) of the theorem. 
%For convenience and without loss of generality, we assume that each of these $n$ machines has $k$ ($k\geq 1$) work tapes.\footnote{One can always add ``dummy'' (never used) work tapes to a given HPM and thus increase the number of work tapes without affecting any other relevant characteristics of the machine.}
Below we describe an  HPM \ ${\mathcal M}$  such that ${\mathcal M} $ wins $F$ under the same interpretation $^*$. It is important to note that our {\em construction} of ${\mathcal M}$  does not depend on $^*$,  $\mathfrak{a},\mathfrak{s},\mathfrak{t}$ and hence on our assumption that  
conditions (i)-(iv) are satisfied; only our {\em claim} that ${\mathcal M} $  wins $F$, and our further claims about its complexities, do.   

As always, we let our machine  ${\mathcal M} $, at the beginning of the play, wait till Environment selects a constant --- call it $c_x$ --- for each free variable $x$ of $F$. Let us fix $e$ as the valuation  that sends every free variable $x$ of $F$ to the denotat of the corresponding constant $c_x$, and (arbitrarily) sends all other variables  to $0$. 

We describe the work of ${\mathcal M} $ afterwards at a high level. A more detailed description and analysis would be neither feasible (since it would be prohibitively long and technical) nor necessary. From our description it will be immediately clear that  $\mathcal M$ is constructed effectively, so this issue will not be explicitly addressed.

To understand the idea, let us first consider the simple case where $\mathcal K$ never makes any replicative moves in the antecedent of $E_1,\ldots,E_n\intimpl F$. 
  The main part of the work of ${\mathcal M} $  consists in continuously monitoring its run tape to see if Environment has made any new moves, combined with  simulating, in parallel, a  play of $E_1,\ldots,E_n\intimpl F$ by the machine $\mathcal K$ and --- for each $i\in\{1,\ldots,n\}$ --- a play of $E_i$ by the machine ${\mathcal N}_i $. 
During simulation, ${\mathcal M} $ ``imagines'' that, at the beginning of the play, for each free variable $x$ of the corresponding formula or sequent, 
the adversary of each machine has chosen the same constant $c_x$ as $\mathcal M$'s adversary did.  After the above initial moves by the real and imaginary adversaries, each of the $n+ 2$ games \[G\ \in\ \{ E_1,\ \ \ldots,\ \ E_n, \ \ E_1,\ldots,E_n\intimpl F,\ \ F\}\] that we consider here (the first $n+1$ games imaginary and the last one real) will be brought down to $e[G]$ but, for readability and because $e$ is fixed, we shall usually omit $e$ and write simply $G$ instead of $e[G]$. 

Since we here assume that $\mathcal K$ never makes any replications in the antecedent of $E_1,\ldots,E_n$ $\intimpl F$, playing this game essentially means simply playing 
%\marginpar{feb18a}
\begin{equation}\label{feb18a}
E_1\mlc \ldots\mlc E_n\mli F,
\end{equation}
so, in what follows, we will talk about (\ref{feb18a}) instead of $E_1,\ldots,E_n\intimpl F$. 

As earlier in similar situations, we assume that, in the real play of $F$, Environment does not make any illegal moves (the Clean Environment Assumption).    We can also safely assume that the simulated machines do not make any illegal moves in the corresponding games, or else our assumptions about their winning those games would be wrong.\footnote{Since we  need to construct ${\mathcal M} $ no matter whether those assumptions are true or not, we can let ${\mathcal M} $ simply retire as soon as some illegal/unexpected behavior is detected.}

If so, what ${\mathcal M} $ does in the above mixture of the real and  simulated plays is that it applies copycat between $n+ 1$ pairs of (sub)games, real or imaginary. Namely, it mimics, in (the real play of) $F$,  $\mathcal K$'s moves made
in the consequent of  (the imaginary play of)  (\ref{feb18a}),  and vice versa: uses   Environment's moves made in the real play of $F$ as $\mathcal K$'s (imaginary) adversary's moves in the consequent of (\ref{feb18a}).  Further, for each   $i\in\{1,\ldots,n\}$,  ${\mathcal M} $ uses the moves made by ${\mathcal N}_i $ in $E_i$  as $\mathcal K$'s imaginary adversary's 
moves in the $E_i$ component of (\ref{feb18a}), and vice versa: uses the moves made by $\mathcal K$ in that component as $ {\mathcal N}_i $'s imaginary adversary's  moves in $E_i$.

Therefore, the final positions   hit by the $n+ 2$ imaginary and real plays of
\[\mbox{$E_1,\ \ldots,\ E_n,\ \ E_1\mlc\ldots\mlc E_n\mli F$ \ and \ $F$}\]
will retain the above forms, i.e., will be
\[\mbox{$E'_1,\ \ldots,\ E'_n,\ \ E'_1\mlc\ldots\mlc E'_n\mli F'$ \ and \ $F'$}\]
for some $E'_1,\ldots,E'_n,F'$. Our assumption that the machines ${\mathcal N}_1  ,\ldots,{\mathcal N}_n  $ and ${\mathcal K}$ win the games $E_1, \ldots, E_n$ and $F_1\mlc\ldots\mlc F_n\mli F$ implies that each $G\in\{E'_1,\ \ldots,\ E'_n,\ E'_1\mlc\ldots\mlc E'_n\mli F'\}$ is $\pp$-won, in the sense that $\win{G}{}\seq{}=\pp$. It is then obvious that so should be  $F'$. Thus, the (real) play of $F$ brings it down to the $\pp$-won  $F'$, meaning that ${\mathcal M} $ wins $F$.

Let us fix $\mathfrak{b}$ as the total number of occurrences of choice operators $\adc,\add,\ada,\ade$ in the $\ada$-closure $\ada(\ref{feb18a})$ of (\ref{feb18a}). Note that  no legal run of $\ada(\ref{feb18a})$ or of any of the games $\ada E_1,\ldots,\ada E_n,\ada F$ will contain more than $\mathfrak{b}$ labmoves.

 Consider an arbitrary play (computation branch) of ${\mathcal M} $, and an arbitrary clock cycle $c$ in the context of that play. Let $\ell$ be the background of $c$.  In the simulations of $\mathcal K$ and ${\mathcal N}_1 $, \ldots, ${\mathcal N}_n  $, every move made by the imaginary adversary of one of these machines is a copy\footnote{Here and later in similar contexts, ``copy'' should be understood in a generous sense, referring only the ``actual meanings'' of  moves, with their ``addressing'' prefixes otherwise possibly modified/readjusted.  For instance, if Environment made the move $\alpha$ in the real play of $F$,  the corresponding ``copy'' move made by ${\mathcal M} $ in the imaginary play of (\ref{feb18a}) for the imaginary adversary of $\mathcal K$ will be $1.\alpha$ rather than $\alpha$, with the prefix ``$1.$'' merely indicating that the move $\alpha$ is made in the consequent. It is important to note that such prefix readjustments in no way affect the magnitudes of the ``copied'' moves. This is one of the reasons why
we chose to understand magnitude as in Convention \ref{decc10a} rather than as raw size.}  of either a move made by Environment in the real play, or a move made by one of the machines ${\mathcal K}$, ${\mathcal N}_1 $, \ldots, ${\mathcal N}_n $ during simulation.  
Let $\beta_1,\ldots,\beta_m$ be the moves by simulated machines that ${\mathcal M} $ detects by time $c$, arranged according to the times of their detections. Let  $${\mathcal G}_1,\ldots,{\mathcal G}_m\in\{{\mathcal K},{\mathcal N}_1  ,\ldots,{\mathcal N}_n  \}$$ be the machines that made these moves, respectively.  Remembering that all machines run in amplitude $\mathfrak{a}$, the magnitude of $\beta_1$ cannot exceed $\mathfrak{a}(\ell)$. That is because, by the time when    ${\mathcal G}_1$ made the move $\beta_1$,  all (if any) moves by ${\mathcal G}_1$'s imaginary adversary were copies of moves made by Environment in the real play rather than moves made by some other simulated machines, and hence the background of $\beta_1$ in the simulated play of ${\mathcal G}_1$ did not exceed $\ell$. For similar reasons, with $\mathfrak{a}(\ell)$ now acting in the role of $\ell$ and ${\mathcal G}_2$ in the role of ${\mathcal G}_1$, the magnitude  of $\beta_2$ cannot exceed $\mathfrak{a}(\mathfrak{a}(\ell))$. Similarly, the magnitude of $\beta_3$ cannot exceed $\mathfrak{a}(\mathfrak{a}(\mathfrak{a}(\ell)))$, etc. Also, by our selection of the constant $\mathfrak{b}$,  at most $\mathfrak{b}$ moves can be made altogether by the simulated machines, 
 so  $m\leq \mathfrak{b}$. Thus, by time $c$, the magnitude of no move made in this mixture of real and simulated plays can exceed $\mathfrak{a}^{\mathfrak{b}}(\ell)$. This means that ${\mathcal M} $, which merely mimics some of $\mathcal K$'s moves, runs in amplitude $\mathfrak{R}=\mathfrak{a}^{\mathfrak{b}}(\ell)$ as promised. 

The next thing to clarify is why the space and time complexities of ${\mathcal M} $ are also as promised. Here we need to look into more details of the work of ${\mathcal M} $. First of all, let us agree for the rest of this section --- or, rather, reiterate our earlier general convention --- that, in the (often only implicit) context of a given computation step of 
${\mathcal M} $, as in the preceding paragraph, $\ell$ stands for the background of that step. As we saw in the preceding paragraph, the magnitude of no real or imaginary move will ever exceed $\mathfrak{R}=\mathfrak{a}^{\mathfrak{b}}(\ell)$ for that very $\ell$. So, while $\ell$ is the ``current'' background in the context of ${\mathcal M} $, the ``current'' background in the context of a simulated machine can instead be as great as $\mathfrak{R}$. Therefore, for both simplicity and safety, in what follows we prefer to talk in terms of (the implicitly ``current'' value of)  $\mathfrak{R}$ rather than $\ell$. 
 
A legitimate question that we have never addressed so far  is the one about the number of work tapes that $\mathcal M$ has. This is our arrangement. Per each simulated machine ${\mathcal G}\in\{{\mathcal K},{\mathcal N}_1,\ldots,{\mathcal N}_n\}$,  $\mathcal M$ has: 
\begin{itemize}
  \item As many work tapes as $\mathcal G$ has. At any time during simulation, $\mathcal M$ keeps the content of each such tape identical to the imaginary content of the corresponding work tape of $\mathcal G$. Furthermore, $\mathcal M$ positions the  scanning head of such a tape also exactly where the  scanning head of the corresponding work tape of $\mathcal G$ would be. In this way, the two work tapes (including their scanning heads) --- one real and one imaginary --- are  perfectly synchronized.  
  \item One work tape to maintain the imaginary content of $\mathcal G$'s run tape. Again, $\mathcal M$ keeps the contents and the scanning head locations of the two tapes identical; the synchronization is perfect just like in the preceding case, except for 
occasional interrupts during which $\mathcal M$ is updating the content of this tape through adding a new labmove to it. 
  \item One work tape to maintain the imaginary content of $\mathcal G$'s buffer. As we may guess, again, the contents of such a tape and such a buffer  are synchronized, with the scanning head of $\mathcal M$'s work tape always located at the leftmost blank cell. As in the preceding case, the synchronization is disrupted only occasionally. Namely, once $\mathcal G$ enters a move state and thus instantaneously clears its buffer, $\mathcal M$ takes a number  of steps to correspondingly blank out its work tape.\footnote{Since we agreed in Section \ref{icp} that HPMs never write \blank\ on their work tapes, deleting or ``blanking out'' here could be understood as changing the content of each cell to some special and never-used-elsewhere (``pseudo-blank'') symbol.}  
\end{itemize}  

\noindent Of course, $\mathcal M$ also needs to keep track of certain other pieces of information. Namely, it needs to remember the state of each simulated machine, as well as remember the (never changing) transition functions of those machines. This information, due to its amount's being bounded (not exceeding a certain constant), can be and is kept track of via $\mathcal M$'s state memory rather than work tape memory, thus contributing nothing to $\mathcal M$'s space complexity. 

Thus, at any time, $\mathcal M$'s overall work-tape space consumption consists of three components: (1) the space needed to remember the work tape contents of the simulated machines; (2) the space needed to remember the run tape contents of the simulated machines; and (3) the space needed to remember the buffer contents of the simulated machines.
 
Each simulated machine contributes to component (1) the amount that is equal to its own space consumption $O\bigl(\mathfrak{s}(\mathfrak{R})\bigr)$.  Since there is only a constant number ($n+1$) of such machines, the overall cost of component (1)  remains   $O\bigl(\mathfrak{s}(\mathfrak{R})\bigr)$.

The run tape of each of the $n+1$ simulated machines can contain at most $\mathfrak{b}$ labmoves, and the size of each such labmove is bounded by $O(\mathfrak{R})$. So, the overall cost of component (2) is $O\bigl((n+1)\times \mathfrak{R}\bigr)$, i.e. $O(\mathfrak{R})$, which, in view of condition (iii) of our theorem, 
can be generously rewritten as $O\bigl(\mathfrak{s}(\mathfrak{R})\bigr)$. 

In view of Lemma \ref{sep1}, we  may   assume that the machines\label{0reas} 
${\mathcal N}_1,\ldots,{\mathcal N}_n$ are provident solutions of the corresponding games.  Therefore, applying clause 1 of Lemma \ref{sep2} to ${\mathcal N}_1,\ldots,{\mathcal N}_n$ and applying clause 2 of the same lemma to (the well behaved and minimal amplitude) $\mathcal K$, we find that  
 representing the   buffer contents of any of the $n+1$ simulated machines  takes at most $O\bigl(\mathfrak{a}(\mathfrak{R})\bigr)$ space. In fact, with a more careful analysis,  this quantity can be lowered to $O(\mathfrak{R})$, which then, as in the preceding paragraph, can be rewritten as the target 
$O\bigl(\mathfrak{s}(\mathfrak{R})\bigr)$. This is so because, as it can be seen with a little thought, due to providence (or well-behavedness which implies providence), whenever the buffer of a given 
simulated machine is nonempty on a given cycle (meaning that the machine is going to move at some later time), the maximum possible  background of that 
cycle is $\mathfrak{a}^{\mathfrak{b}-1}(\ell)$ rather than  $\mathfrak{a}^{\mathfrak{b}}(\ell)$ established earlier for the general case; and if so, Lemma \ref{sep2} tells us that the buffer content can be recorded in space $O\bigl(\mathfrak{a}(\mathfrak{a}^{\mathfrak{b}-1}(\ell))\bigr)$, i.e. $O\bigl(\mathfrak{a}^{\mathfrak{b}}(\ell)\bigr)$, i.e. $O(\mathfrak{R})$, rather than  the earlier claimed $O\bigl(\mathfrak{a}(\mathfrak{R})\bigr)$. From here, as in the cases of components (1) and (2), we conclude that 
 the overall cost of component (3) does not exceed $O\bigl(\mathfrak{s}(\mathfrak{R})\bigr)$.

Summarizing the above four paragraphs, the overall space consumption of $\mathcal M$ --- and thus (``even more so'') the space consumption of any particular work tape of $\mathcal M$ --- is indeed $O\bigl(\mathfrak{s}(\mathfrak{R})\bigr)$, as desired. 

$\mathcal M$ initializes its simulation routine by copying Environment's initial moves (in the ``$\oo 1.$''-prefixed form) onto its work tape that keeps track of the content of $\mathcal K$'s run tape.  After that it keeps its run-tape head at the leftmost blank cell and, every time a new move $\alpha$ is detected made by either player $\xx$, further adds the string $\xx 1.\alpha$ to the content of the above-mentioned work tape. $\mathcal M$ similarly copies the moves made by the simulated machines onto its work tapes representing the run tapes of those machines and their adversaries. An exception here is the case of ``$1.$''-prefixed moves made by $\mathcal K$; each such move $\alpha$, after (or simultaneously with) copying it on its work tape representing $\mathcal K$'s run tape, $\mathcal M$ further copies  into its buffer, and then makes $\alpha$ as a move in the real play.\footnote{At this point one may notice that, despite our having qualified $\mathcal M$'s strategy as applying copycat,   the run $\Gamma$ generated in the real play of $F$ may be not exactly the same as the run $\Delta$ generated in the consequent of (\ref{feb18a}). For instance, if  Environment made a move $\beta$ while $\mathcal M$ was constructing the above $\alpha$ in its buffer, the labmove $\oo\beta$ will appear {\em before} $\pp\alpha$ in $\Gamma$ but {\em after} $\pp\alpha$ in $\Delta$. Is this a problem? Not at all. It is not hard to see that, even if the orders of labmoves in $\Gamma$ and $\Delta$ are not exactly the same,  both runs are legal and they bring $F$ down to the same position, in the sense that $\seq{\Gamma}F=\seq{\Delta}F$, and this is all that eventually matters. Using the terminology of  the end of Section \ref{nncg}, $\Gamma$ is a $\pp$-delay of $\Delta$, which, taking into account that the game $F$ is static, is just as good as if the two runs were identical. Namely, $\Delta$'s being $\pp$-won implies that $\Gamma$ is also $\pp$-won. Similarly, by the Clean Environment Assumption, $\Gamma$ is $\oo$-legal; but $\Gamma$'s being a $\pp$-delay of $\Delta$ implies that $\Delta$ is a $\oo$-delay of $\Gamma$; this, by the definition of static games, further implies that $\Delta$ is $\oo$-legal. Thus, the assumption that $\Delta$ was a $\oo$-legal run of $F$, on which (whether the reader noticed or not) we have  implicitly relied in our argument,   was legitimate.} 
 Since the number of moves that will be made in the mixture of the real and simulated plays does not exceed the constant $\mathfrak{b}$ and the magnitude of each move does not exceed $\mathfrak{R}$, the above move-copying work, altogether, contributes only $O(\mathfrak{R})$ to $\mathcal M$'s overall time consumption. In view of condition (iv) of our theorem, asymptotically this quantity adds nothing to our target time  $O\bigl(\mathfrak{s}(\mathfrak{R})\bigr)$ and hence can be ignored. 
 
The above move-copying events can be viewed as exceptions occurring occasionally during the otherwise ``normal''  simulation process. The latter consists in following the $n+1$ simulated machines, in parallel and step by step, via keeping correspondingly updating (within the same, single, step) the corresponding work tape contents and head locations. That is, during the ``normal'' simulation process, $\mathcal M$ works at exactly the same pace as the simulated machines, spending a single clock cycle of its own on tracing one step of all simulated machines simultaneously.    
 Taking into account that each of the simulated machines runs in time $O\bigl(\mathfrak{t}(\mathfrak{R})\bigr)$ and hence consumes at most $O\bigl(\mathfrak{t}(\mathfrak{R})\bigr)$ steps before responding with a move, it is not hard to see that  
 the number of steps performed by $\mathcal M$ during the ``normal'' simulation process since Environment's  last move (or since step $0$ if there is no such move) each time before making a new move of its own is $O\bigl(\mathfrak{t}(\mathfrak{R})\bigr)$. Thus, $\mathcal M$ indeed runs in time $O\bigl(\mathfrak{t}(\mathfrak{R})\bigr)$, as desired. 

Whatever we have said so far was about the simple case where $\mathcal K$ makes no replicative moves in the antecedent of $E_1,\ldots,E_n\intimpl F$. How different is the general case, where $\mathcal K$ can make replications?
 Not very different. The overall work of ${\mathcal M} $ 
 remains the same, with the only difference that, every time $\mathcal K$ replicates one of $E_i$ (more precisely, to whatever game a given copy of $E_i$ has evolved by that time), ${\mathcal M} $ splits the corresponding simulation of ${\mathcal N}_i $ into two identical copies (using two different sets of its work tapes), with the same past but possibly diverging futures. This increases the number of simulated plays and the corresponding number of to-be-synchronized (by the copycat routine) pairs of games, as well as the number of work tapes that $\mathcal M$ utilizes,   but otherwise ${\mathcal M} $ continues working as in the earlier described scenario. ${\mathcal M} $ is guaranteed to win for the same  reasons as before. Furthermore, the  complexity analysis that we provided earlier still remains valid as long as we appropriately readjust (increase) the value of the parameter $\mathfrak{b}$. As we remember, $\mathfrak{b}$ was chosen to be the total number of occurrences of choice (``move-inducing'') operators in the $\ada$-closure of the sequent. The only reason for this choice of $\mathfrak{b}$ was to make sure that it was a bound on the lengths of (numbers of moves in) legal runs of the game represented by the sequent. If, however, $\mathcal K$ makes some replicative moves in the antecedent, then each occurrence of a choice operator in the antecedent of the original sequent may eventually give rise to many moves rather than only one move. But how many moves? Luckily, as we remember,  $\mathcal K$ is well behaved, so that, even if it makes replicative moves, it does so only a certain bounded (constant) number of times that only depends on $\mathbb{P}$. This means that the lengths of the runs of the $\ada$-closure of  $E_1,\ldots,E_n\intimpl F$ generated by $\mathcal K$ still remain bounded. Setting the new value of $\mathfrak{b}$ to that bound, $\mathfrak{b}$ thus remains constant. 
And this is all that is necessary for our earlier complexity analysis to remain valid.

\section{Second preservation theorem}\label{sssp}
%\marginpar{sssp}

The conditions of the following theorem essentially differ from those of Theorem \ref{feb9dt} only in that they replace the ``at least linear'' requirement (condition (iii)) on the space complexity of the premise-solving machines by the weaker ``at least logarithmic'' condition. The price to pay is an increase of the time complexity of the conclusion-solving machine relative to the time complexities of the premise-solving machines.  
 
\begin{thm}\label{feb9ds}  
%\marginpar{feb9ds}
 There is an  effective procedure that takes an arbitrary $\cltw$-proof \ $\mathbb{P}$ of an arbitrary sequent {\em $E_1,\ldots,E_n\intimpl F$}, arbitrary HPMs ${\mathcal N}_1,\ldots,{\mathcal N}_n$
 and constructs an HPM ${\mathcal M}$ satisfying the condition below:

\begin{quote}
Assume $^*$ is an interpretation and $\mathfrak{a},\mathfrak{s},\mathfrak{t}$ are unary arithmetical functions such that:
\begin{description}
  \item[(i)] For each $i\in\{1,\ldots,n\}$,  \ ${\mathcal N}_i$ is an $\mathfrak{a}$ amplitude, $\mathfrak{s}$ space and $\mathfrak{t}$ time solution of $E_{i}^{*}$.
  \item[(ii)] For any $x$, $\mathfrak{a}(x)\geq \max\{x,\mathfrak{c}\}$, where  $\mathfrak{c}$ is the native magnitude of 
{\em $E_1,\ldots,E_n\intimpl F$}.  
  \item[(iii)]   For any $x$,  $\mathfrak{s}(x)\geq \log(x)$.
  \item[(iv)] For any $x$,  $\mathfrak{t}(x)\geq x$ \ and \ $\mathfrak{t}(x)\geq \mathfrak{s}(x)$.\footnote{While this condition is always automatically satisfied in the traditional complexity theory, in our case this is not so. After all, think of a scenario where the machine consumes a lot of space but makes no moves.}
\end{description} 
Then there are numbers $\mathfrak{b},\mathfrak{d}$ which only depend on $\mathbb{P}$ such that, with $\mathfrak{R}$ abbreviating $\mathfrak{a}^{\mathfrak{b}}(\ell)$,  \ ${\mathcal M}$ is a solution of $F^*$ that runs in amplitude $\mathfrak{R}$, space $O\bigl(\mathfrak{s}(\mathfrak{R})\bigr) $ and time 
$O\bigl((\mathfrak{t}(\mathfrak{R}))^{\mathfrak{d}}\bigr)$. 
\end{quote}
\end{thm}

\noindent The rest of the present section is exclusively devoted to a proof of the above theorem. 

 Consider an arbitrary sequent $E_1,\ldots,E_n\intimpl F$ together with a $\cltw$-proof $\mathbb{P}$ of it. Let $\mathfrak{c}$ be the native magnitude of 
$E_1,\ldots,E_n\intimpl F$, and let $\mathfrak{a},\mathfrak{s},\mathfrak{t}$ be functions satisfying conditions (ii), (iii) and (iv) of the theorem. 
By Theorem \ref{feb9c}, there is a well behaved logical solution  $\mathcal K$ of $E_1,\ldots,E_n\intimpl F$ which runs in $\max\{\ell,\mathfrak{c}\}$ amplitude, constant space and linear time,  and such a $\mathcal K$ --- fix it --- can be effectively constructed from $\mathbb{P}$. In view of conditions (ii)-(iv), $\mathcal K$ automatically also runs in amplitude 
$\mathfrak{a}(\ell)$, space $O\bigl(\mathfrak{s}(\ell)\bigr)$ and time $O\bigl(\mathfrak{t}(\ell)\bigr)$.  Assume that, in the context of a given arbitrary 
interpretation (which we notationally suppress as before), ${\mathcal N}_1,\ldots,{\mathcal N}_n$ are HPMs satisfying condition (i) of our theorem. 
Below we describe an  HPM ${\mathcal M}$  such that ${\mathcal M}  $ wins $F$. 
From the description it will be immediately clear that $\mathcal M$ is constructed effectively.

Since the present case is considerably more challenging than the case dealt with in Section \ref{sstm}, we will make a simplifying assumption that the formulas of $E_1,\ldots,E_n\intimpl F$ have no free occurrences of variables. This special case rather easily (in the style of Section \ref{sstm}) generalizes to all cases, and we leave making such a generalization to the reader if necessary. Furthermore, as we did at the beginning of the proof of Theorem \ref{feb9dt}, we will adopt another simplifying assumption, according to which $\mathcal K$ never makes any replicative moves in the antecedent of the sequent, so that the game it plays is (essentially)  
%\marginpar{feb18aa}
\begin{equation}\label{feb18aa}
E_1\mlc \ldots\mlc E_n\mli F.
\end{equation}
Again, this case then generalizes to all cases in the same way and for similar reasons as at the end of Section \ref{sstm}, so the issue of such a generalization will no longer be explicitly addressed in the present proof. 
%For further convenience, we will assume that each of the machines ${\mathcal N}_i$ has a single work tape. Generalizing this case to all cases does not present any problem and is not worth paying our attention. 
Finally, as before, we assume that neither ${\mathcal M}  $'s environment nor any of the machines that it simulates ever make illegal moves.

As in the proof of Theorem   \ref{feb9dt}, the idea for the strategy of ${\mathcal M}  $ is to let $\mathcal K$ (through simulation)  
play against ${\mathcal N}_1  $, \ldots, ${\mathcal N}_n  $ in the corresponding $n$ components of the antecedent of
(\ref{feb18aa}), and mimic, in the real play of $F$, the play of $\mathcal K$ in the consequent of (\ref{feb18aa}). Then, for reasons pointed out in 
the proof of Theorem   \ref{feb9dt}, ${\mathcal M}  $ is guaranteed to win $F$. Furthermore, for exactly the same reasons as before, 
${\mathcal M}  $ is also guaranteed to run in amplitude\label{0amcom} $\mathfrak{R}=\mathfrak{a}^{\mathfrak{b}}(\ell)$ as desired, where, as
in Section \ref{sstm},   $\mathfrak{b}$ is the number of occurrences of choice operators in the $\ada$-closure of (\ref{feb18aa}).\footnote{In the general case, i.e., in the case where $\mathcal K$ may make replications in the antecedent, the number $\mathfrak{b}$ will be greater, but will still remain a constant, depending on the $\cltw$-proof $\mathbb{P}$ of the sequent but not on any moves made in the play.}

So, we only need to understand  how to achieve --- through appropriately (re)arranging the details of ${\mathcal M}  $'s work --- that   
the space and time complexities of ${\mathcal M}  $ are also as promised in the present theorem. In the context of a given computation step of
${\mathcal M}  $, as before, $\ell$ will stand for the background of that (``current'') step.  As noted in Section \ref{sstm}, then the 
``current'' background of any simulated machine can be at most $\mathfrak{R}$. Therefore, as before, in our complexity analysis we will typically think and talk in terms of (the implicitly ``current'' value of)  $\mathfrak{R}$ rather than $\ell$. 

In our present case  ${\mathcal M}  $ can no longer afford to perform simulation in the same way as it did in the preceding section, because the amount $\mathfrak{s}(\mathfrak{R})\geq \log(\mathfrak{R})$ of work-tape space available (asymptotically) to it may be less than the previously available amount  $\mathfrak{s}(\mathfrak{R})\geq \mathfrak{R}$. What caused ${\mathcal M}  $'s high space consumption in Section \ref{sstm}  was the fact that it had represented the contents of the run tapes and the buffers of the simulated machines on its work tapes, which generally takes $O(\mathfrak{R})$ rather than $O\bigl(\mathfrak{s}(\mathfrak{R})\bigr)$ space. In the present case, when dealing with a $j$th computation step of a machine 
${\mathcal G}\in\{{\mathcal K},{\mathcal N}_1  ,\ldots,{\mathcal N}_1  \}$,  we let  ${\mathcal M}  $ in its work-tape memory only keep representations of the other (and some additional, previously redundant)  components of the corresponding configuration of ${\mathcal G}$. 
 Namely, with ``current'' below referring to an arbitrary given $j$th computation step of ${\mathcal G}$, ${\mathcal M}  $ maintains  the following pieces of information --- call them together the {\bf sketch}\label{0sketch} of the $j$th configuration (computation step) of ${\mathcal G}$:
\begin{description}
  \item[1st component:] The current state of ${\mathcal G}$. This only takes a constant amount of space.\footnote{Of course, as in the preceding section, $\mathcal M$ can just as well keep track of this piece of information through its state memory rather than work-tape memory.}
  \item[2nd component:] The current contents of the work tapes of ${\mathcal G}$. This piece of information can be represented with $O\bigl(\mathfrak{s}(\mathfrak{R})\bigr)$ space, because  ${\mathcal G}$ runs in space $\mathfrak{s}$ and the magnitude of no (real or imaginary) move exceeds $\mathfrak{R}$.
  \item[3rd component:]  The current  locations of the work-tape heads of ${\mathcal G}$. The amount of space needed for this obviously does not exceed the preceding amount $O\bigl(\mathfrak{s}(\mathfrak{R})\bigr)$ --- in fact, it is  $O\bigl(\log(\mathfrak{s}(\mathfrak{R}))\bigr)$.   
  \item[4th component:] The current  location of the run-tape head of ${\mathcal G}$. This,  in view  of the fact that the magnitude of no move exceeds $\mathfrak{R}$ and that there is a constant bound (namely, $\mathfrak{b}$) on the maximum number of moves that can emerge on the run tape of ${\mathcal G}$, obviously takes $O\bigl(\log(\mathfrak{R})\bigr)$ space. But, in view of condition (iii) of the theorem, $\mathfrak{s}(\mathfrak{R})\geq \log(\mathfrak{R})$. So,  the present component can be remembered with $O\bigl(\mathfrak{s}(\mathfrak{R})\bigr)$ space. 
\item[5th component:] The number of moves that ${\mathcal G}$ has made so far (at steps $\leq j$) in the play. This number can never exceed $\mathfrak{b}$, so holding it in memory only takes a constant amount of space. 
\item[6th component:] The current number of symbols in the buffer of ${\mathcal G}$. For the same reasons as those relied upon on page \pageref{0reas} when
proving  Theorem 
\ref{feb9dt}, we find that the size of the buffer content of $\mathcal G$  is  $O(\mathfrak{R})$. Hence, recording  this quantity only takes  
 $O\bigl(\log(\mathfrak{R})\bigr)$ space and therefore, as in the case of the 4th component, $O\bigl(\mathfrak{s}(\mathfrak{R})\bigr)$ space.
\item[7th component:] The (possibly empty) string $\alpha$ that has been added to  the buffer of $\mathcal G$ when it made a transition to the $j$th step from the preceding, $(j-1)$th, step; here we assume that if $j=0$, i.e., if there is no preceding step, then the string $\alpha$ is empty. In either case, recording such an $\alpha$, of course, only takes a constant amount of space.  
\item[8th component:] In addition, if ${\mathcal G}$ is $\mathcal K$, the sketch has a record of the first few (bounded number of) symbols from the current content of $\mathcal G$'s  buffer sufficient to determine in which of the $n+1$ components of (\ref{feb18aa}) the move that is being constructed in that buffer is meant. For instance, if $n=2$, then remembering the first four or even three symbols would be sufficient. That is because every (legal) move that $\mathcal K$ makes will have either the prefix $0.0.$ indicating that the move is being made in $E_1$, or the prefix $0.1.$ indicating that the move is being made in $E_2$, or the prefix $1.$ indicating that the move is being made in $F$.
%\footnote{If the first few symbols do not potentially match any of these, then the move that is being constructed in the buffer is illegal and, in view of our assumptions, will never be made; in this case, what is in the buffer is inconsequential, so we may pretend that such a pathological case never occurs.} 
  \end{description}
Summing up all of the above quantities, we find that maintaining the sketch for any step of any given simulated machine ${\mathcal G}$ takes $O\bigl(\mathfrak{s}(\mathfrak{R})\bigr)$ space.\label{0ske}

Unfortunately, the sketch of a given computation step $j$ of ${\mathcal G}$ alone is not sufficient to fully trace the subsequent steps of ${\mathcal G}$ and thus successfully conduct simulation. 
One reason is that, in order to compute (the sketch of) the $(j+1)$th step of ${\mathcal G}$, one needs to know the content of the cell scanned by the run-tape head of ${\mathcal G}$. However, sketches do not keep track of what is on the run tape, and that information --- unless residing on the run tape of ${\mathcal M}  $ itself by good luck\footnote{Namely, the ``lucky'' case is when ${\mathcal G}$ is $\mathcal K$ and it is trying to read some move made by its adversary in the consequent of (\ref{feb18aa}).} --- is generally forgotten. Another reason is that, when ${\mathcal G}$ makes a move on a given step,
${\mathcal M}  $ is unable to act accordingly (copy that move in the real play or use it as another machine's imaginary adversary's move)  because it does not remember the content of the buffer of ${\mathcal G}$ --- this information, just like information on the content of the run tape, is not present in the sketch.  

We handle the above difficulty by letting the simulation routine  recompute the missing information every time such information is needed. This is done  through recursive calls to the routine itself. Required space efficiency here is achieved at the expense of time:  the same computations will generally be performed many times over and over because the procedure, to save space, keeps forgetting certain crucial and reusable results of its previous computations. 

Properly materializing the above general idea requires quite some care though, namely when it comes to the details of how the simulations of different machines should be synchronized. Among the crucial conditions for our recursive procedure to work within the required complexity limits is to make sure that the depth of the recursion stack never exceeds a certain constant bound. 
For simplicity, we design  $\mathcal M$ as a single-work-tape machine,  even though with many work tapes $\mathcal M$ could have achieved a ``somewhat'' better time efficiency. 

Below, by an {\bf agent}\label{0agent} we shall mean either Environment (that is, ${\mathcal M}  $'s real adversary in its play of $F$) or any one of the $n+1$ machines ${\mathcal K}$, ${\mathcal N}_1  ,\ldots,{\mathcal N}_n  $.

We let ${\mathcal M}  $, in addition to the sketches for the simulated steps of the simulated machines, maintain (one single/common copy of) what we call the {\bf global history}.\label{0gh} The latter is a list of all moves made by all $n+2$ agents throughout the mixture of the real and imaginary plays ``so far''. More precisely, this is not a list of moves themselves, but rather entries with certain partial information on those moves. Namely, the entry for each  
 move $\alpha$ 
does not indicate the actual content of $\alpha$ (which would require $O(\mathfrak{R})$ space, thus possibly exceeding the available amount $O(\mathfrak{s}(\mathfrak{R}))$), but rather only the  size of $\alpha$, i.e.  the number of symbols in $\alpha$. 
Recording this quantity only takes $O\bigl(\log(\mathfrak{R})\bigr)$ and --- in view of condition (iii) of the theorem --- $O\bigl(\mathfrak{s}(\mathfrak{R})\bigr)$ space. The entry for $\alpha$ also indicates which of the $n+2$ agents has made (is the {\bf author}\label{0author} of) the move; in addition, if the author of the move is $\mathcal K$, there is  a record of in which of the $n+1$ components of (\ref{feb18aa}) the move was made. Since representing the last two pieces of information only takes a constant amount of space, it does not add anything to the asymptotic size of an entry, so the entry for each move $\alpha$ in the global history takes $O\bigl(\mathfrak{s}(\mathfrak{R})\bigr)$ space. Further, remembering where the number $\mathfrak{b}$ came from, we see that the number of entries in the global history can never exceed $\mathfrak{b}$. Since $\mathfrak{b}$ is constant, we find that ${\mathcal M}  $ only spends the same  $O\bigl(\mathfrak{s}(\mathfrak{R})\bigr)$ amount of space on maintaining the overall global history.\label{0gloh} While a move $\alpha$ is not the same as the entry for it in the global history, in the sequel we will often terminologically identify these two. 

What do we need the global history for? As noted earlier, during its work, ${\mathcal M}  $ will often have to resimulate some already simulated portions of the work of one or another machine ${\mathcal G}$. To make such a resimulation possible, it is necessary to have information on the times at which ${\mathcal G}$'s adversary has made its moves in the overall scenario that we are considering and re-constructing. Recording the actual move {\em times} as they were detected during the initial simulation, however, could take us beyond our space limits (think of a situation where, say, $\mathcal K$ waits ``very long'' before its environment makes a move). So, instead, we only keep track --- via the global history --- of the {\em order} of moves. Then we neutralize the problem of not remembering the ``actual'' times of ${\mathcal G}$'s adversary's moves by simply assuming that ${\mathcal G}$'s adversary always makes its moves instantaneously in response to ${\mathcal G}$'s moves. The point is that, if ${\mathcal G}$ wins its game, it does so in all scenarios, including the above scenario of an instantaneously moving adversary. This is our preliminary and very rough attempt to explain the use of the global history, of course, and more can be understood after seeing the details of ${\mathcal M}  $'s work below.

What follows is a high-level yet relatively detailed description of the work of ${\mathcal M}  $. The latter relies on the three subprocedures called \transition,  \call\ and \main. We start with \transition.

\subsection{Procedure \transition}\label{0us} In the context of a given global history $\mathbb{H}$, this procedure  
takes the sketch $\mathbb{S}^{{\mathcal G}}_{j}$ of a given computation step $j$ of a given machine ${\mathcal G}\in\{{\mathcal K},{\mathcal N}_1  ,\ldots,{\mathcal N}_n  \}$, and returns the sketch $\mathbb{S}^{{\mathcal G}}_{j+1}$ of the next computation step $j+1$ of the same  machine. 

 Let $m$ be the 5th component of $\mathbb{S}^{{\mathcal G}}_{j}$. The number $m$ tells us how many moves ${\mathcal G}$ had made by time $j$. 
In most cases, \transition\ will be used while re-constructing some past episode of ${\mathcal G}$'s work. It is then possible that the global history contains an $(m+1)$th  move by ${\mathcal G}$. If so, then such a move, together with all subsequent moves by whichever agents, are ``future moves'' from the perspective of the $j$th step of ${\mathcal G}$ that \transition\ is currently dealing with.
 This means that, when ``imagining'' the situation at the $j$th step of 
 ${\mathcal G}$, those moves should be discarded. So, let $\mathbb{H}'$ be the result of deleting from the global history $\mathbb{H}$ the $(m+1)$th move of ${\mathcal G}$ and all subsequent moves (if there are no such moves, then simply $\mathbb{H}'=\mathbb{H}$). Next, 
{\em relevant} to ${\mathcal G}$ are only the moves from $\mathbb{H}'$ that are made either by ${\mathcal G}$, or by ${\mathcal G}$'s imaginary adversary. For instance, if ${\mathcal G}$ is ${\mathcal N}_1  $, then the only relevant moves are those made either by ${\mathcal N}_1  $ or by $\mathcal K$ in the $E_1$ component of (\ref{feb18aa}). Correspondingly, let $\mathbb{H}''$ be the result of further deleting from $\mathbb{H}'$ all moves that are not relevant to $\mathcal G$.\footnote{Note that, if ${\mathcal G}={\mathcal K}$, then all moves are relevant to it, so $\mathbb{H}'=\mathbb{H}''$.} Thus, the moves of $\mathbb{H}''$ prefixed  with corresponding labels are exactly the labmoves that $\mathcal G$ would see --- in the same order as they appear in $\mathbb{H}''$ --- on its run tape at step $j$, except that, if $\mathcal G$ is $\mathcal K$, each move of $\mathbb{H}''$ should additionally take an {\em addressing prefix}\label{0adpref} --- a string indicating in which of the $n+1$ components of (\ref{feb18aa}) the move is made. 
For instance, if the author of the move is ${\mathcal N}_1$, then on its run tape $\mathcal K$ will see $\oo 0.0.\alpha$ rather than just $\oo \alpha$. Here ``$0.0.$'' is the addressing prefix of the move,  indicating that $\alpha$ is made in the first conjunct of the antecedent of (\ref{feb18aa}).

The information contained in $\mathbb{S}^{{\mathcal G}}_{j}$ is ``almost'' sufficient for \transition\ to calculate the sought value of $\mathbb{S}^{{\mathcal G}}_{j+1}$. The only missing piece of information is the symbol $s$ scanned by the run-tape head of ${\mathcal G}$ on step $j$.  \transition\ thus needs, first of all, to figure out what that symbol $s$ is. To do this, using the move sizes recorded in the global history, \transition\ computes the 
length $p$ of the ``active'' content of ${\mathcal G}$'s run tape; that is, $p$ is the ordinal number of the rightmost non-blank cell of the run tape of $\mathcal G$ on its $j$th step.  
Next, let $q$ (found in the 4th component of $\mathbb{S}^{{\mathcal G}}_{j}$) be the number indicating the location of the run-tape head of ${\mathcal G}$ 
on step $j$. \transition\ compares $q$ with $p$. If $q>p$, it concludes that $s$ is \blank. Otherwise, $s$ should be a symbol of some labmove 
$\lambda$ residing on $\mathcal G$'s run tape. 
%Let $\mathcal A$ be the author of that $\lambda$ as recorded in the global history. 
Below, by the {\em preamble}\label{0preamble} of 
$\lambda$ we shall mean $\lambda$'s label and addressing prefix,\footnote{$\lambda$ has a (nonempty) addressing prefix if and only if ${\mathcal G}={\mathcal K}$.} and we shall refer to the rest of $\lambda$ as (the move) $\alpha$.  

In a rather similar manner, based on the above $q$ and using the relevant information contained in $\mathbb{H}$,  \transition\  figures out whether $s$ is in $\lambda$'s preamble or in $\lambda$'s (``main'') $\alpha$  part. If $s$ resides in the preamble,  \transition\ directly figures out what particular symbol $s$ is. Otherwise, if $s$ resides within $\alpha$,   \transition\ 
 finds two integers $X$ and $Y$. Here $X$ is the number of moves made by $\alpha$'s author $\mathcal A$ before this agent made the move $\alpha$. And $Y$ is the ordinal number of the sought symbol $s$ within $\alpha$.
If $\mathcal A$ is (the real) Environment,\footnote{And, hence, ${\mathcal G}$ is $\mathcal K$.} using $X$ and $Y$, \transition\ finds $s$ on the run tape of 
${\mathcal M}$. In any other  case, \transition\ calls the below-described procedure \call\ on $({\mathcal A},X,Y)$. As will be seen later, \call\  then returns the sought symbol $s$. Thus, in any case, \transition\ now knows the symbol $s$ read by the run-tape head of ${\mathcal G}$ on step $j$. Omitting  details, we just  want to once again point out that 
doing all of the above --- except executing the \call\ subroutine --- only requires some straightforward logarithmic space arithmetic in the style seen 
in the preceding paragraph.

Keeping the above $s$ in mind, \transition\ now additionally consults $\mathbb{S}^{{\mathcal G}}_{j}$ (namely, the 2nd and the 3rd components of it) for 
the symbols scanned by the work-tape heads of ${\mathcal G}$ on step $j$, as well as for the state of ${\mathcal G}$ on that step (the 1st component of 
$\mathbb{S}^{{\mathcal G}}_{j}$). Using this information and its knowledge of ${\mathcal G}$'s transition function, \transition\ is now able to find the 
state of ${\mathcal G}$ on step $j+1$, the directions in which its scanning heads moved when transitioning from step $j$ to step $j+1$, the symbols
 by which the old scanned symbols were replaced on the work tapes, the string added to the buffer on the transition, the new size of (number of symbols 
in) the buffer, the new number of ${\mathcal G}$-authored moves (the latter will be $m$ if the state on step $j+1$ is not a move state, and $m+1$ otherwise), and 
the first few symbols of the buffer if required. To summarize omitting the straightforward details of these calculations, \transition\ finds 
(all 8 components of) the sought sketch $\mathbb{S}^{{\mathcal G}}_{j+1}$.

\subsection{Procedure \call}\label{sis}
%\marginpar{sis}
 In the context of a given global history $\mathbb{H}$, this procedure takes a machine ${\mathcal G}\in\{{\mathcal K},{\mathcal N}_1  ,\ldots,{\mathcal N}_n  \}$ and two numbers $X,Y$, where $X$ is smaller than the number of ${\mathcal G}$'s moves in the global history,
and $Y$ is a number not exceeding the length of ${\mathcal G}$'s $(X+1)$th move there. The goal of \call\ is to return, through rerunning ${\mathcal G}$, the $Y$th symbol of the $(X+1)$th move of ${\mathcal G}$.\footnote{Here, if ${\mathcal G}={\mathcal K}$, the addressing prefix of the move should be ignored when counting its symbols.}

To achieve the above goal, \call\ creates a variable $\mathbb{S}$ to hold a sketch of ${\mathcal G}$, and sets the initial value of $\mathbb{S}$ to the {\bf initial sketch}\label{0isk} of ${\mathcal G}$. By the latter we mean the sketch of the initial configuration of ${\mathcal G}$, i.e. the configuration where ${\mathcal G}$ is in its start state, the buffer and the work tapes are empty,\footnote{As for the run tape, what is on it is irrelevant because a sketch has no record of the run-tape content anyway.} and all scanning heads are looking at the leftmost cells of their tapes. 

After the above initialization step, \call\ performs the following subprocedure:

\begin{enumerate}
  \item Remember the 5th and 6th components --- call them $Z$ and $T$, respectively --- of $\mathbb{S}$, and then perform \transition\ on $\mathbb{S}$. Let $\mathbb{S}'$ be the resulting sketch, and let $\sigma$ be the 7th component of $\mathbb{S}'$. Below, as always, $|\sigma|$ means the length of (number of symbols in) $\sigma$. Also, let $\mathfrak{r}$ be the length of the addressing prefix of the $(X+1)$th $\mathcal G$-authored move recorded in $\mathbb{H}$; it is understood here that  $\mathfrak{r}=0$ if ${\mathcal G}\not={\mathcal K}$. 
  \item If $Z=X$ and $T<Y+\mathfrak{r} \leq T+|\sigma|$, then return the $(Y+\mathfrak{r}-T)$th symbol of $\sigma$. Otherwise, update (the value of) 
 $\mathbb{S}$  to $\mathbb{S}'$, destroy your memory of  $Z$ and $T$ to recycle space, and go back to step 1. 
\end{enumerate}

 \subsection{Procedure \main}\label{0mh} For the convenience of description, below we assume that the number $n$ from (\ref{feb18aa}) is at least $1$. The case of $n=0$ is simpler if not trivial.

The procedure \main\ takes a global history $\mathbb{H}$ as an argument  and, treating $\mathbb{H}$ as a variable that may undergo updates,  acts according to the following prescriptions: 
\begin{description}
  \item[Stage $1$:] Create  variables 
$\mathbb{S}^{\mathcal K}$, $\mathbb{S}^{{\mathcal N}_1}$, \ldots, $\mathbb{S}^{{\mathcal N}_n}$, each one to hold a sketch of the corresponding simulated machine from the list 
${\mathcal K}$, ${\mathcal N}_1  $, \ldots, ${\mathcal N}_n  $. Initialize these variables to the initial sketches (see Subsection \ref{sis}) of the corresponding machines. Proceed to Stage $2$. 
\item[Stage $2$:] See if Environment has made a new move (this can be done, say, by counting Environment's moves on ${\mathcal M}  $'s run tape, and comparing their number with the number of Environment-authored  moves recorded in the global history). If yes, update the global history $\mathbb{H}$ by adding to it a record for that move, and repeat \main. If not, go to Stage $3$. 
\item[Stage $3$:]  \ 
\begin{description}
  \item[(a)] Perform \transition\ on $\mathbb{S}^{\mathcal K}$. Let $\mathbb{T}^{\mathcal K}$ be the resulting sketch.
  \item[(b)] If $\mathcal K$ did not make a {\bf globally new}\label{0gnm} move on its transition from $\mathbb{S}^{\mathcal K}$ to $\mathbb{T}^{\mathcal K}$,\footnote{Here and later in similar contexts, we terminologically identify sketches with the corresponding steps of the corresponding machines.} change the value of the variable $\mathbb{S}^{\mathcal K}$ to $\mathbb{T}^{\mathcal K}$, and  proceed to Stage $4$.  
Here and later in similar contexts, by a ``globally new'' move we mean a move not recorded in the global history $\mathbb{H}$. Figuring out whether $\mathcal K$ made a globally new move is easy. Technically, $\mathcal K$ made a globally new move if and only if, firstly, it {\em did} make a move, i.e., the 1st component of $\mathbb{T}^{\mathcal K}$ is a move state; and secondly, such a move is not recorded in $\mathbb{H}$, meaning that the 5th component of $\mathbb{T}^{\mathcal K}$ exceeds the total number of ${\mathcal K}$'s moves recorded in $\mathbb{H}$. 
  \item[(c)] If $\mathcal K$ made a globally new move and the move was in one of the $E_i$ components of (\ref{feb18aa}), update $\mathbb{H}$ by adding to it a record for that move, and repeat \main. 
  \item[(d)] Suppose now $\mathcal K$ made a globally new move --- call it $\alpha$ --- in the $F$ component of (\ref{feb18aa}). Let $X$ be the 5th component of 
 $\mathbb{S}^{\mathcal K}$, and $Y$ be the 6th component of $\mathbb{S}^{\mathcal K}$. Thus, $Y$ is the size of the string $1.\alpha$, and $X$ is the number of moves that $\mathcal K$ had made before it made the move $\alpha$. In this case, call \call\ $Y-2$ times: first on $({\mathcal K},X,3)$, then on $({\mathcal K},X,4)$, \ldots, finally on $({\mathcal K},X,Y)$. Copy to the buffer (of ${\mathcal M}  $) each of the $Y-2$ symbols returned by these calls.
% except for the first two symbols comprising the string ``$1.$''. 
Once all calls have been made, go to a move state. Notice that this results in ${\mathcal M}  $ making the move $\alpha$  in the real play. Now update the global history $\mathbb{H}$ by adding to it a record for the move $\alpha$ made by $\mathcal K$, and repeat \main. 
\end{description}
\item[Stage ($3+i$) ($1\leq i\leq n$):] \ 
\begin{description} 
  \item[(a)] Perform \transition\ on $\mathbb{S}^{{\mathcal N}_i}$. Let $\mathbb{T}^{{\mathcal N}_i}$ be the resulting sketch. 
  \item[(b)] If ${\mathcal N}_i  $ did not make a globally new move (in the sense explained for $\mathcal K$ in Stage 3) on its transition from $\mathbb{S}^{{\mathcal N}_i}$ to $\mathbb{T}^{{\mathcal N}_i}$,  then update the value of $\mathbb{S}^{{\mathcal N}_i}$ to $\mathbb{T}^{{\mathcal N}_i}$. After that, if $i<n$, go to Stage $3+i+1$, and if $i=n$, go to Stage $2$. 
  \item[(c)] If ${\mathcal N}_i  $ made a globally new move, then update the global history $\mathbb{H}$ by adding to it a record for that move, and repeat \main.    
\end{description}  
\end{description}

\subsection{The overall strategy and an example of its run} The overall strategy followed by ${\mathcal M}  $ consists in creating the variable $\mathbb{H}$, initializing it to the empty global history, and then (forever) running \main. 

Note that \main\ will be iterated at most $\mathfrak{b}$ times, because every iteration increases the number of moves in $\mathbb{H}$, and that number, as already observed, cannot exceed $\mathfrak{b}$. Since \main\  is restarted only finitely many times, the last iteration of it never terminates. 

Let us look at an example scenario to make sure we understand the work of  ${\mathcal M}  $.  For simplicity, assume $n=1$, so that (\ref{feb18aa}) is $E_1\mli F$. At the very beginning of its work,  ${\mathcal M}  $ creates the variable $\mathbb{H}$ and sets its value to the empty global history $\seq{}$. The rest of its work just consists in running \main. So, in what follows, we can use ``${\mathcal M}  $'' and ``\main'' as synonyms.

During Stage $1$, \main\ creates 
two variables $\mathbb{S}^{\mathcal K}$ and $\mathbb{S}^{{\mathcal N}_1}$, and sets their values to the initial sketches of ${\mathcal K}$ and ${\mathcal N}_1  $, respectively. 
The result of this step reflects the start situation,  where ``nothing has happened yet'' in the mixture of the real play of
 $F$ by ${\mathcal M}  $ and the simulated plays of $E_1\mli F$ and  $E_1$  by the machines $\mathcal K$ and ${\mathcal N}_1  $.  

Now \main\ starts performing, over and over,  Stages $2$ through $4$. The work in those stages can be characterized as ``global simulation''. This is a routine that keeps updating, in turn and one step at a time, the two sketches $\mathbb{S}^{\mathcal K}$ (Stage $3$) and $\mathbb{S}^{\mathcal N}_1$ (Stage $4$) to the sketches of the ``next configurations'' of the corresponding machines in the scenario where the adversaries of those machines have made no moves; simultaneously, \main\ keeps checking (Stage $2$) the run tape of ${\mathcal M}  $ to see if Environment has made a move. This will continue until either Environment or one of the two simulated machines is detected to make a move. In our example, let us imagine that Environment was the first of the three agents to make a move, and such a move was $\alpha_1$. What happens in this case? 

\main\ simply restarts the global simulation by resetting (in Stage 1) the two sketches $\mathbb{S}^{\mathcal K}$ and 
$\mathbb{S}^{{\mathcal N}_1}$ to the initial sketches of ${\mathcal K}$ and ${\mathcal N}_1  $. The earlier-described ``Stage $2$ through Stage $4$ over and over'' routine will be repeated, with the only difference that the global history $\mathbb{H}$ is now showing the presence of the Environment-authored $\alpha_1$. This means that the simulation of the machine $\mathcal K$ will now proceed in the scenario where, at the very beginning of the play, $\mathcal K$'s adversary had made the move $\alpha_1$ in the $F$ component. So,  every time the simulated $\mathcal K$ tries to read one of the symbols of $\alpha_1$ on its imaginary run tape, \main --- ${\mathcal M}  $, that is --- looks that symbol up on its own run tape. As for ${\mathcal N}_1  $, 
its (re)simulation  will proceed exactly as before (during the first iteration of \main), for its imaginary adversary has not yet made any moves. By switching to this new scenario, \main, in fact, deems the previous scenario (the scenario where none of the agents had made any moves) invalid, and simply forgets about it.
This new,  2nd attempt of global simulation (the second iteration of \main, that is) will continue until one of the three agents, again, is detected to make a move. 

Let us say it is again Environment, which makes move $\alpha_2$. Then \main\ again correspondingly updates $\mathbb{H}$, deems the previous global simulation scenario invalid, forgets it and restarts global simulation for the scenario where, from the very beginning, $\mathcal K$'s adversary had made the two moves $\alpha_1$ and $\alpha_2$ (and the adversary of ${\mathcal N}_1  $,
again, had made no moves).
This 3rd attempt of global simulation will continue until, again, a move is detected by one of the three agents. 

Let us say this time it is $\mathcal K$, which makes move $\beta$ in the consequent of $E_1\mli F$. In this event, through invoking \call\ as many times as the length of $\beta$, \main\ --- ${\mathcal M}  $, that is --- assembles $\beta$ in its buffer symbol-by-symbol, and then makes the move $\beta$ in the real play. After that, as always when a move by one of the agents is detected, the global simulation restarts. Now the global history $\mathbb{H}$ is showing (records for) the sequence $\seq{\alpha_1,\alpha_2,\beta}$ of three moves. In the present, 4th attempt of global simulation, ${\mathcal N}_1  $ is (re)simulated again in the scenario where no moves had been made by ${\mathcal N}_1  $ or its adversary. Similarly, as in the previous case, 
$\mathcal K$ is resimulated in the scenario where, at the very beginning, its adversary had made the two moves $\alpha_1$ and $\alpha_2$. The only difference between the present attempt of global simulation and the previous one is that, once $\mathcal K$ is detected to make the expected move $\beta$, nothing special happens. Namely, the global history is not updated (as $\beta$ is already there), the move $\beta$ is not made in the real play (because it has already been made), and the global simulation continues in the ordinary fashion rather than restarts. 
The present attempt of global simulation, again, will be interrupted if and when one of the agents is detected to make a globally new move, i.e. a move not recorded in the global history. 

Let us say it is again Environment, which makes move $\alpha_3$. As always, a record for $\alpha_3$ is added to $\mathbb{H}$ and the global simulation restarts. For ${\mathcal N}_1  $ everything will be the same as before. As for $\mathcal K$, its resimulation will start in the scenario where, at the beginning of the play, its adversary had made the moves $\alpha_1$ and $\alpha_2$. We already know that, in this scenario, sooner or later, $\mathcal K$ will make its previously detected move $\beta$. Once this event is detected, ${\mathcal K}$'s simulation continues for the scenario where its adversary responded by the move $\alpha_3$ {\em immediately} after $\mathcal K$ made the move $\beta$. 

Let us now imagine that the above, fifth attempt of global simulation detects that $\mathcal K$ has made a move $\gamma$ in the $E_1$ component of $E_1\mli F$. As always, $\mathbb{H}$ is correspondingly updated and the global simulation restarts. For $\mathcal K$, it will be in the same scenario as before (the adversary had made the moves $\alpha_1$ and $\alpha_2$ in the $F$ component right at the beginning of the play, and made the third move $\alpha_3$ there right after $\mathcal K$ moved $\beta$), with the only difference that detecting $\mathcal K$'s expected move $\gamma$ does not cause \main\ to restart. As for ${\mathcal N}_1  $, this time it will be simulated in the scenario where, at the very beginning of the play, its adversary had made the move $\gamma$. So, every time the simulated ${\mathcal N}_1  $ tries to read a symbol of $\gamma$ from its imaginary run tape, \call\ is called to get that symbol and feed it to the simulation. 

Imagine that the final globally new move detected was one by ${\mathcal N}_1  $. Call that move $\delta$.\footnote{It does not matter whether the event of ${\mathcal N}_1  $ making the move $\delta$ was detected before or after the simulated $\mathcal K$ already made both of its expected moves $\beta$ and $\gamma$.}  The global simulation again restarts with the correspondingly updated $\mathbb{H}$. 
${\mathcal N}_1  $ is simulated exactly as in the previous scenario, with the only difference that nothing happens (namely, the 
global simulation does not restart) when ${\mathcal N}_1  $ makes its (expected) move $\delta$. As for $\mathcal K$, it is also simulated 
as in the previous case, with the only difference that, after $\mathcal K$ makes its move $\gamma$, the simulation continues its work ``imagining'' that the adversary immediately responded with the move $\delta$ in the $E_1$ component. So, every time $\mathcal K$ is reading a symbol of $\delta$ on its 
imaginary run tape, that symbol is found calling \call. In contrast, when reading a symbol of $\alpha_1$, $\alpha_2$ or $\alpha_3$, that symbol 
is found on the run tape of ${\mathcal M}  $. 

The last attempt of global simulation (the one that never got discarded/reconsidered) corresponds to the ``ultimate'' scenario that determined  ${\mathcal M}  $'s real play. Namely, in our present example, the ``ultimate'' scenario in which ${\mathcal N}_1  $ was simulated was that, at the very beginning of the play, ${\mathcal N}_1  $'s adversary had made the move $\gamma$, to which ${\mathcal N}_1  $ later responded by $\delta$, and no moves were made ever after. As for $\mathcal K$, the ultimate scenario for it was that, at the very beginning, $\mathcal K$'s adversary had made the moves $\alpha_1$ and $\alpha_2$ in the $F$ component of $E_1\mli F$, to which $\mathcal K$ later responded by the move $\beta$ in the same component; to $\beta$, the adversary instantaneously responded by the move $\alpha_3$ in the $F$ component; after that, at some point, $\mathcal K$ made the move $\gamma$ in the $E_1$ component, to which the adversary instantaneously responded by the move $\delta$ in the same component, and no further moves were ever made in the game. Since it is our assumption that ${\mathcal N}_1  $ wins $E_1$, the two-move run (consisting of $\gamma$ and $\delta$) that took place in the antecedent of $E_1\mli F$ is won by ${\mathcal N}_1  $ and hence lost by ${\mathcal K}$. So, since $\mathcal K$ wins the overall game $E_1\mli F$, the four-move run (consisting of $\alpha_1,\alpha_2,\beta,\alpha_3$) that took place in the consequent of $E_1\mli F$ is won by $\mathcal K$. But the same run was generated in the real play of $F$, which makes ${\mathcal M}  $ the winner. 

Why do we need to restart the global simulation every time a globally new move is detected? The reason is that otherwise we generally would not be able to rely on calls to \call\ for obtaining required symbols.  Going back to our example, imagine we did not restart the global simulation (\main) after the moves $\alpha_1$, $\alpha_2$ and $\alpha_3$ were made by Environment. Perhaps (but not necessarily), as before, $\mathcal K$ would still make its move $\beta$ sometime between $\alpha_2$ and $\alpha_3$, and then the move $\gamma$ after  $\alpha_3$. Fine so far. But the trouble starts when, after that event, ${\mathcal N}_1  $ tries to read some symbol of $\gamma$ from its imaginary run tape. A way to provide such a symbol is to invoke \call, which will resimulate $\mathcal K$ to find that symbol. However, in order to properly resimulate $\mathcal K$ up to the moment when it made the move $\gamma$ (or, at least, put the sought symbol of the latter into its buffer), we need to know when (on which computation steps of $\mathcal K$), exactly, the moves $\alpha_1$, $\alpha_2$ and $\alpha_3$ emerged on $\mathcal K$'s run tape. Unfortunately,  we do not remember this piece of information, because, as noted earlier, remembering the exact times (as opposed to merely remembering the order) of moves may require more space than we possess. So, instead, we assume that the moves $\alpha_1$ and $\alpha_2$ were made right at the beginning of $\mathcal K$'s play, and then $\alpha_3$ was made right after $\mathcal K$ made its move $\beta$. This assumption, however, disagrees with the scenario of the original simulation, where $\alpha_1$ was perhaps only made at step $100$, $\alpha_2$ at step $200$, and $\alpha_3$ perhaps $9999$ steps after $\beta$. Therefore, there is no guarantee that $\mathcal K$ will still generate the same move $\gamma$ in response to $\alpha_3$. Restarting the global simulation --- as we did --- right after $\alpha_1$ was made,  then restarting it again after $\alpha_2$ was detected, and then restarting it again after Environment made the move $\alpha_3$ in response to $\beta$, neutralizes this problem. If $\mathcal K$ made its move $\gamma$ after $\alpha_3$ in this new scenario (the scenario where its imaginary adversary always acted instantaneously), then every later resimulation, no matter how many times \main\ is restarted, will again take us to the same move $\gamma$ made after $\alpha_3$, because the global history, which ``guides''  resimulations, will always be showing the first four moves in the order $\alpha_1,\alpha_2,\beta,\alpha_3$. To see this, note that all updates of the global history only add some moves to it, and otherwise do not affect the already recorded moves or their order.

We also want to understand one remaining issue. As we should have noticed, \call\ always calls \transition, and the latter, in turn, may again call \call. Where is a guarantee that infinitely many or ``too many'' nested calls will not occur? Let us again appeal to our present example, and imagine we (\transition, that is) are currently simulating a step of $\mathcal K$ sometime after ${\mathcal N}_1  $ already has made its move $\delta$. Whenever $\mathcal K$ tries to read a symbol of $\delta$, \call\ is called to resimulate  ${\mathcal N}_1  $ and find that symbol. While resimulating ${\mathcal N}_1  $, however, we may find that, at some point, its run-tape head is trying to read a symbol of $\gamma$. To get that symbol, \call\ will be again called  to resimulate $\mathcal K$ (the author of $\gamma$) and find that symbol. Can this process of mutual resimulations go on forever? Not really. Notice that, when \call\ is called on $\mathcal K$ to find the sought symbol of $\gamma$, \call, guided by the global history, will resimulate $\mathcal K$ only up to the moment when it made the move $\gamma$. But during that episode of $\mathcal K$'s work, the move $\delta$ was not yet on its run tape. So, \call\ will not have to be further called on ${\mathcal N}_1  $. Generally, as we are going to see in the next subsection, at any time there can be at most $\mathfrak{b}$ nested invocations of \call\ or \transition.

\subsection{Complexity analysis}

Let $\mathbb{H}$ be a global history, $m$ a natural number, and $\mathcal G$ one of the machines ${\mathcal K}$, ${\mathcal N}_1  $, \ldots, ${\mathcal N}_n  $. We define the {\bf $\mathbb{H}$-index} of the pair $({\mathcal G},m)$ as the number of moves in $\mathbb{H}'$, where $\mathbb{H}'$ is the result of deleting from $\mathbb{H}$ the $(m+1)$th move of ${\mathcal G}$ and all subsequent moves by any agents; if here $\mathbb{H}$ does not contain more than $m$ moves authored by $\mathcal G$, then $\mathbb{H}'$ is simply $\mathbb{H}$. Next, where 
$\mathbb{S}^{\mathcal G}$ is a sketch of ${\mathcal G}$, we define the {\bf $\mathbb{H}$-index} of $\mathbb{S}^{\mathcal G}$  as the $\mathbb{H}$-index of $({\mathcal G},m)$, where $m$ is (the value of) the 5th component of $\mathbb{S}^{\mathcal G}$. We extend the concept of $\mathbb{H}$-index to particular runs/iterations  of \transition\ and \call\ in the process of performing \main. Namely, \transition\ is always run on a sketch $\mathbb{S}^{\mathcal G}$ of a machine $\mathcal G$, and we define the {\bf $\mathbb{H}$-index} of that  run of \transition\ as the $\mathbb{H}$-index of $\mathbb{S}^{\mathcal G}$. Similarly, 
\call\ is always called on a triple $({\mathcal G},X,Y)$ for some machine $\mathcal G$ and numbers $X$ and $Y$, and we define the {\bf $\mathbb{H}$-index} of such a  call/run of \transition\ as the $\mathbb{H}$-index of the pair $({\mathcal G},X)$ ($Y$ is thus irrelevant here). If $\mathbb{H}$ is fixed or clear from the  context, as it always is when we talk about a given iteration of \main,\footnote{Remember that each iteration of \main\ deals with one single value of $\mathbb{H}$ given at the very beginning of the iteration; the value of $\mathbb{H}$ only changes on a transition from one iteration of 
\main\ to another.} \ we may omit ``$\mathbb{H}$-'' and simply say ``{\bf index}''.\label{0index}

 \begin{lem}\label{mmm}
%\marginpar{mmm}
In the process of any given iteration of \main\ with argument $\mathbb{H}$, we have: 
\begin{enumerate}[label=\arabic*.]
\item The $\mathbb{H}$-index of any run of \transition\  does not exceed $\mathfrak{b}$.

\item Whenever a given  run of \transition\ calls \call, the $\mathbb{H}$-index of the callee is strictly smaller than  that of the caller.  

\item Whenever given run of \call\ calls \transition, the $\mathbb{H}$-index of the callee does not exceed 
that of the caller.
\end{enumerate}
\end{lem}

\begin{proof} Clause 1 is immediate from the obvious fact that an index can never exceed the number of moves in the global history, and the latter, in view of our assumption that no agents ever make illegal moves, cannot exceed $\mathfrak{b}$. Clauses 2 and 3 can be verified through a rather straightforward (albeit perhaps somewhat long) analysis of the two procedures \transition\ and \call;  details of such an analysis are left to the reader. \end{proof}

That the amplitude complexity of ${\mathcal M}  $ is as required was observed earlier in the present proof on page \pageref{0amcom}. 

Next, we examine space complexity. The space consumption of ${\mathcal M}  $ comes from the need to simultaneously maintain the global history and various sketches of the simulated machines. As observed earlier, maintaining the global history consumes $O\bigl(\mathfrak{s}(\mathfrak{R})\bigr)$ space (page \pageref{0gloh}), and each sketch also consume $O\bigl(\mathfrak{s}(\mathfrak{R})\bigr)$ space (page \pageref{0ske}). At any given time, the global history is kept in memory in a single copy. So, to show that the overall space consumption is $O\bigl(\mathfrak{s}(\mathfrak{R})\bigl)$, we need to show that, at any given time, the number of sketches simultaneously kept in the memory of ${\mathcal M}  $ does not exceed a certain constant. But this is indeed so. Looking 
back at the work of \main, we see that,  at any time, its top level simultaneously maintains the constant number $n+1$ of sketches: $\mathbb{S}^{\mathcal K},\mathbb{S}^{{\mathcal N}_1},\ldots,\mathbb{S}^{{\mathcal N}_n}$, one for each  simulated machine. It also keeps going through these sketches and updating them through \transition, one after one and one step at a time. Since updates are done sequentially rather than in parallel, space used for them can be recycled, so that space consumptions for updating different sketches (this includes not only the top-level $n+1$ sketches of \main, but also many additional sketches that will emerge during calls to \call\ when updating each individual sketch) do not add together unless those sketches happen to be on a same branch of nested recursive calls that \transition\ and \call\ make to each other. In view of Lemma \ref{mmm}, however, the depth of recursion (the height of the recursion stack at any time) is bounded, because the index of \transition\ in the topmost level of recursion does not exceed $\mathfrak{b}$, and every pair of successor levels of recursion strictly decreases the index of the corresponding call of \transition.

Finally, we look at time complexity. Remember that our goal here is to show that ${\mathcal M}  $ runs in time polynomial  in
$\mathfrak{t}(\mathfrak{R})$ --- namely, time $O\bigl((\mathfrak{t}(\mathfrak{R}))^\mathfrak{d}\bigr)$ for some constant number $\mathfrak{d}$ that only depends on the proof $\mathbb{P}$ of the sequent. Our discourse throughout the rest of this section should be understood in the context of some arbitrary but fixed computation branch of $\mathcal M$.

During the entire work of ${\mathcal M}  $, \main\ will be iterated at most $\mathfrak{b}$ times. This is so because each iteration 
strictly increases the number of moves in the global history, which, as already observed, can never exceed $\mathfrak{b}$. The last iteration of 
\main\ will run forever, but ${\mathcal M}  $ will not be billed for that time because it makes no moves during that period. Likewise, ${\mathcal M}  $ will not be billed for the time spent on an iteration of \main\ that was interrupted at Stage 2, because a move by Environment resets $\mathcal M$'s time counter to $0$. Call all other (other than the above two) sorts of iterations of \main\ {\bf time-billable}.\label{0tmb} So, it 
is  sufficient for us to understand how much time a single time-billable iteration of \main\ takes. Pick  any such iteration and fix it throughout the 
context of the rest of this subsection, including the forthcoming Lemmas \ref{mmmm} and \ref{nnnn}. We 
let $\ell$ denote the background of the last clock cycle of that iteration and, as before, use $\mathfrak{R}$ as an abbreviation of $\mathfrak{a}^{\mathfrak{b}}(\ell)$. 

\begin{lem}\label{mmmm} \ 
%\marginpar{mmmm}
\begin{enumerate}
\item The time consumed by any single run of \transition\ of index $i$  is  $O\bigl((\mathfrak{t}(\mathfrak{R}))^{i+1}\bigr)$.

\item The time consumed by any single run of \call\ of index $i$  is
  $O\bigl((\mathfrak{t}(\mathfrak{R}))^{i+2}\bigr)$.
\end{enumerate}
\end{lem}
\begin{proof} We verify this lemma by induction on the index of the corresponding call/run of \transition\ or \call. 

Assume  the index of a given run of \transition\ is  $i\geq 0$. Looking back at our description of \transition, we see that this routine makes at most one call of \call. First, assume no such call is made.  With some analysis it is clear that \transition\ in this case spends $O\bigl(\mathfrak{s}(\mathfrak{R})\bigr)+O(\mathfrak{R})$ time. In view of condition (iv) of the theorem, this quantity does not exceed $O\bigl(\mathfrak{t}(\mathfrak{R})\bigr)$. So, the time consumption is $O\bigl(\mathfrak{t}(\mathfrak{R})\bigr)$ and hence, of course, also $O\bigl((\mathfrak{t}(\mathfrak{R}))^{i+1}\bigr)$. 
Now assume \transition\ {\em does} call \call. 
By clause 2 of Lemma \ref{mmm}, the index $j$ of such a call is less than $i$. Hence, by the induction hypothesis, the time taken by the call 
 is 
$O\bigl((\mathfrak{t}(\mathfrak{R}))^{j+2}\bigr)$. Since $j<i$, we may just as well say $O\bigl((\mathfrak{t}(\mathfrak{R}))^{i+1}\bigr)$ instead. In addition to this, \transition\ only spends the same amount $O\bigl((\mathfrak{t}(\mathfrak{R}))^{i+1}\bigr)$ of time to complete its work as in the preceding case. Thus, in either case,  the time consumption of \transition\ is  $O\bigl((\mathfrak{t}(\mathfrak{R}))^{i+1}\bigr)$.

 Now consider a run of \call, and let $i\geq 0$ be its index. By clause 3 of Lemma \ref{mmm}, the index of any call of \transition\ that the given run of \call\  makes is at most $i$. By the induction hypothesis,  each such call of \transition\ consumes at most $O\bigl((\mathfrak{t}(\mathfrak{R}))^{i+1}\bigr)$ time.  Processing any such  call (doing additional work related to it), in turn, obviously takes at most $O\bigl(\mathfrak{t}(\mathfrak{R})\bigr)$ time. So, each call of \transition\ costs our run of \call\ at most 
$O\bigl((\mathfrak{t}(\mathfrak{R}))^{i+1}\bigr)$ time. How many such calls of \transition\ will \call\ make? Remembering our assumption that each machine runs in time $\mathfrak{t}$, with a little thought one can see that the number of calls of \transition\ is at most $O\bigl(\mathfrak{t}(\mathfrak{R})\bigr)$. So, the overall time cost of the run of \call\ is $O\bigl(\mathfrak{t}(\mathfrak{R})\bigr)\times O\bigl((\mathfrak{t}(\mathfrak{R}))^{i+1}\bigr)= O\bigl((\mathfrak{t}(\mathfrak{R}))^{i+2}\bigr)$. 
\end{proof}

\begin{lem}\label{nnnn}
%\marginpar{nnnn}
No single run of \transition\ or \call\  consumes more than  $O\bigl((\mathfrak{t}(\mathfrak{R}))^{\mathfrak{b}+1}\bigr)$ time. 
\end{lem}

\begin{proof} As we already know from clause 1 of Lemma \ref{mmm}, the index of no run of \transition\ may ever exceed $\mathfrak{b}$. So, by clause 1 of Lemma \ref{mmmm}, no run of \transition\ consumes more than $O\bigl((\mathfrak{t}(\mathfrak{R}))^{\mathfrak{b}+1}\bigr)$ time. As for \call, this procedure is only run when \transition\ calls it. The maximum possible index of the caller, as just noted, is $\mathfrak{b}$. Therefore, by clause 2 of Lemma \ref{mmm}, the maximum possible index of any given run of \call\ is $\mathfrak{b}-1$. Then, by clause 2 of Lemma \ref{mmmm}, such a run consumes at most $O\bigl((\mathfrak{t}(\mathfrak{R}))^{(\mathfrak{b}-1)+2}\bigr)=O\bigl((\mathfrak{t}(\mathfrak{R}))^{\mathfrak{b}+1}\bigr)$ time. \end{proof}

We are now ready to look at the time consumption of the single time-billable iteration of \main\ fixed earlier. 

Stage 1 of \main\ obviously takes a constant amount of time, and this stage is iterated only once. So, asymptotically, it contributes nothing to the overall time consumption of the procedure.  

Stage 2 of a time-billable iteration of \main\ obviously takes $O(\mathfrak{R})$ time and hence, in view of condition (iv) of the theorem, not more than  $O\bigl(\mathfrak{t}(\mathfrak{R})\bigr)$ time. Namely, checking out the run tape may require moving the run-tape head of ${\mathcal M}  $ (at most) from one end  of (the non-trivial, i.e. non-blank part of) the tape to the other end, and the length of that segment of the tape is at most $\mathfrak{b}$ times $\mathfrak{R}$. 
%Additionally, the global history needs to be updated, but  this can be dove even faster.  

Stage 3 starts with performing \transition\ (substage (a)), and this, by Lemma \ref{nnnn}, takes $O\bigl((\mathfrak{t}(\mathfrak{R}))^{\mathfrak{b}+1}\bigr)$ time. With a little thought, the time taken by substages (b) and (c) of Stage 3 can be seen to be at most linear  in  $\mathfrak{R}$ and hence in 
$\mathfrak{t}(\mathfrak{R})$. So is the time taken by substage (d) without counting the calls of \call\ that it performs. Each call of \call, by Lemma \ref{nnnn}, additionally takes $O\bigl((\mathfrak{t}(\mathfrak{R}))^{\mathfrak{b}+1}\bigr)$ time, and altogether substage (d) performs at most $O(\mathfrak{R})$  and thus $O\bigl(\mathfrak{t}(\mathfrak{R})\bigr)$ calls. So, the overall time consumption of substage (d) is $O\bigl((\mathfrak{t}(\mathfrak{R})^{\mathfrak{b}+2})\bigr)$.

A similar but simpler analysis applies to each stage $3+i$ ($1\leq i\leq n$), after which we find that the time consumption of such a stage is $O\bigl((\mathfrak{t}(\mathfrak{R}))^{\mathfrak{b}+1}\bigr)$. 

To summarize, none of the $3+n$ stages of the iteration of \main\ takes more than $O\bigl((\mathfrak{t}(\mathfrak{R}))^{\mathfrak{b}+2}\bigr)$ time. Stage 1 is repeated only once, and the remaining stages are repeated at most $O\bigl(\mathfrak{t}(\mathfrak{R})\bigr)$ times as can be seen with a little thought, keeping in mind that the iteration of \main\ that we are dealing with is a time-billable one and that each simulated machine runs in time $\mathfrak{t}$, including the machine whose move interrupted the iteration. If so, the overall time consumption is  $O\bigl((\mathfrak{t}(\mathfrak{R}))^{\mathfrak{b}+3}\bigr)$. Taking $\mathfrak{d}=\mathfrak{b}+3$ completes our proof of the claim of the theorem regarding the time complexity of ${\mathcal M}  $.

\section{Final remarks}

The following is an immediate corollary of Theorem  \ref{feb9ds}:

\begin{cor}\label{feb9e} 
%\marginpar{feb9e}
Whenever a formula $F$ is a logical consequence of formulas $E_1,\ldots,E_n$ and the latter have polynomial amplitude, logarithmic space and polynomial time solutions under a given interpretation $^*$,  so does the former. Such a solution can be  effectively constructed from a $\cltw$-proof of {\em $E_1,\ldots,E_n\intimpl F$} and the solutions of $E_{1}^{*},\ldots,E_{n}^{*}$.  
\end{cor}

To see the import of the above corollary, imagine we want to construct a clarithmetical theory $S$ in the style of  systems {\bf CLA4}-{\bf CLA7}  discussed in Section 1, such  that the number-theoretic problem expressed by any formula $F$ provable in $S$ has a polynomial amplitude, logarithmic space and polynomial time solution. Call this property of formulas the {\em Polynomial-Logarithmic-Polynomial} property.  To achieve this, it would be sufficient to just make sure that every nonelementary (containing some choice operators) axiom  of $S$ enjoys the Polynomial-Logarithmic-Polynomial property, and that every nonlogical inference rule of $S$ --- if such are present --- preserves the Polynomial-Logarithmic-Polynomial property of formulas. As for the elementary (choice-operator-free) axioms of $S$,  they just need to be true, in which case they also automatically enjoy the Polynomial-Logarithmic-Polynomial property because the problems they express are ``solved'' by an HPM that does nothing and hence ``runs'' in $0$ amplitude, $0$ space and $0$ time.  Then we have a guarantee that 
every theorem of $S$ also enjoys the Polynomial-Logarithmic-Polynomial property as desired. This is so because the only logical inference rule of $S$ (just as of any other $\cltw$-based applied theory) is Logical Consequence, which, according to Corollary \ref{feb9e}, preserves the Polynomial-Logarithmic-Polynomial property. As an aside, 
note that we would have full flexibility in selecting or varying the set of elementary axioms of $S$ as long as those axioms are true, with such variations having no effect on the soundness of $S$ with respect to the Polynomial-Logarithmic-Polynomial property. Typically we would want the elementary axioms of $S$ to be Peano axioms, but nothing holds us from choosing any stronger or weaker collection of elementary axioms if and when a need arises. 
 
It is easily understood that Corollary \ref{feb9e} is but one of a series of nice corollaries of Theorems \ref{feb9dt} and \ref{feb9ds}. Various corollaries in the same style will be dealing with other natural and important complexity triples, such as Linear-Logarithmic-Polynomial,  \ \ Polynomial-Polylogarithmic-Polynomial,  \ \ Linear-Polynomial-Exponential, \ \ Polynomial-Polynomial- Exponential, etc. --- you name it!
This means that $\cltw$ is an adequate and very scalable common logical basis for a wide class of complexity-oriented or complexity-sensitive applied systems. $\cltw$ --- more precisely, the associated rule of Logical Consequence --- is adequate because, on one hand, by Theorems \ref{feb9dt}, \ref{feb9ds} and their corollaries in the style of Corollary \ref{feb9e},   it is sound for such systems, and, on the other hand, by Theorem \ref{feb9b}  and/or Thesis \ref{thesis} (feel free to also throw the discussion of Subsection \ref{comment} into the mix), it is as strong as a logical rule of inference could possibly be. 

We want to close this paper with a couple of terminological conventions for future use.  
By the {\bf unarification}\label{xunf} of an $n$-ary ($n\geq 0$) arithmetical function $h$ we shall mean the unary arithmetical function $h'$ such that, for any number $\ell$, \ $h'(\ell)=h(\ell,\ldots,\ell)$, with $n$ occurrences  $\ell$ in ``$(\ell,\ldots,\ell)$''. 
%Note that, if here $h$ is unary, then $h=h'$.

The following convention conservatively generalizes Definition \ref{deftcs} from unary functions to functions of any arities. Such a generalization is of no conceptual interest, but it may offer some technical convenience  in many treatments.

\begin{conv}\label{ioo} Assume $h$ is a (not-necessarily-unary) arithmetical function, $\mathcal M$ is an HPM, and $A$ is a constant game. By saying that $\mathcal M$ plays $A$ in time $h$ we shall mean that $\mathcal M$ plays $A$ (in the sense of Definition \ref{deftcs}) in time  $h'$, where $h'$ is the unarification of $h$. Similarly for space and amplitude. 
\end{conv}

\begin{conv}\label{iii} Assume $H$ is a set of arithmetical functions, $\mathcal M$ is an HPM, and $A$ is a constant game. By saying that $\mathcal M$ plays $A$ in time $H$ we shall mean that $\mathcal M$ plays $A$ (in the sense of convention \ref{ioo}) in time  $h$ for some $h\in H$.  Similarly for ``$\mathcal M$ is an $H$ time solution of $A$'', ``$\mathcal M$ runs in time $H$'', ``$\mathcal M$ is an $H$ time machine'', etc. Similarly for space and amplitude.
\end{conv}

Finally, for brevity's sake, we also agree on the following: 

\begin{conv}\label{iop} Assume $h_1,h_2,h_3$ are arithmetical functions, $\mathcal M$ is an HPM, and $A$ is a constant game. By saying that $\mathcal M$ plays $A$ in {\bf tricomplexity} $(h_1,h_2,h_3)$ we shall mean that $\mathcal M$ plays $A$ (in the sense of Convention \ref{ioo}) in amplitude $h_1$, space $h_2$ and time $h_3$. Similarly for ``$\mathcal M$ is an $(h_1,h_2,h_3)$ tricomplexity solution of $A$'', ``$\mathcal M$ runs in $(h_1,h_2,h_3)$ tricomplexity'', ``$\mathcal M$ is an $(h_1,h_2,h_3)$ tricomplexity machine'', etc. Similarly for $(H_1,H_2,H_3)$ instead of $(h_1, h_2,h_3)$, where $H_1,H_2,H_3$ are sets of arithmetical functions. 
%Next, assume $H_1,H_2,H_3$ are sets of arithmetical functions. By saying that $\mathcal M$ plays $A$ in tricomplexity $(H_1,H_2,H_3)$ we shall mean that $\mathcal M$ plays $A$ in tricomplexity $(h_1,h_2,h_3)$ for some $(h_1,h_2,h_3)\in H_1\times H_2\times H_3$. Similarly for ``$\mathcal M$ is a $(H_1,H_2,H_3)$ tricomplexity solution of $A$'', ``$\mathcal M$ runs in $(H_1,H_2,H_3)$ tricomplexity'', ``$\mathcal M$ is a $(H_1,H_2,H_3)$ tricomplexity machine'', etc. 
\end{conv}

\appendix
\section{Proof of Lemma \ref{sep1}}\label{sapa}
%\marginpar{sapa}

Pick an 
 arbitrary HPM $\mathcal N$  together with an arbitrary formula $E$. We may assume that $E$ is closed, or else consider $\ada E$ instead. Let us fix  $\mathfrak{d}$ as the greatest possible number of labmoves in any  legal run of $E$.\footnote{Such a number $\mathfrak{d}$ is said to be the {\em depth} of $E$.}  
 We want to (see how to) construct an HPM $\mathcal Q$ which is as promised in the lemma.   As usual, in both  our construction of such a machine $\mathcal Q$ and our further analysis of its work, we --- mostly implicitly --- rely on the Clean Environment Assumption.

We let the machine $\mathcal Q$  have the same number of work tapes as $\mathcal N$ has. As  mentioned in the proof idea given on page \pageref{pid}, $\mathcal Q$ works through  simulating 
$\mathcal N$.   During this process, $\mathcal Q$ maintains a  list $L=\seq{n_1,\ldots,n_k,m}$ of integers, where $0\leq n_1<\ldots < n_k\leq m\leq \mathfrak{d}$.  Intuitively,  $k$ is  the number of moves made by  $\mathcal Q$ ``so far'';  each $n_i\in\{n_1,\ldots,n_{k}\}$  is the  number of labmoves  residing on $\mathcal N$'s run tape throughout the time period during which $\mathcal N$ was ``thinking on''  (constructing) its $i$th move; and $m$ is the  number of labmoves  residing on $\mathcal N$'s run tape at the beginning of the ``final'' episode of simulation --- the episode following $\mathcal N$'s   $k$th move.  
It is important to point out that $\mathcal Q$ remembers the list $L$ through its {\em state memory} rather than work-tape memory. Namely, 
since there is only a bounded number $b$ of all possible values $L_1,\ldots,L_b$ of $L$, we can think of each state of $\mathcal Q$ as being indexed by one of such values. Then, $\mathcal Q$'s being in an $L_j$-indexed state can be understood as that $L_j$ is the ``current'' value of $L$ in $\mathcal Q$'s state memory.     

In a similar fashion, we let $\mathcal Q$, through its state memory, be always conscious of the ordinal number of the labmove that it is currently scanning on the run tape. We further let $\mathcal Q$   keep track of the greatest value that such an ordinal number has ever achieved --- that is, keep track of the overall number of labmoves it has ever seen ``so far'' on its run tape.  This quantity will be referred to as the {\bf historical move count}.\label{0hmc}

After initializing the  values of both $k$ and $m$ to $0$, the 
simulation routine of $\mathcal Q$ consists of the two subroutines   described below, starting from Subroutine 1. In our description of those subroutines, whenever we talk about simulating $\mathcal N$, it is to be understood in the sense of an ``almost perfect simulation'' --- namely, a simulation 
during which the  work-tape contents of $\mathcal Q$ fully coincide with the (corresponding) work-tape contents of the simulated configuration of $\mathcal N$, and so do the locations of the work-tape and run-tape scanning heads. Furthermore,  the position spelled on the imaginary run tape  of $\mathcal N$ at any time is also either the same as or, at least, an initial segment of, the position spelled on $\mathcal Q$'s run tape. This synchronization, however, does not extend to the buffer contents of the two machines: unless (as in stage (b) of Subroutine 2) explicitly specified otherwise, $\mathcal Q$ puts no nonempty strings into its buffer during simulation, regardless of what $\mathcal N$ does in this respect. In fact, for the exception of stage (b) of Subroutine 2, $\mathcal Q$'s buffer remains empty throughout the entire process.\vspace{7pt} 

{\em Subroutine 1}: With the current value of the list $L$ being $\seq{n_1,\ldots,n_k,m}$, assume\footnote{This assumption/condition will always be automatically satisfied on each actual iteration of Subroutine 1.} there are at least $m$ labmoves on $\mathcal Q$'s run tape,   exactly $k$ of those labmoves are $\pp$-labeled, and $\mathcal Q$'s buffer is empty. If $m=\mathfrak{d}$, $\mathcal Q$ retires.  
%``does nothing'' --- just loops forever in a non-move state without consuming space and without putting any nonempty strings into the buffer. 
Otherwise:

(a) $\mathcal Q$ simulates $\mathcal N$ in the scenario where, from the  very beginning of the play and up to 
(not including) the step $t_1$ at which $\mathcal N$ makes its $1$st move $\alpha_1$,\footnote{Here and later: if and when this event actually occurs, of course; however, as it happens, the expected event will indeed always occur.}   
the position spelled  on $\mathcal N$'s run tape consists of exactly the first $n_1$ labmoves (order respecting) found on $\mathcal Q$'s run tape; then, starting from $t_1$ and up to   the step $t_2$ at which $\mathcal N$ makes its $2$nd move $\alpha_2$, the position spelled  on $\mathcal N$'s run tape consists of exactly the first $n_2$ labmoves  found on $\mathcal Q$'s run tape;
and so on; finally,    starting from $t_{k-1}$ and up to   the step $t_k$ at which $\mathcal N$ makes its $k$th move $\alpha_k$, the position spelled  on $\mathcal N$'s run tape consists of exactly the first $n_k$ labmoves  found on $\mathcal Q$'s run tape. Here, if $k=0$, there is no work to be done during the present stage (a) and, for the purposes of the following stage (b), we let $t_k$ be $0$.

(b) $\mathcal Q$ continues the above simulation\footnote{Or, rather, {\em starts} simulation if $k=t_k=0$. The same comment applies to stage (b) of Subroutine 2.} in the scenario where, starting from  step $t_k$, at any (``current'') time, the content of $\mathcal N$'s run tape  is fully identical to the (``current'') content of $\mathcal Q$'s own run tape. If and when, at some time $t_{k+1}$ during this process, $\mathcal Q$ is detected to make its $(k+1)$th move $\alpha_{k+1}$, $\mathcal Q$ acts as follows, with $h$ below standing for the  historical move count at time $t_{k+1}$: 

(b.1) If $m<h$, $\mathcal Q$ sets the value of $m$ to be equal to $h$,  and repeats Subroutine 1. 

(b.2) If $m=h$, $\mathcal Q$ goes to Subroutine 2.
 
({\em Technical note:} It is understood that, in either one of the above cases (b.1) or (b.2), as well as in stage (b) of Subroutine 2 below, before exiting a given iteration of the subroutine, $\mathcal Q$ makes sure to delete everything on  its work tapes, and move all scanning heads to the leftmost cells of the corresponding tapes.)\vspace{4pt} 

{\em Subroutine 2} (with the list $L=\seq{n_1,\ldots,n_k,m}$ inherited from the preceding iteration of Subroutine 1): 

(a) $\mathcal Q$ resimulates $\mathcal N$ exactly as in stage (a) of Subroutine 1, from the very beginning of the play up to the step $t_k$ at which   $\mathcal N$ is found to make its $k$th move $\alpha_k$. As before, if $k=0$,  there is no work to be done during the present stage, and we simply declare the value of $t_k$ to be $0$.  
 
(b) $\mathcal Q$ continues simulation from   the above step $t_k$ up to the step $t_{k+1}$ at which $\mathcal Q$ makes its (expected) $(k+1)$th move $\alpha_{k+1}$, in the scenario where, throughout the entire episode,    $\mathcal N$'s run tape contains precisely the first $m$ labmoves of $\mathcal Q$'s run tape. Also,  unlike the   earlier simulation modes, now  $\mathcal Q$ puts into its buffer   every string that $\mathcal N$ is putting into its imaginary buffer. When, at time $t_{k+1}$,  $\mathcal N$ eventually makes its $(k+1)$th move $\alpha_{k+1}$, $\mathcal Q$ enters a move state,  making the same move $\alpha_{k+1}$ in the real play. After that,  $\mathcal Q$ updates the  list $L$ from $\seq{n_1,\ldots,n_k,m}$ to $\seq{n_1,\ldots,n_k,m,m}$ (meaning that the old value of $k$ is now incremented by $1$, with the value of the ``new $n_k$'' set to $m$ while the values of all other --- old --- elements of the list left unchanged), and goes to Subroutine 1.\vspace{7pt}

Consider an arbitrary interpretation $^*$ and unary arithmetical functions $\mathfrak{a},\mathfrak{s},\mathfrak{t}$.  
Assume that   $\mathcal N$ is an $\mathfrak{a}$ amplitude, $\mathfrak{s}$ space and $\mathfrak{t}$ time solution of $E^*$.    
We want to show that, as promised in the lemma, 
$\mathcal Q$  is a provident, $\mathfrak{a}$ amplitude, $O(\mathfrak{s}) $ space and $O(\mathfrak{t})$ time solution of $E^*$. 

Looking back at our description of the work of $\mathcal Q$, we see that $\mathcal Q$ constructs a nonempty move in its buffer only during   Subroutine 2. Next, $\mathcal Q$'s work within Subroutine 2 always terminates ($\mathcal Q$ returns to Subroutine 1). This is so because, during the preceding iteration of Subroutine 1, $\mathcal N$ was detected to make a new, $(k+1)$th move $\alpha_{k+1}$. But Subroutine 2 resimulates  $\mathcal N$'s work in exactly the same scenario in which it was simulated during the last iteration of Subroutine 1. So, while (re)simulating $\mathcal N$ within Subroutine 2, $\mathcal Q$ has a guarantee that, sooner or later, $\mathcal N$ will be (re)detected to make its $(k+1)$th move $\alpha_{k+1}$. But, in this case, as prescribed in the  (b) part of Subroutine 2, $\mathcal Q$ also enters a move state, thus  making a move and  emptying its buffer. To summarize, every time some nonempty string appears (during Subroutine 2) in $\mathcal Q$'s buffer, sooner or later the buffer will be emptied. This means nothing but that, as desired, $\mathcal Q$ plays $E^*$  providently. 

With some further analysis, which is left to the reader, one can see that every run generated by $\mathcal Q$ is also a run generated by $\mathcal N$. This means that $\mathcal Q$ inherits, from $\mathcal N$, the property of being an  $\mathfrak{a}$ amplitude solution of $E^*$.  

 It remains to analyze the running space and time of $\mathcal Q$. Before that, it is worth noting that, 
in view of the Clean Environment Assumption, all runs generated by $\mathcal Q$ are (not only $\oo$-legal but also simply) legal. This is so because, as pointed out in the preceding paragraph, all  runs generated by $\mathcal Q$ are also runs generated by $\mathcal N$, and the latter, as a winning strategy for $E^*$, never makes illegal moves. 
%If so, it is obvious that there is a constant bound on the maximum values of the $k$ and $n_i$ ($0\leq i\leq k$) parameters of $L$.  [!!!]

In view of simulating $\mathcal N$ in the earlier noted ``almost perfect fashion'', on each work tape $\mathcal Q$ obviously consumes exactly the same amount of space as $\mathcal N$ does on the corresponding work tape. 
For the time complexity of $\mathcal Q$, let us first try to understand how many steps a single iteration of Subroutine 2 takes. With the value of $L$ being $\seq{n_1,\ldots,n_k,m}$, an iteration of Subroutine 2 simulates $\mathcal N$ from the very beginning of the play up   to the point when it makes its $(k+1)$th move in the scenario where $\mathcal N$'s imaginary adversary always acted instantaneously, that is, never took any time to ``think'' before making its moves. Under these circumstances the simulated $\mathcal N$, which runs in time $\mathfrak{t}$, spends at most $k\mathfrak{t}(\ell)$\footnote{Where, according to our already established notational practice, $\ell$ is the background  of the then-current computation step of $\mathcal Q$.} time altogether. Then so does 
$\mathcal Q$ while ``almost perfectly'' simulating $\mathcal N$ within Subroutine 2 because,   as we already know, its tape contents and scanning head locations are synchronized with those of $\mathcal N$, meaning (or, at least, making it possible) that the simulation proceeds at exactly the same pace as $\mathcal N$ works. Of course, at the end of Subroutine 2, $\mathcal Q$ additionally needs to clean up its work tapes and move all of its scanning heads to the leftmost cells of the corresponding tapes, but asymptotically this does not add anything to its time consumption. Thus, the time cost of each iteration of Subroutine 2 is $O\bigl(\mathfrak{t}(\ell)\bigr)$.   

A very similar analysis reveals that every terminating iteration of Subroutine 1 also costs $\mathcal Q$ at most $O\bigl(\mathfrak{t}( \ell) \bigr)$ steps 
 as long as Environment makes no moves while $\mathcal Q$ is performing that iteration. If Environment moves while $\mathcal Q$ is performing a terminating iteration of Subroutine 2,  the duration of the iteration could be greater than $O\bigl(\mathfrak{t}( \ell)\bigr)$;\footnote{Namely, this would happen due to Environment   taking ``too long'' before moving.}  however, the portion of that duration following the time $t$ of Environment's (last) move   remains $O\bigl(\mathfrak{t}( \ell) \bigr)$ for the same reasons as in the above, ``normal'' cases; as for the   (``excessively long'') portion of the iteration time preceding step $t$, its duration is irrelevant  because,  at step $t$,   $\mathcal Q$'s time counter is reset to $0$.

Unlike Subroutine 2, one of the iterations of Subroutine 1 ---  the last iteration ---   never terminates. However, $\mathcal Q$ will not be billed for any time spent within that iteration because, after it reaches such an iteration, it no longer makes any moves. 

Putting it all together, the time bill of $\mathcal Q$ for each move it makes consists of (at most) the time spent within the iteration of Subroutine 2 in which the move was made ($\mathcal Q$ only moves within Subroutine 2), plus the time spent within the preceding --- and hence terminating ---  iterations of Subroutines 1 and 2 starting from the time of Environment's last move (or time $0$ if there is no such move).  We have observed above that this quantity  does not exceed $O\bigl(\mathfrak{t}( \ell) \bigr)$ steps per iteration. So, 
how many (terminating) iterations of Subroutines 1 and 2 may occur before any given iteration of Subroutine 2? Note that  every 
such iteration --- or, at least, every pair of consecutive iterations --- increases the value of either the $k$ or the $m$ parameter of the list $L=\seq{n_1,\ldots,n_k,m}$. And neither value can ever exceed $\mathfrak{d}$. Thus, there is a constant bound on the number of iterations,  implying that the running time of $\mathcal Q$ is  $O\bigl(\mathfrak{t}(\ell)\bigr)$  as desired.

\begin{rem}\label{rem7}
%\marginpar{rem7}
Looking back at our proof of Lemma \ref{sep1}, we see that it goes through   not only for games of the form $\ada E^*$ where $E$ is a formula, but, in fact, for all  constant games $G$ satisfying the following condition for some integer $\mathfrak{d}$:
%\marginpar{33}
\begin{equation}\label{33}
\mbox{\em No legal run of $G$ contains more than $\mathfrak{d}$ labmoves.}
\end{equation}
Furthermore, the machine $\mathcal Q$ constructed from $\mathcal N$ not only wins the game(s) $G$ won by $\mathcal N$, but in fact generates the same runs (of relevance) as  $\mathcal N$ does.  Specifically,  Lemma \ref{sep1} holds in the following, strong,  form:  

\begin{quote}{\em 
There is an effective procedure that takes an 
 arbitrary HPM $\mathcal N$, together with an arbitrary integer $\mathfrak{d}$,  and constructs an HPM ${\mathcal Q}$  such that, for any constant game $G$ satisfying condition (\ref{33})    and any unary arithmetical functions 
$\mathfrak{a},\mathfrak{s},\mathfrak{t}$,  if $\mathcal N$ is an   
$\mathfrak{a}$ amplitude, $\mathfrak{s}$ space and $\mathfrak{t}$ time solution of $G$, then  
${\mathcal Q}$  is a provident,  $\mathfrak{a}$ amplitude, $O(\mathfrak{s})$ space and $O(\mathfrak{t})$ time solution of $G$. Such a ${\mathcal Q}$ has the same number of work tapes as $\mathcal N$ does.  Furthermore, for any 
run $\Gamma$ containing at most $\mathfrak{d}$ labmoves, $\Gamma$ is a run generated by $\mathcal Q$ if and only if $\Gamma$ is a run generated by $\mathcal N$.}
\end{quote}
\end{rem}

\section{Proof of Lemma \ref{sep2}}\label{sapb}
%\marginpar\label{sapb}
For clause 1, pick  any    unary arithmetical function $\mathfrak{a}$,  formula $E$, interpretation $^*$ and HPM $\mathcal Q$ such that $\mathcal Q$ is a provident,  $\mathfrak{a}$ amplitude solution  of $E^*$.  
  Consider any $\oo$-legal play of $E$ by  $\mathcal Q$ and any clock cycle $t$. 

Let $\ell$ be the background of $t$, $\Phi$   the   position   spelled on $\mathcal Q$'s run tape at time $t$, and  $\alpha$   the string spelled in $\mathcal Q$'s buffer at time $t$. Assume $\alpha$ is nonempty (otherwise its size does not exceed $0$ and the case is trivial). We may further assume that, in the play that we are considering,  $\mathcal Q$'s adversary makes no moves at any time $\geq t$. Since  $\mathcal Q$ is provident, there is a (smallest) clock cycle $t'>t$ such that $\mathcal Q$ makes a move $\alpha'$ at time $t'$, where $\alpha$ is a prefix of $\alpha'$. In view of our assumption that $\mathcal Q$ wins $E^*$, such an $\alpha'$ should be a legal move of $E$ by $\pp$ in position $\Phi$. Let $m$ be the magnitude of $\alpha'$. 

Obviously there are two (smallest) finite sets $A$ and $B$ of strings such that  any   move of any legal run of  $E$, including $\alpha'$,   satisfies one of the following two conditions:
\begin{description}
  \item[(i)]  $\alpha'$ does not contain the symbol ``$\#$'', and $\alpha'$ is an element of $A$;
  \item[(ii)] $\alpha'$ is $\beta\# d$ for some element $\beta\#$ of $B$ and some constant $d$.
\end{description}
For instance, if   $E$ is   $(G_0\add G_1)\mlc \ade x H(x)$, then $A=\{0.0,0.1\}$ and $B=\{1.\#\}$. 
Let $c$ be the length of  a longest string of $A\cup B$.\footnote{In the pathological case of $A\cup B=\emptyset$, $c$ can be taken to be $0$.} Note that ($A$, $B$ and hence) $c$ only depends on $E$, not on $^*$, $\mathcal Q$ or $\mathfrak{a}$. Obviously the length of $\alpha'$ does not exceed $m+c$. 
 But, by our assumption,  $\mathcal Q$ runs in amplitude $\mathfrak{a}$, meaning that $m\leq \mathfrak{a}(\ell)$. Thus,  the length of $\alpha'$ does not exceed $\mathfrak{a}(\ell)+c$. Then the same holds  for $\alpha$, because $\alpha$ is a prefix of $\alpha'$.  

Clause 2, with a sequent $X$ instead of a formula $E$, is handled in a rather similar way. Indeed, by assumption, $\mathcal Q$ is a well behaved solution of $X$. This, by condition 4 of Definition \ref{wb}, implies that $\mathcal Q$ is also a provident solution of $X$. So, here we can rely on the   providence of $\mathcal Q$    in the same way as we did when dealing with clause 1. Also, by condition 1 of Definition \ref{wb}, $\mathcal Q$  makes only a bounded number of replications in $X$'s antecedent, which, for our purposes, makes $X$ essentially behave as if it was a formula rather than a sequent. Namely, as before, we have two finite sets $A$ and $B$ such that, in the context of any  legal play of $X$ by $Q$, any       move on $\mathcal Q$'s run tape 
 satisfies the above conditions (i) and (ii). This said, the only minor difference between the present case and the previous one is that $c$ now should be the length of the longest string of  $A\cup B$ {\em plus} the native magnitude of $X$, rather than just the length of the longest string of $A\cup B$. On the assumption that $\mathcal Q$'s runs in minimal amplitude, we  then have a guarantee that the length of ($\alpha'$ and hence of) $\alpha$ does not exceed $\ell+c$, as desired.

\section{Proof of the completeness of \texorpdfstring{$\cltw$}{CL12}}\label{scompleteness}
%\marginpar{scompleteness}
 
In this appendix, as promised, we reproduce the proof of the completeness theorem for $\cltw$ from \cite{lbcs}. 

\begin{thm}\label{feb9b}
%\marginpar{feb9b}
Every sequent with a logical solution  is provable in $\cltw$.
\end{thm}

\begin{proof} Assume $X$ is a sequent not provable in $\cltw$. Our goal is to show that $X$ has no logical solution (let alone an efficient  and/or well behaved logical solution).

Here we describe a {\em counterstrategy},\label{0counterstrategy} i.e., Environment's strategy, against which any particular HPM (in the usual role of $\pp$) loses $X^*$ for an appropriately selected interpretation $^*$. In precise terms, as a mathematical object, our counterstrategy --- let us call it $\mathcal C$ --- is a (not necessarily effective) function that prescribes, for each possible content of the run tape that may arise during the process of playing the game, a (possibly empty) sequence of moves that Environment should make on the corresponding clock cycle. In what follows, whenever we say that $\mathcal C$ wins or loses, we mean that so does $\bot$ when it acts according to such prescriptions. $\mathcal C$ and $\bot$ will be used interchangeably, that is.

By a {\em variables-to-constants mapping} --- or {\bf vc-mapping}\label{0vc-mapping} for short --- for a sequent $Y$ we shall mean a function whose domain is some finite set of variables that contains all (but not necessarily only) the free variables of $Y$ and whose range is some set of constants not occurring in $Y$,  such that to any two (graphically) different variables are assigned (graphically) different constants. When $e$ is a vc-mapping for $Y$, by $e[Y]$ we shall mean the result of replacing in $Y$ each free occurrence of every variable with the constant assigned to that variable by $e$.

At the beginning of the play, $\mathcal C$ chooses different constants for (all) different free variables of $X$, also making sure that none of these constants are among the ones that occur in $X$. Let $g$ be the corresponding vc-mapping for $X$. This initial series of moves brings $X$ (under whatever interpretation) 
down to the constant game $g[X]$ (under the same interpretation).   
 
The way $\mathcal C$ works after that can be described recursively.  At any time, $\mathcal C$ deals with a pair $(Y,e)$, where  $Y$ is a $\cltw$-unprovable sequent and $e$  is a vc-mapping for 
$Y$, such that $e[Y]$ is the game to which the initial $\ada X$ has been brought down ``by now''.   The  initial value of $Y$ is $X$, and the initial value of $e$ is the above vc-mapping $g$.  How $\mathcal C$ acts on $(Y,e)$ depends on whether $Y$ is stable or not. 

{\em CASE 1}: $Y$ is stable.  Then there should be a $\cltw$-unprovable sequent $Z$ satisfying one of the following conditions, for otherwise $Y$ would be derivable 
by Wait. $\mathcal C$  selects one such $Z$ (say, lexicographically the smallest one), and acts according to the corresponding prescription as given below. 

{\em Subcase 1.1:} $Y$ has the form $\vec{E}\intimpl F[G_0\adc G_1]$, and $Z$ is $\vec{E}\intimpl F[G_i]$ ($i=0$ or $i=1$). In this case, $\mathcal C$ makes the move that brings $Y$ down to $Z$ (more precisely, $e[Y]$ down to $e[Z]$), and calls itself on $(Z,e)$. 

{\em Subcase 1.2:} $Y$ has the form $\vec{E},F[G_0\adc G_1],\vec{K}\intimpl H$, and $Z$ is $\vec{E},F[G_i],\vec{K}\intimpl H$. This subcase is similar to the previous one. 
 
{\em Subcase 1.3:} $Y$ has the form $\vec{E}\intimpl F[\ada xG(x)]$, and $Z$ is $\vec{E}\intimpl F[G(y)]$, where $y$ is a variable not occurring in $Y$. In this case, $\mathcal C$ makes a move that brings $Y$ down to $\vec{E}\intimpl F[G(c)]$ for some (say, the smallest) constant $c$ such that $c$ is different from any constant occurring in $e[Y]$.  After this move, $\mathcal C$  calls itself on $(Z,e')$, where $e'$ is the vc-mapping for $Z$ that sends $y$ to $c$ and agrees with $e$ on all other variables. 

{\em Subcase 1.4:} $Y$ has the form $\vec{E},F[\ada xG(x)],\vec{K}\intimpl H$, and $Z$ is $\vec{E},F[G(y)],\vec{K}\intimpl H$, where $y$ is a variable not occurring in $Y$. This subcase is similar to the previous one.

$\mathcal C$ repeats the above until (the continuously updated) $Y$ becomes unstable. This results is some finite series of moves made by $\mathcal C$. We assume that all these moves are made during a single clock cycle (remember that there are no restrictions in the HPM model on how many moves Environment can make during a single cycle).   

{\em CASE 2}:  $Y$ is unstable.  ${\mathcal C}$ does not make any moves, but rather waits until its adversary makes a move. 

{\em Subcase 2.1}: The adversary never makes a move. Then the run of $e[Y]$ that is generated is empty.  As $Y$ is unstable, $\elz{Y}$ and hence $\elz{e[Y]}$ is not classically valid. That is, $\elz{e[Y]}$ is false in some classical model. But classical models are nothing but our interpretations restricted to elementary formulas.   So, $\elz{e[Y]}$ is false under some interpretation $^*$. This, in view of Lemma \ref{new1}, implies that $\win{(e[Y])^*}{}\seq{}=\oo$ and hence $\mathcal C$ is the winner.

{\em Subcase 2.2}: The adversary makes a move $\alpha$. We may assume that such a move is legal, or else $\mathcal C$ immediately wins.  There are two further subcases to  consider here:

{\em Subsubcase 2.2.1}: $\alpha$ is a move in the succedent, or a  nonreplicative move in one of the components of the antecedent, of $Y$. With a little thought, it can be seen that then 
$\alpha$ brings $e[Y]$ down to $e'[Z]$, where $Z$ is a sequent from which $Y$ follows by one of the four Choose rules, and $e'$ is a certain vc-mapping for $Z$. In this case, $\mathcal C$ calls itself on $(Z,e')$.  
 
{\em Subsubcase 2.2.2}: $\alpha$ is a replicative move in one of the components of the antecedent of $Y$. Namely, assume $Y$   (after disabbreviating $\intimpl$) is 
the game (\ref{fff11a2}) of Subsection \ref{sep10bb3}, and the replicative move is made in its $\st E$ component. This brings $e[Y]$ down to $e[(\ref{fff11b})]$. The latter, however, is ``essentially the same as'' $e[Z]$, where $Z$ abbreviates the game (\ref{fff11a1}). So, $\mathcal C$ can pretend that $e[Y]$ has been brought down to $e[Z]$, and call itself on $(Z,e)$. The exact meaning of ``pretend'' here is that, after calling itself on $(Z,e)$, $\mathcal C$ modifies its behavior --- by ``reinterpreting'' moves --- in the same style as machine $\mathcal M$ modified $\mathcal N$'s behavior in Subsection \ref{sep10bb3}.

This completes our description of the work of $\mathcal C $.
 
Assume a situation corresponding to Subsubcase 2.2.2 occurs only finitely many times. Note that all other cases, except Subcase 2.1, strictly  decrease the complexity of $Y$. So, the play finally stabilizes in a situation corresponding to Subcase 2.1 and, as was seen when discussing that subcase, $\mathcal C$ wins.  

Now, assume a situation corresponding to Subsubcase 2.2.2 occurs infinitely many times, that is, $\mathcal C$'s adversary makes infinitely many replications in the antecedent. And, for a contradiction, assume that
%\marginpar{ffeb12a}
\begin{equation}\label{ffeb12a}
\mbox{\em $\mathcal C$ loses the play of $\ada X^*$ for every interpretation $^*$.}
\end{equation}
Let $F$ be the (constant/closed) game/formula to which the succedent of the original $g[X]$ is eventually brought down. Similarly, let $\mathcal A$ be the set of all (closed) formulas to which various copies of various formulas of the antecedent of $g[X]$ are eventually brought down.  With a little thought and with Lemma \ref{new1} in mind, it can be seen that (\ref{ffeb12a}) implies the following:

%\marginpar{ffeb12b}
\begin{equation}\label{ffeb12b}
\mbox{\em The set $\{\elz{E}\ |\ E\in{\mathcal A}\}\cup\{\elz{\gneg F}\}$ is unsatisfiable (in the classical sense).}
\end{equation}
By the  compactness theorem for classical logic, (\ref{ffeb12b}) implies that, for some {\em finite} subset ${\mathcal A}'$ of $\mathcal A$, we have:

%\marginpar{ffeb12c}
\begin{equation}\label{ffeb12c}
\mbox{\em The set $\{\elz{E}\ |\ E\in{\mathcal A}'\}\cup\{\elz{\gneg F}\}$ is unsatisfiable (in the classical sense).}
\end{equation}
Consider a step $t$ in the work of $\mathcal C$ such that, beginning from $t$ and at every subsequent step, the antecedent of (the then current) $e[Y]$ contains all formulas of ${\mathcal A}'$. It follows easily from (\ref{ffeb12c}) that, beginning from $t$, (the continuously updated) $Y$ remains stable. This means that $\mathcal C$ deals only with CASE 1. But, after making a certain finite number of moves as prescribed by CASE 1, $Y$ is brought down to a stable sequent that contains no surface occurrences of $\adc,\ada$ in the succedent and no surface occurrences of $\add,\ade$ in the antecedent. Every such sequent follows from the empty set of premises by Wait, which is a contradiction because, as we know, the sequent $Y$ at any step of the work of $\mathcal C$ remains $\cltw$-unprovable. \end{proof}

\makeatletter
\renewenvironment{theindex}
               {\section*{\indexname}%
                \@mkboth{\MakeUppercase\indexname}%
                        {\MakeUppercase\indexname}%
                \thispagestyle{plain}\parindent\z@
                \parskip\z@ \@plus .3\p@\relax
                \columnseprule \z@
                \columnsep 35\p@
                \let\item\@idxitem}
               {}
\makeatother
\twocolumn
\begin{theindex}
\item {address} \pageref{0address}
\item {addressing prefix} \pageref{0adpref}
\item {agent} \pageref{0agent}
\item {amplitude (complexity)} \pageref{0ampl3},\pageref{0amplitude}
\item {antecedent (of sequent)} \pageref{0antecedent}
\item {arithmetical function} \pageref{0aff}
\item {arity of function} \pageref{0argf} 
\item {arity of  game} \pageref{0arg}
\item {arity of letter} \pageref{0arp} 
\item {atomic formula} \pageref{0af}
\item {author (of move)} \pageref{0author}
\indexspace

\item {background} \pageref{0background}
\item {background parameter} \pageref{0bp}
\item {Big-O notation} \pageref{xbigo}
\item {binary numeral} \pageref{0binnum} 
\item {\blank} \pageref{0blank}
\item {blank symbol} \pageref{0blank}
\item {blind existential quantification} \pageref{0cle},\pageref{0cle2}
\item {blind universal quantification} \pageref{0cla},\pageref{0cla2}
\item {bound (for complexity)} \pageref{0bound}
\item {branching recurrence} \pageref{0st}
\item {buffer} \pageref{0buffer}
\indexspace

\item {canonical constant} \pageref{0constant}
\item {cell (of tape)} \pageref{0cell}
\item {choice conjunction} \pageref{0adc1},\pageref{0adc2}
\item {choice disjunction} \pageref{0add1},\pageref{0add2}
\item {choice existential quantification} \pageref{0ade},\pageref{0ade2}
\item {choice universal quantification} \pageref{0ada},\pageref{0ada2}
\item {Choose ($\add$-,$\adc$-,$\ade$-,$\ada$-$\sim$)} \pageref{0choose}
\item {cirquent} \pageref{0cirquent}
\item {\bf CL12} \pageref{0cl12a},\pageref{0cl12b},\pageref{cl12c}
\item {\bf CLA5} - {\bf CLA10} \pageref{0clar}
\item {clarithmetic} \pageref{0clarithmetic}
\item {Clean Environment Assumption} \pageref{0cea}
\item {clock cycle} \pageref{0cc}   
\item {closed formula} \pageref{0closedformula}
\item {closed sequent} \pageref{0closedsequent}
\item {closure: $\ada$-$\sim$} \pageref{0adaclosure}
\item {CoL} \pageref{0col}
\item {computable game} \pageref{0computable}
\item {computation branch} \pageref{0cb}
\item  {computation step} \pageref{0cc}
\item {computational problem} \pageref{0compp}
\item {computational resource} \pageref{0resource1},\pageref{0resource2}
\item {compute (game)} \pageref{0cg}
\item {conclusion (of rule)} \pageref{0ruleconclusion}
\item {configuration} \pageref{0configuration}
\item {constant} \pageref{0constant},\pageref{0constant2}
\item {constant function} \pageref{0constantfunction}
\item {constant game} \pageref{0constantgame}
\item {$\constants$} \pageref{0consntme}
\item {counterstrategy} \pageref{0counterstrategy}
\indexspace

\item {delay} \pageref{0delay}
\item {denotat (of constant)} \pageref{0denotat} 
\item $\mbox{\em Denotation}^*$ \pageref{000den}
\item {denotation (of interpretation)} \pageref{0OIKOL} 
\item {denotation  (of universe)} \pageref{0nf}
\item {depend (on  variable)} \pageref{0depend},\pageref{0dependf} 
\item $\mbox{\em Domain}^*$ \pageref{0xcs}
\item {domain (of interpretation)} \pageref{0OIKOL} 
\item {domain (of universe)} \pageref{xxdomain1}
\indexspace

\item {$\epsilon$} \pageref{0epsilon}
\item {elementarization} \pageref{0elz},\pageref{0elz2}
\item {elementary game} \pageref{0elgame3},\pageref{0elemgamea},\pageref{0elemgameb}
\item {elementary formula} \pageref{0elformula}
\item {elementary sequent} \pageref{0elsequent}
\item {empty run} \pageref{0emptyrun}
\item {Environment} \pageref{0environment3},\pageref{0environment}
\item {extensional understanding of predicates} \pageref{0euop}
\indexspace

\item {\call} \pageref{sis}
\item {finitary function} \pageref{0finitaryf}
\item {finitary game} \pageref{0finitary} 
\item {focused} \pageref{0focused}
\item {formula} \pageref{0formula}
\item {function (on universe)} \pageref{0fun}
\item {function letter} \pageref{0fl}
\indexspace

\item {game (on universe)} \pageref{0game3},\pageref{0constantgame},\pageref{0game}
\item {global history} \pageref{0gh}
\item {globally new move}  \pageref{0gnm}
\indexspace

\item {head (scanning $\sim$)} \pageref{0sch}
\item {historical move count} \pageref{0hmc}
\item {HPM} \pageref{0HPM}
\indexspace

\item {ideal universe} \pageref{0ideal}
\item {illegal move} \pageref{0illegmove}
\item {illegal run} \pageref{0illegrun}
\subitem {$\xx$-illegal} \pageref{0pillegal}
\item {index ($\mathbb{H}$-$\sim$)} \pageref{0index}
\item {individual (of universe)} \pageref{0individual}
\item {initial configuration} \pageref{0initial configuration}
\item {initial legal (lab)move} \pageref{0ilm}
\item {initial sketch} \pageref{0isk}
\item {instance (of game)} \pageref{0instance}
\item {instance (of rule)} \pageref{0ruleinstance} 
\item {interpret} \pageref{0interpret} 
\item {interpretation} \pageref{0int}
\indexspace

\item  {$\ell$} \pageref{0ell}
\item {label} \pageref{0label}
\item {labmove (labeled move)} \pageref{0labmove}
\item {leaf} \pageref{0leaf}
\item {legal move} \pageref{0legmove}
\item {legal play}: $\oo$-$\sim$ of \ldots by \ldots  \pageref{0lpob} 
\item {legal run} \pageref{0legrun} 
\subitem {$\xx$-legal} \pageref{0plegal}
\item {logical consequence (relation)} \pageref{0logcon3},\pageref{0lcr}
\item {Logical Consequence (rule)}   \pageref{0logcon3},\pageref{0lcl}
\item {logical solution} \pageref{0uniformsolution}
\item {lost run} \pageref{0lostrun}
\item {$\legal{A}{}$} \pageref{0lr}
\item {$\legal{A}{e}$} \pageref{0lre}
\indexspace

\item {Machine} \pageref{0machine3},\pageref{0machine}
\item {\main} \pageref{0mh}
\item {magnitude} \pageref{0magnitudenew},\pageref{0magnitude}
\item {minimal amplitude logical solution} \pageref{0minamp}  
\item {move} \pageref{0move}
\item {move state} \pageref{0movestate}
\indexspace

\item {name (of individual)} \pageref{0name}
\item {naming (of universe)} \pageref{0naming}
\item {native magnitude} \pageref{0nm}
\item {negation}  \pageref{0gneg1},\pageref{0gneg2}
\item {nonreplicative move} \pageref{0nonrep}
\indexspace

\item {$\xx$} \pageref{0xx}
\item {parallel conjunction} \pageref{0mlc1},\pageref{0mlc2}
\item {parallel disjunction} \pageref{0mld1},\pageref{0mld2}
\item {play} \pageref{0play}
\item {position} \pageref{0position},\pageref{0posa}
\item {preamble} \pageref{0preamble}
\item {predicate letter}:
\subitem {nonlogical} \pageref{0predicatelettern}; {logical} \pageref{0predicateletterl}
\item {prefixation} \pageref{0prefixation}
\item {premise (of rule)} \pageref{0premise}
\item {proposition} \pageref{0proposition}
\item {provident branch} \pageref{0prvb}
\item {provident solution} \pageref{0prvs}
\item {providently (play $\sim$)} \pageref{0prvp}
\indexspace

\item {reduction} \pageref{0reduction}
\item {Replicate} \pageref{0replicate}
\item {replicative move} \pageref{0rep}
\item {retire} \pageref{x16}
\item {revisit (labmove)} \pageref{0revisit}
\item {rule (of inference)} \pageref{0rule}  
\item {run} \pageref{0run}
\item {run generated by machine} \pageref{0rgb}
\item {run spelled by computation branch} \pageref{0run spelled by a computation branch}
\item {run tape} \pageref{0runtape}
\indexspace

\item {sequent} \pageref{0sequent},\pageref{0sequent2}
\item {sketch} \pageref{0sketch} 
\item {solution} \pageref{0solution3},\pageref{0sol}
\item {solve (problem, game)} \pageref{0cg}
\item {space (complexity)} \pageref{0space}
\item {spacecost} \pageref{0space}
\item {stable formula} \pageref{0stable}
\item {start state} \pageref{0startstate}
\item {state (of HPM)} \pageref{0state}
\item {static game} \pageref{0static}
\item {strict reduction} \pageref{0mli2}
\item {subextensional understanding of predicates or functions} \pageref{0iuop}
\item {substitution of variables} \pageref{0dash}
\item {succedent} \pageref{0succedent}
\item {surface occurrence} \pageref{0surface occurrence}
\indexspace

\item {tape symbol} \pageref{0tapesymbol}
\item {term} \pageref{0term}
\item {time (complexity)} \pageref{0time}
\item {time-billable} \pageref{0tmb}
\item  {timecost} \pageref{0timecost}
\item {transition function} \pageref{0tf}
\item {tree of games} \pageref{0treeofgames}
\item {tricomplexity} \pageref{iop}
\item {Turing reduction} \pageref{0intimpl}
\indexspace

\item {unarification} \pageref{xunf}
\item {unfocused} \pageref{0unfocused}
\item {uniform solution} \pageref{0uniformsolution}
\item {unistructural game} \pageref{0unistructural}
\item {$\mbox{\em Universe}^*$} \pageref{00unhg}
\item {universe (of discourse)} \pageref{0universe}
\item {universe (of interpretation)} \pageref{00unhg}
\item  {unstable formula} \pageref{0unstable}
\item {\transition} \pageref{0us}
\indexspace

\item {valuation} \pageref{0valuation}
\item {$\valuations$} \pageref{0valu}
\item {$\mbox{\em var}_i$} \pageref{0vari}
\item {variable} \pageref{0variable},\pageref{0variable2}
\item {$\variables$} \pageref{0variablee}
\item {vc-mapping} \pageref{0vc-mapping}
\indexspace

\item {Wait} \pageref{0wait}
\item {well behaved logical solution} \pageref{0wbs}
\item {win} \pageref{0win}
\item {$\win{A}{}$} \pageref{0wn}
\item {$\win{A}{e}$} \pageref{0wne}
\item  {winning strategy} \pageref{0sol}
\item {won run} \pageref{0wonrun}
\item {work tape} \pageref{0worktape}
\indexspace

\item {yield} \pageref{0yield}

\indexspace

\indexspace

\item {$\vdash$} \pageref{0stopor}\vspace{3pt}
\item {$\twg$ (as  game)} \pageref{0twg}\vspace{3pt}
\item {$\twg$ (as  player)} \pageref{0machine3},\pageref{0pp}\vspace{3pt}
\item {$\tlg$ (as  game)} \pageref{0tlg}\vspace{3pt}
\item {$\tlg$ (as player)} \pageref{0environment3},\pageref{0oo}\vspace{3pt}
\item {$\gneg$ (as  operation on games)} \pageref{0gneg1},\pageref{0gneg2}\vspace{3pt}
\item {$\gneg$ (as  operation on players)} \pageref{0pneg}\vspace{3pt}
\item {$\gneg$ (as  operation on runs)} \pageref{0rneg}\vspace{3pt}
\item {$\mlc$} \pageref{0mlc1},\pageref{0mlc2}\vspace{3pt}
\item {$\mld$} \pageref{0mld1},\pageref{0mld2}\vspace{3pt}
\item {$\mli$} \pageref{0mli1},\pageref{0mli2}\vspace{3pt}
\item {$\cla$} \pageref{0cla},\pageref{0cla2}\vspace{3pt}
\item {$\cle$} \pageref{0cle},\pageref{0cle2}\vspace{3pt}
\item {$\adc$} \pageref{0adc1},\pageref{0adc2}\vspace{3pt}
\item {$\add$} \pageref{0add1},\pageref{0add2}\vspace{3pt}
\item {$\ada$} \pageref{0ada},\pageref{0ada2},\pageref{0adaclosure}\vspace{3pt}
\item {$\ade$} \pageref{0ade},\pageref{0ade2}\vspace{3pt}
\item {$\pst$} \pageref{0pst}\vspace{3pt}
\item {$\st$} \pageref{0st}\vspace{3pt}
\item \hspace{-2pt}{$\intimpl$} \pageref{0intimpl00},\pageref{0intimpl}\vspace{3pt}
\item {$\circ$} \pageref{0circ}\vspace{3pt}
\item {$\#$} \pageref{0pound}\vspace{3pt}
\item $\equiv$ \pageref{0equiv}\vspace{3pt}
\item {$\emptyrun$} \pageref{0emptyrun}\vspace{3pt}

\item {$\preceq$} \pageref{0preceq}\vspace{3pt}
\item {$\models$} \pageref{0models}\vspace{3pt}
\item {$|\ldots |$} \pageref{0|1},\pageref{0|2}\vspace{3pt} 
\item {$\elz{\ldots}$} \pageref{0elz},\pageref{0elz2}\vspace{3pt}  
\item {$F[E]$} \pageref{0fe}\vspace{3pt}   
\item {$e[A]$} \pageref{0ea}\vspace{3pt}
\item {$\seq{\Phi}A$} \pageref{0pr}\vspace{3pt}
\item {$A(f_1/x_1,\ldots,f_n/x_n)$} \pageref{0dash}\vspace{3pt}
\item {$\Gamma^\alpha$} \pageref{0apr2}\vspace{3pt}
\item {$\Gamma^{\preceq v}$} \pageref{0susu}\vspace{3pt}
\end{theindex}
\vspace{-20 pt}
\end{document}